\documentclass[acmsmall]{acmart}

\usepackage{textcomp}
\usepackage{xcolor, framed}
\usepackage{soul}
\usepackage{url}

\usepackage{graphicx}
\usepackage{balance} 

\usepackage{hyperref}
\usepackage{color}
\usepackage{multirow}
\usepackage{subcaption}
\usepackage[export]{adjustbox}
\usepackage[normalem]{ulem}
\useunder{\uline}{\ul}{}

\usepackage{bbm}
\usepackage{makecell}
\usepackage[most]{tcolorbox}

\newcommand{\stitle}[1]{\noindent\textup{\textbf{#1}}}

\newcommand{\eat}[1]{}
\newcommand{\sprint}{\textsc{{SPRinT}}\xspace}
\newcommand{\sprintv}{\textsc{{SPRinT-V}}\xspace} 
\newcommand{\sprintc}{\textsc{{SPRinT-C}}\xspace}

\newcommand{\squishlist}{
 \begin{list}{$\bullet$}
  { \setlength{\itemsep}{0pt}
     \setlength{\parsep}{3pt}
     \setlength{\topsep}{3pt}
     \setlength{\partopsep}{0pt}
     \setlength{\leftmargin}{1.5em}
     \setlength{\labelwidth}{1em}
     \setlength{\labelsep}{0.5em} } }

\newcommand{\squishlisttwo}{
 \begin{list}{$\bullet$}
  { \setlength{\itemsep}{0pt}
    \setlength{\parsep}{0pt}
    \setlength{	opsep}{0pt}
    \setlength{\partopsep}{0pt}
    \setlength{\leftmargin}{2em}
    \setlength{\labelwidth}{1.5em}
    \setlength{\labelsep}{0.5em} } }

\newcommand{\squishend}{
  \end{list}  }

\newtcolorbox{responsebox}{
  colback=gray!5!white,    
  colframe=gray!40!white,  
  boxrule=0.4pt,           
  arc=2pt,                 
  left=4pt, right=4pt, top=4pt, bottom=4pt, 
  fontupper=\normalsize,   
  before skip=8pt, after skip=8pt
}

\usepackage{algorithmicx}
\usepackage{algorithm}
\usepackage{algpseudocode}
\usepackage{xspace}
\usepackage{makecell}
\usepackage[italicdiff]{physics}

\newtheorem{theorem}{Theorem}

\AtBeginDocument{%
  }

\setcopyright{cc}
\setcctype{by}
\acmJournal{PACMMOD}
\acmYear{2026} \acmVolume{4} \acmNumber{1 (SIGMOD)} \acmArticle{58} \acmMonth{2} \acmPrice{}\acmDOI{10.1145/3786672}

\begin{document}

\title{On Efficient Approximate Aggregate Nearest Neighbor Queries over Learned Representations} 
\author{Carrie Wang}
\affiliation{%
  \institution{The
     University of Hong Kong}
  \city{Hong Kong}
  \country{Hong Kong}
}
\email{carrie07@connect.hku.hk}

\author{Sihem Amer-Yahia}
\affiliation{%
  \institution{CNRS, Univ. Grenoble Alpes}
  \city{Saint Martin D'Hères}
  \country{France}
}
\email{sihem.amer-yahia@univ-grenoble-alpes.fr}

\author{Laks V. S. Lakshmanan}
\affiliation{%
  \institution{The
     University of British Columbia}
  \city{Vancouver}
  \country{Canada}
}
\email{laks@cs.ubc.ca}

\author{Reynold Cheng}
\affiliation{%
  \institution{The
     University of Hong Kong}
  \city{Hong Kong}
  \country{Hong Kong}
}
\email{ckcheng@cs.hku.hk}

\renewcommand{\shortauthors}{Wang et. al}

\begin{abstract}
We study Aggregation Queries over Nearest Neighbors (AQNN), which compute aggregates over the learned representations of the neighborhood of a designated query object. For example, a medical professional may be interested in {\em the average heart rate of patients whose representations are similar to that of an insomnia patient}. Answering AQNNs accurately and efficiently is challenging due to the high cost of generating high-quality representations (e.g., via a deep learning model trained on human expert annotations) and the different sensitivities of different aggregation functions to neighbor selection errors. We address these challenges by combining high-quality and low-cost representations to approximate the aggregate. We characterize value- and count-sensitive AQNNs and propose the \textit{Sampler with Precision-Recall in Target} (\sprint), a query answering framework that works in three steps: (1) sampling, (2) nearest neighbor selection, and (3) aggregation. We further establish theoretical bounds on sample sizes and aggregation errors. Extensive experiments on five datasets from three domains (medical, social media, and e-commerce) demonstrate that \sprint achieves the lowest aggregation error with minimal computation cost in most cases compared to existing solutions. \sprint's performance remains stable as dataset size grows, confirming its scalability for large-scale applications requiring both accuracy and efficiency.
\end{abstract}



\received{July 2025}
\received[revised]{October 2025}
\received[accepted]{November 2025}

\maketitle

\section{Introduction}\label{sec:intro}
Many database applications require computing efficient aggregates of the neighborhood of a designated query object, where the neighborhood is over learned representations. For example, medical professionals may explore {\em the average heart rate of patients whose representations are similar to that of an insomnia patient} \cite{DBLP:journals/isci/RodriguesGSBA21}. 
In online education, instructors may be interested in analyzing performance metrics over groups of students with similar learning trajectories, where similarity is over an embedding space to better capture complex behavioral patterns. 
Such queries are increasingly common, where the neighborhood is determined, not based on raw data, but on \textit{learned representations}, which are multi-dimensional vectors or embeddings derived from machine learning models. We call these queries AQNNs, or \textit{Aggregation Queries over Nearest Neighbors}. AQNNs extend the Fixed-Radius Near Neighbor (FRNN) queries~\cite{FRNNSurvey} by operating on {\em learned representations} and applying standard {\em aggregation functions} (e.g., \(\mathtt{AVG}\), \(\mathtt{SUM}\), percentage \(\mathtt{PCT}\), \(\mathtt{COUNT}\), and \(\mathtt{VAR}\)) over an attribute value of objects in the resulting neighborhood. The evaluation of AQNNs can be costly, as it relies on generating high-quality representations of data objects using an expensive deep learning model, a.k.a. an oracle \cite{DBLP:conf/sigmod/LaiHLZ0K21, DBLP:journals/pvldb/KangGBHZ20, DujianPQA} in order to determine the neighborhood of the input object and compute an aggregate result. 

\begin{figure}[h!]
    \centering \includegraphics[width=0.7\linewidth]{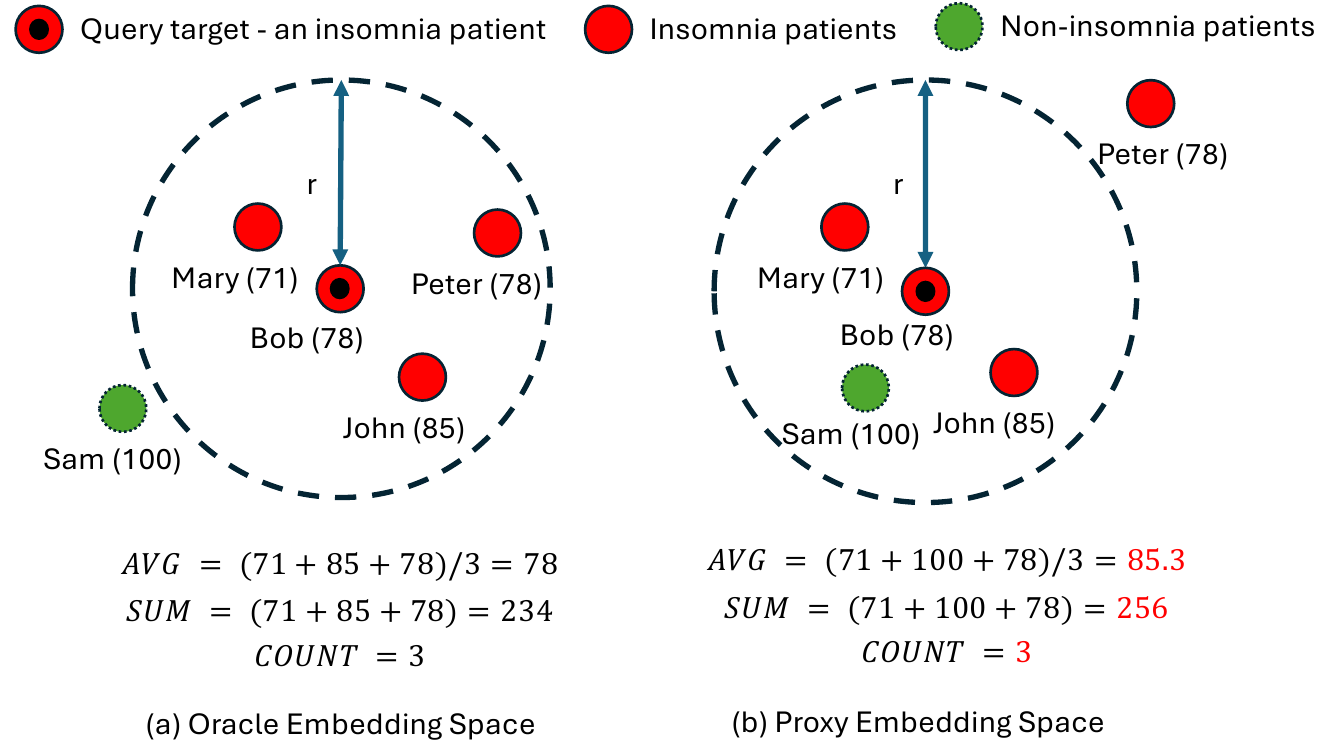}
    \caption{Nearest neighbors in oracle and proxy embedding spaces and their impact on AQNN results. Numbers in parentheses indicate patients' heart rates (bpm).}
    \label{fig:p_r_example}
\end{figure}


In this paper, we study the evaluation of AQNNs by seeking an efficient approximation of the aggregate result. Our approach leverages both an expensive oracle and a cheaper representation learning model, namely a proxy \cite{DBLP:conf/icde/AndersonCRW19, DBLP:conf/sigmod/LuCKC18, DujianPQA, DBLP:journals/pvldb/KangEABZ17}, to determine the embeddings and hence the neighborhood for aggregation.

\begin{figure*}[t!]
    \centering \includegraphics[width=\linewidth]{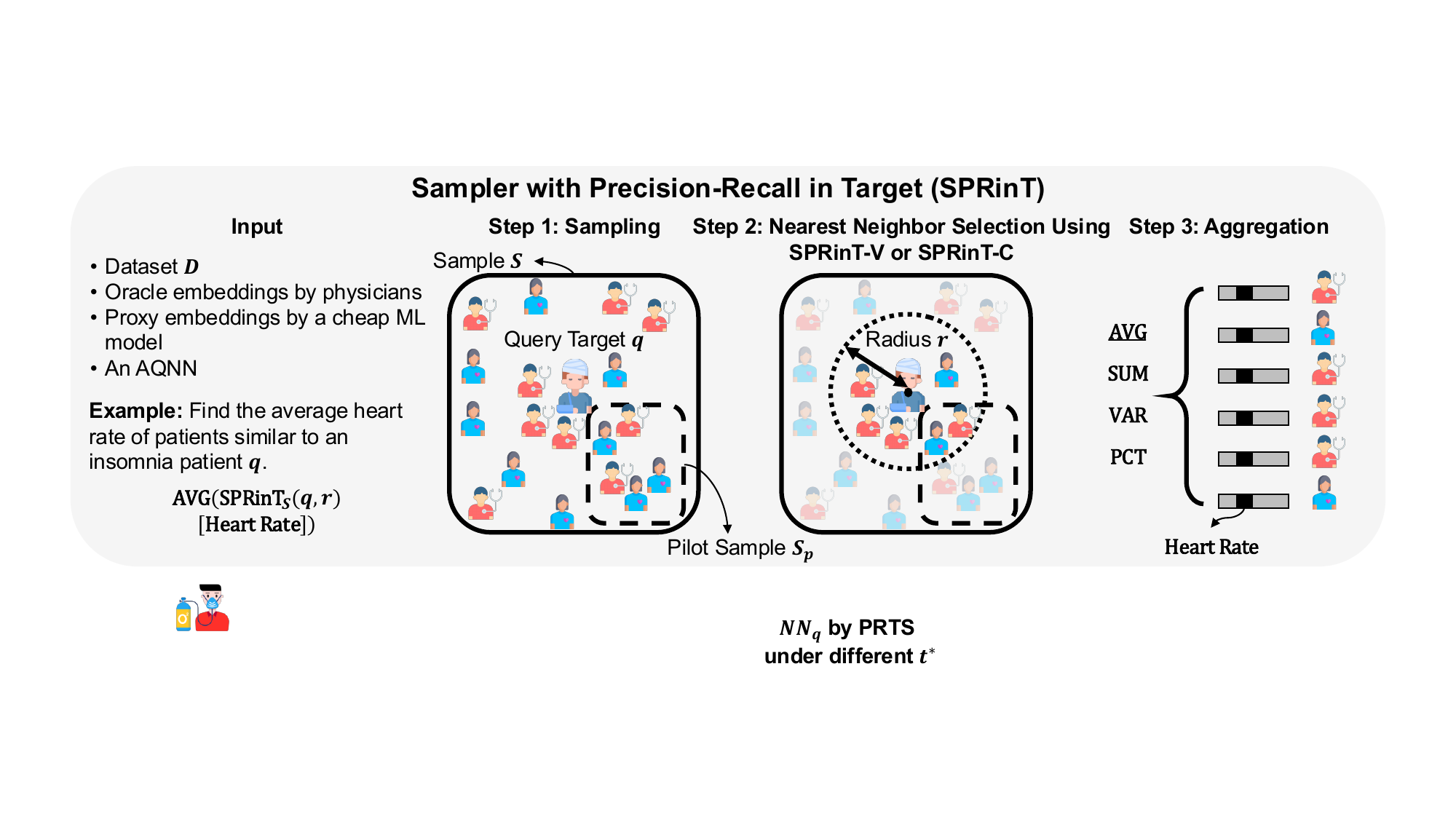}
    \caption{Overview of the \sprint framework for answering AQNNs.}
    \label{fig:framework}
\end{figure*}

\stitle{Objectives and Challenges.} Effectively answering an AQNN over learned representations requires addressing two key objectives: (\textbf{O1}) reducing computational costs, which include {\em embedding generation} and {\em query execution costs}, and (\textbf{O2}) minimizing the aggregation error between the approximate and true results. A key challenge for \textbf{O1} is that learned representations often need to be recomputed at query time, particularly in settings where data evolves rapidly. In healthcare, a patient' physiological state changes frequently, necessitating on-demand embedding updates to ensure accurate neighbor retrieval and attribute aggregation. In online education, student embeddings should be updated with learning progress and interaction history.
Using stale embeddings risks inaccurate nearest neighbor (NN) retrieval and unreliable estimation of query answers. Hence, the computational bottleneck lies in generating \emph{oracle embeddings} using high-capacity models such as deep neural networks trained on expert annotations, which can require tens to hundreds of GPU-hours according to throughput benchmarks reported in ~\cite{reddi2020mlperf}. Recent work has recognized this issue and proposed the use of \emph{proxy embeddings}, which are less accurate but significantly cheaper to generate for efficient approximate nearest neighbor (ANN) search with probabilistic guarantees~\cite{DBLP:conf/sigmod/LaiHLZ0K21, DBLP:conf/sigmod/LuCKC18, DBLP:journals/pvldb/KangGBHZ20, DBLP:journals/pvldb/KangEABZ17}. Empirically, generating high-quality oracle embeddings is \(2-10\times\) slower than generating proxy embeddings according to existing benchmarks~\cite{DBLP:conf/nips/WangW0B0020, reddi2020mlperf}. We will empirically confirm this trend in \S~\ref{exp:framework}.



However, existing probabilistic ANN algorithms that leverage both proxy and oracle embeddings are typically optimized for either \textit{precision} or \textit{recall}, without carefully considering how neighbor selection errors affect the aggregate result. As illustrated in Fig.~\ref{fig:p_r_example}, the neighborhood retrieved in the proxy space may include false positives (e.g., the non-insomnia patient Sam) or omit true positives (e.g., the insomnia patient Peter). These neighbor selection errors propagate to the aggregation step, potentially leading to large aggregation error (\textbf{O2}). Moreover, different aggregation functions respond differently to neighbor selection errors. For example, \(\mathtt{AVG}\) is affected by the attribute values of selected neighbors while \(\mathtt{COUNT}\) and \(\mathtt{PCT}\) are more sensitive to the cardinality of the selected neighborhood, regardless of the attribute values. \(\mathtt{SUM}\), instead, can be affected by both extreme attribute values and the over- or under-estimation of the neighborhood cardinality. Mitigating such errors can be achieved by jointly controlling the precision and recall of neighbor selection based on the aggregation function’s aggregation error sensitivity. Therefore, {\em our overarching goal is to strike a balance between cost and aggregation error by optimizing precision and recall in a manner that is tailored to different AQNNs.}

\stitle{Our Contributions.} In this work, we address the challenges and make the following contributions:
\squishlist 
    \item We introduce and formalize Aggregation Queries over Nearest Neighbors (AQNNs), a novel query abstraction that unifies learned representation-based NN retrieval with standard SQL aggregation functions (\S~\ref{sec:def}).
    \item We propose \textit{Sampler with Precision-Recall in Target} (\sprint), a framework for answering AQNNs, in three steps (Fig.~\ref{fig:framework}): (1) sampling, (2) NN selection, and (3) aggregation (\S~\ref{sec:sprint}).
    \item We analyze aggregation sensitivity and develop two NN selection algorithms: \sprintv, which maximizes F1-score for value-sensitive queries (e.g., \(\mathtt{AVG}, \mathtt{VAR}\)), and \sprintc, which balances precision and recall for count-sensitive ones (e.g., \(\mathtt{PCT}, \mathtt{COUNT}\)). For \(\mathtt{SUM}\), which is both value- and count-sensitive, we heuristically combine \sprintv and \sprintc (\S~\ref{sec:sprint}, \S~\ref{sec:SUM_algo}).
    \item We establish theoretical bounds on the aggregation error for most aggregation functions, deriving minimum sample and pilot sample sizes needed to achieve a target aggregation error with high probability (\S~\ref{sec:theory}).
    \item Extensive experiments on five datasets across three domains (medical, social media, and e-commerce) show that \sprint achieves significant speedup in terms of embedding generation cost and end-to-end cost, lower aggregation error, and scales efficiently compared to existing methods, corroborating our theoretical analysis. We further showcase how AQNNs enable downstream tasks, particularly one-sample hypothesis testing, which requires neighborhood-level statistics over learned representations (\S~\ref{sec:exps}).
\squishend
\S~\ref{sec:relwork} discusses related work, while \S~\ref{sec:conclusion} concludes the paper and outlines directions for future research.

\section{Problem Studied}\label{sec:def}
We first discuss Fixed-Radius Near Neighbor (FRNN) queries and then formalize Aggregation Queries over Nearest Neighbors (AQNNs) that build on them. We will also state our problem.

\subsection{Nearest Neighbor Queries}\label{subsec:FRNN}
Fixed-Radius Near Neighbor (FRNN) queries~\cite{FRNNSurvey} retrieve all NNs of a query target \(q\) within a given radius \(r\) under a distance function \(dist\). More formally:
\[
NN_D(q, r) = \{x \in D \mid dist(x, q) \leq r\},
\]
where \(x\) is any object in \(D\), and \(dist(x, q)\) denotes a standard distance metric such as Euclidean distance.
We build on \emph{generalized} FRNN queries by extending them to learned representation spaces, where distances are computed over embeddings produced by machine learning models (e.g., deep neural networks). This enables richer notions of similarity and supports flexible distance metrics, such as cosine similarity. 
We use \(|NN_D(q,r)|\) to denote the neighborhood size of \(q\) in \(D\). Fig.~\ref{fig:framework} shows an example that given the radius \(r\) and target patient \(q\), patients in the dotted circle are NNs, and the neighborhood size is 6.

Note that in settings where data evolves rapidly, it is often impractical to assume that all oracle or proxy embeddings are precomputed for all objects in \(D\); instead, they are typically generated on demand during query processing.

\subsection{Aggregation Queries over Nearest Neighbors}\label{subsec:AQNN} 
Given an FRNN query target \(q\) in dataset \(D\), a radius \(r\), and an attribute \(\texttt{attr}\), an Aggregation Query over Nearest Neighbors (AQNN) is defined as:
\[ \text{agg}(NN_D(q,r)[\texttt{attr}]) \]
where agg is an aggregation function, such as $\mathtt{AVG}$, $\mathtt{SUM}$, and $\mathtt{PCT}$, and \(NN_D(q,r)[\texttt{attr}]\) denotes the bag of values of attribute \texttt{attr} of all FRNN results of \(q\) within radius \(r\).

We classify AQNNs by their sensitivity to NN selection errors: (i) \textit{value-sensitive} queries (e.g., \(\mathtt{AVG}\), \(\mathtt{VAR}\)), which depend critically on the attribute values of neighbors; and (ii) \textit{count-sensitive} queries (e.g., \(\mathtt{PCT}\), \(\mathtt{COUNT}\)), which are strongly influenced by the cardinality of the selected neighborhood regardless of attribute values. Notably, \(\mathtt{SUM}\), which is both value- and count-sensitive, exhibits sensitivity to both, as errors in either can distort the aggregate result. 

An AQNN expresses aggregation operations to capture key insights about the neighborhood of a query target. For example, \(\mathtt{AVG}\) could reflect the average heart rate of patients in a neighborhood, providing a measure of typical health conditions. \(\mathtt{SUM}\) is useful for assessing cumulative effects, such as the total cost of treatments in the neighborhood, which informs public health policy. \(\mathtt{PCT}\) measures the proportion of patients in the neighborhood relative to the total population in the dataset.

Fig. \ref{fig:framework} illustrates an example of an AQNN: ``\textit{Find the average heart rate of patients similar to an insomnia patient \(q\)}''. The aggregation function is \(\mathtt{AVG}\) and the target attribute of interest is heart rate. Exact query evaluation requires computing embeddings for all patients in \(D\) using high-quality but expensive models (e.g., deep learning models) or expert annotation~\cite{DBLP:conf/sigmod/LuCKC18}, followed by identifying \(q\)'s neighbors within radius \(r\)~\cite{DBLP:journals/isci/RodriguesGSBA21}. We refer to such accurate but computationally expensive models as \textit{oracle models} (\(O\))~\cite{sze2017efficient, DujianPQA, DBLP:journals/pvldb/KangGBHZ20} and those at least \(2\times\) cheaper as \textit{proxy models} (\(P\))~\cite{DBLP:conf/nips/WangW0B0020}. In the example, if the oracle embedding by a physician costs one unit per patient, the proxy embedding by a lightweight machine learning model would cost at most one-half of that. However, relying solely on proxy embeddings to answer the AQNN may lead to inaccurate neighborhood identification, which in turn degrades the aggregate result. Thus, we aim to judiciously generate oracle and proxy embeddings in order to balance computational cost and aggregation error.

Once similar patients are identified, their heart rate values can be averaged and returned as the output. Note that the values of the target attribute \texttt{attr} (e.g., heart rate) are not predicted but are instead known quantities. Moreover, they are excluded from the representation learning process to avoid biasing similarity toward the very outcome we wish to analyze. This ensures that AQNN results reflect how similar objects relate w.r.t. \texttt{attr}, rather than simply aggregating objects that already share similar values.

\subsection{Problem Statement}
Given an AQNN, our goal is to return an approximate aggregate result by leveraging both oracle and proxy embeddings while reducing cost and error.

\begin{table}[t!]
\begin{small}
\centering
\caption{Table of Notations.}
\begin{tabular}{c|c}
\hline
\textbf{Notation} & \textbf{Description} \\ \hline
\( D \) & the dataset \\ 
\( O \) & oracle model \\ 
\( P \) & proxy model \\ 
\( q \) & query target \\ 
\( r \) & query radius \\ 
\( S \) & random sample of \( D \) with size \(s\) \\ 
\( ON_D \) & oracle neighbors in \( D \) \\ 
\( ON_S \) & oracle neighbors in \( S \) \\ 
\( \sprint_S \) & $\sprint$ neighbors in \(S\) \\ 
\( S_{p} \) & pilot sample in \( S \) with size \(s_p\) \\
\( t^* \) & optimal precision target \\ 
\(\text{R}_S\), \(\text{P}_S\) & recall, precision in \(S\) \\
\(\text{F1}_S\) & F1 score in \(S\) \\
\hline
\end{tabular}
\end{small}
\end{table}
\section{Our Solution}\label{sec:algo}
In this section, we first discuss our objectives and challenges (\S~\ref{sec:challenges}), then describe the Sampler with Precision-Recall in Target (\sprint) framework for answering AQNNs (\S~\ref{sec:sprint}). For value- and count-sensitive AQNNs, we introduce \sprintv (\S~\ref{subsec:sprint_v}) and \sprintc (\S~\ref{subsec:sprint_c}), which respectively maximize F1 score and equalize precision and recall with high probability. For \(\mathtt{SUM}\), we design a heuristic strategy, namely Two-Phase, that combines \sprintv and \sprintc (\S~\ref{sec:SUM_algo}). We also theoretically analyze \sprintv, \sprintc, and Two-Phase by deriving their approximation error bounds and the minimal sample and pilot sample sizes required for \(\mathtt{AVG}\), \(\mathtt{VAR}\), \(\mathtt{PCT}\), \(\mathtt{COUNT}\), and \(\mathtt{SUM}\) (\S~\ref{sec:theory} and \S~\ref{sec:SUM_algo}).

\subsection{Objectives and Challenges}
\label{sec:challenges}
As mentioned in \S~\ref{sec:intro}, answering an AQNN involves two key objectives:  (\textbf{O1}) minimize computational cost and (\textbf{O2}) ensure a low approximation error. 

Achieving \textbf{O1} is challenging because learned representations often need to be computed or refreshed at query time rather than being precomputed or cached. In many applications, data evolves rapidly. In healthcare, a patient' physiological state changes frequently, necessitating on-demand embedding updates to ensure accurate neighbor retrieval and attribute aggregation. In online education, student embeddings should be updated with learning progress and interaction history. Using stale embeddings risks inaccurate NN retrieval and unreliable estimation of query answers. Therefore, the computational bottleneck lies in generation \emph{oracle embeddings}, which can require tens to hundreds of GPU-hours according to throughput benchmarks reported in ~\cite{reddi2020mlperf}. Recent work has recognized this issue and proposed the use of \emph{proxy embeddings}, which are less accurate but significantly cheaper to generate for efficient ANN search with probabilistic guarantees~\cite{DBLP:conf/sigmod/LaiHLZ0K21, DBLP:conf/sigmod/LuCKC18, DBLP:journals/pvldb/KangGBHZ20, DBLP:journals/pvldb/KangEABZ17}. Empirically, generating high-quality oracle embeddings is \(2-10\times\) slower than generating proxy embeddings according to existing benchmarks~\cite{DBLP:conf/nips/WangW0B0020, reddi2020mlperf}. We will empirically confirm this trend in \S~\ref{exp:framework}.

However, existing probabilistic ANN algorithms that leverage both proxy and oracle embeddings are typically optimized for either \textit{precision} or \textit{recall}, without carefully considering how neighbor selection errors affect the aggregate result. As shown in Fig. ~\ref{fig:p_r_example}, the neighborhood retrieved in the proxy space may include false positives (e.g., the non-insomnia patient Sam) or omit true positives (e.g., the insomnia patient Peter). Such inaccuracies in neighbor selection directly propagate to the aggregation step, making it challenging to achieve \textbf{O2}. In Fig. ~\ref{fig:p_r_example}(a), the average heart rate (\(\mathtt{AVG}\)) of the selected neighbors in the oracle embedding space is 78, which reflects the true health condition of patients similar to the query target. In contrast, in the proxy embedding space (Fig. ~\ref{fig:p_r_example}(b)), the inclusion of a non-insomnia patient (Sam, with heart rate 100) inflates the computed average to 85.3, introducing a substantial bias. Similarly, for \(\mathtt{SUM}\), the total heart rate increases from 234 to 256 due to this false positive. Interestingly, the \(\mathtt{COUNT}\) remains unchanged at 3 in both spaces because the number of selected neighbors is the same, even though their composition differs. This illustrates that different aggregation functions respond differently to neighbor selection errors.

\textit{Value-sensitive} queries (e.g., \(\mathtt{AVG}\), \(\mathtt{VAR}\)) are prone to distortion from individual attribute values. Hence, both the correctness and completeness of neighbor selection are essential. This requirement can be formally expressed as achieving high precision and recall with high probability. Precision measures the proportion of retrieved neighbors that are correct, while recall measures the proportion of true neighbors that are retrieved, reflecting completeness. Typically, higher precision may exclude true neighbors and hence lower recall, whereas higher recall may include incorrect neighbors and hence reduce precision. Therefore, we aim to maximize both precision and recall as much as possible simultaneously. 

\textit{Count-sensitive} queries (e.g., \(\mathtt{PCT}\), \(\mathtt{COUNT}\)) are less affected by individual false positives or false negatives. Instead, the aggregation error primarily depends on the cardinality of the retrieved neighborhood. This suggests that minimizing error for count-sensitive queries hinges on balancing false positives and false negatives, i.e., equalizing precision and recall. \(\mathtt{SUM}\)-type queries, in contrast, exhibit dual sensitivity. Errors in both neighbor count and attribute values can contribute to aggregation errors.

To achieve both efficiency and accuracy in answering AQNNs, we propose \sprint, a three-step framework that combines random sampling with tailored NN selection strategies that optimize precision and recall in accordance with the query’s sensitivity to aggregation errors. This design significantly reduces the computational cost of generating proxy and oracle embeddings. We detail the framework in \S~\ref{sec:sprint}.

\subsection{The \sprint Framework}
\label{sec:sprint}
Fig. ~\ref{fig:framework} illustrates \sprint, our framework for answering AQNNs. It consists of three steps: (1) sampling, (2) NN selection, and (3) aggregation. First, given a dataset \(D\), instead of directly evaluating AQNNs over it, we draw a random sample \(S \subset D\) of size \(s\) using a random sampler. We then perform neighbor search and aggregation over \(S\). As \(s \ll |D|\) and proxy embeddings are computed only for objects in \(S\), this step significantly reduces proxy computation and alleviates the cost burden of \textbf{O1}.

In the second step, we design NN selection strategies tailored to the aggregation error sensitivity of different AQNNs, addressing \textbf{O2}. For value-sensitive queries, we jointly optimize precision and recall by maximizing the F1 score, which harmonizes the two metrics. For count-sensitive queries, we equalize precision and recall to offset over- and under-counting errors. For queries sensitive to both, such as \(\mathtt{SUM}\), we combine these strategies heuristically. To enable adaptive NN selection from \(S\), we draw a small pilot sample \(S_p \subset S\) of size \(s_p\), on which oracle embeddings are computed. This pilot sample provides lightweight oracle supervision that serves a dual role. First, it enables us to compute precision and recall in \(S_p\), where all objects have both proxy and oracle embeddings. By Hoeffding’s inequality~\cite{hoeffding1994probability, serfling1974probability}, the precision and recall in \(S_p\) provide reliable estimates of these metrics in \(S\). Second, the true neighbors in \(S_p\) can be used to refine NN selection in \(S\) according to each AQNNs' error sensitivity. For example, it can further improve the F1 score for value-sensitive queries or balance precision and recall for count-sensitive queries. Overall, the pilot sample incurs minimal oracle embedding generation cost (\textbf{O2}) since \(s_p < s \ll |D|\), while also playing an important role in controlling the quality of NN selection (\textbf{O1}). 

In the third step, we apply the desired aggregation function to the target attribute of the selected neighbors in \(S\), and return the aggregate result as an approximation to the AQNN. 


\subsubsection{\sprintv} \label{subsec:sprint_v}
For value-sensitive AQNNs, we propose \sprintv to find neighbors in \(S\). \sprintv maximizes the F1 score in \(S\) with high probability. However, directly optimizing \(\text{F1}_S\) is challenging, as it depends non-linearly on precision and recall, which are often at odds.

\begin{figure}[t]
    \centering
    \begin{subfigure}{0.48\linewidth}
        \centering
        \includegraphics[width=0.8\linewidth]{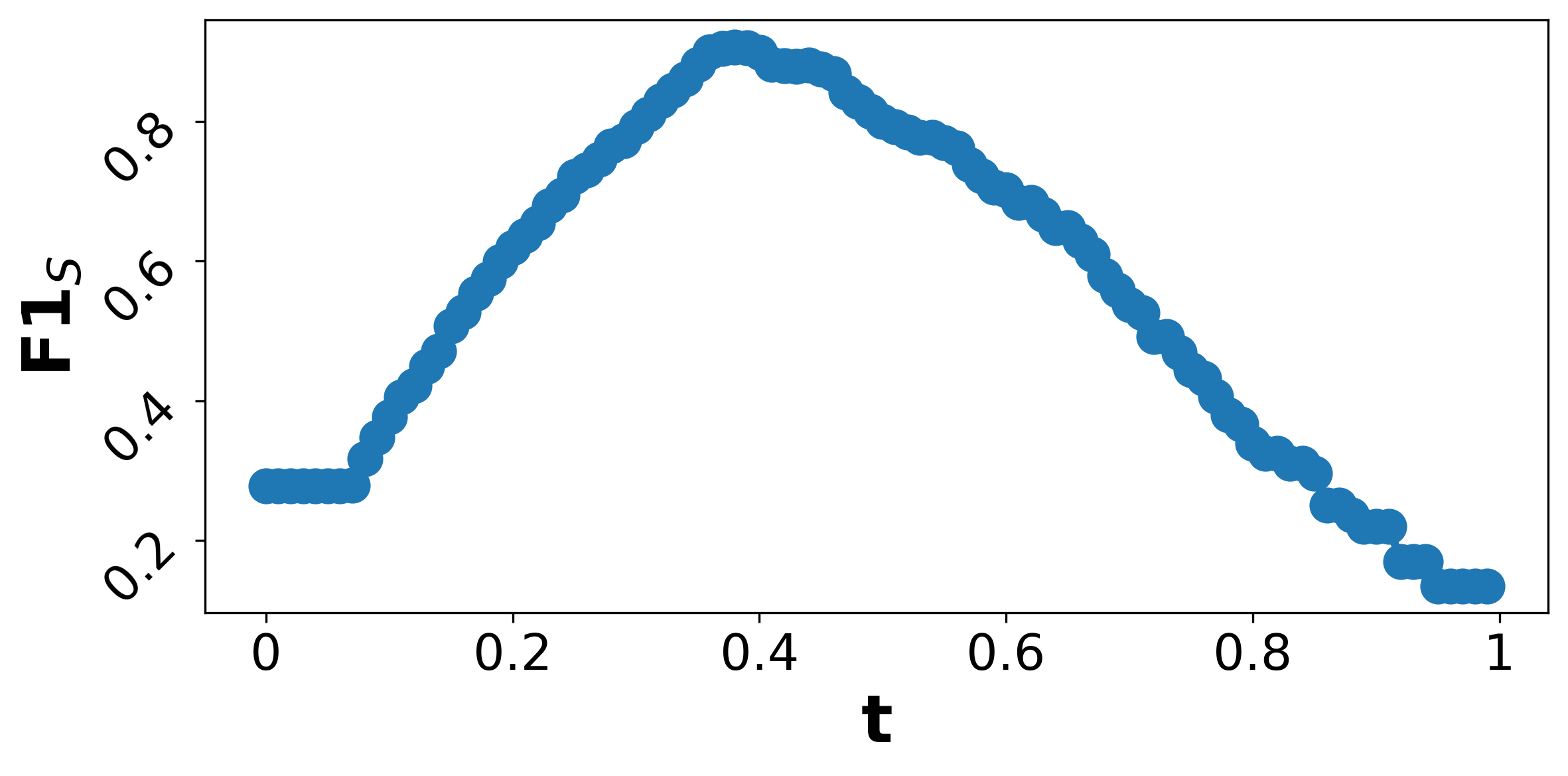}
        \caption{eICU}
        \label{fig:unimodal-eICU}
    \end{subfigure}
    \hfill
    \begin{subfigure}{0.48\linewidth}
        \centering
        \includegraphics[width=0.8\linewidth]{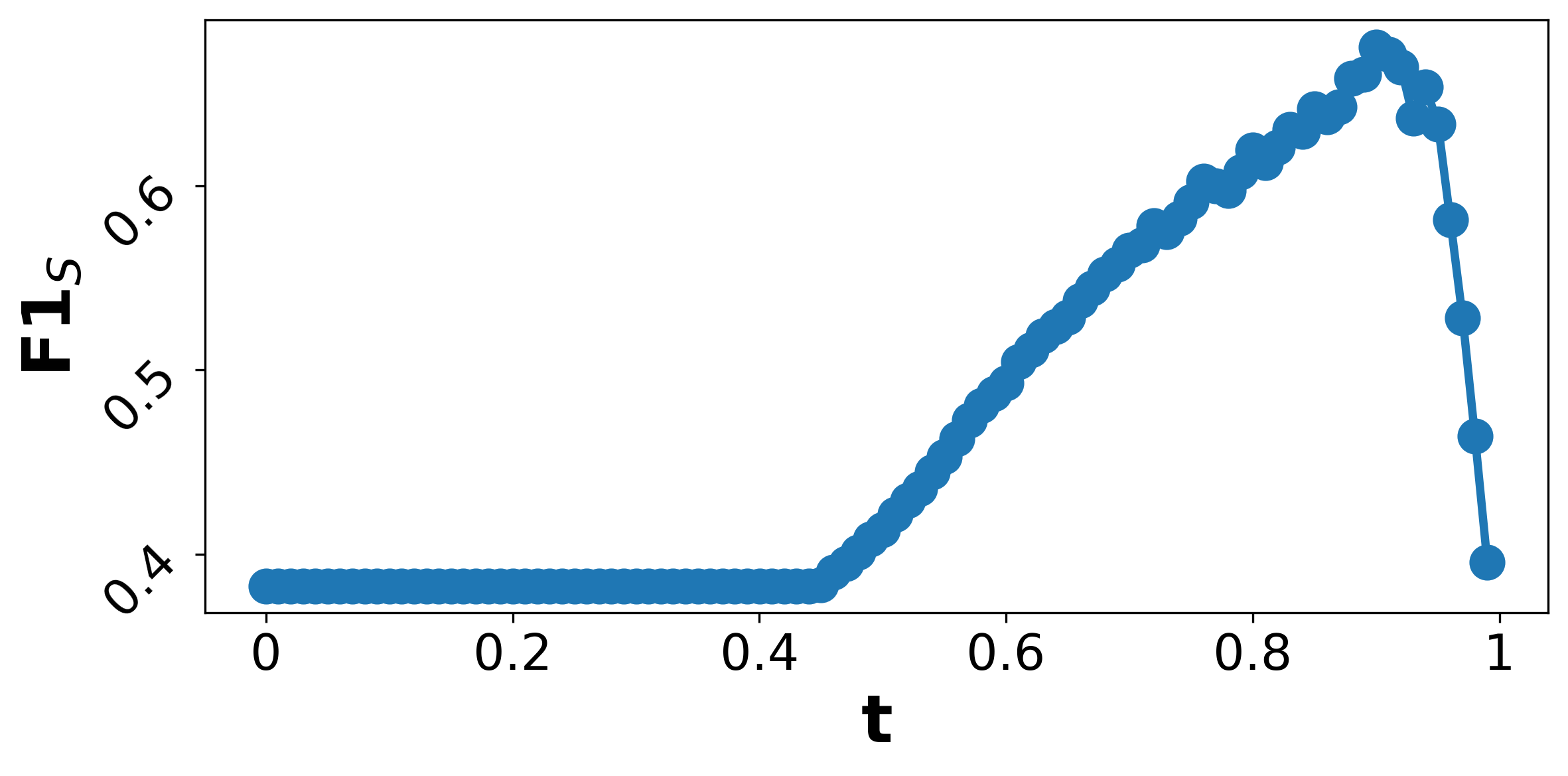}
        \caption{Yelp}
        \label{fig:unimodal-yelp}
    \end{subfigure}
    \caption{F1 score curves w.r.t. the precision target (\(t\)).}
    \label{fig:f1-empirical}
\end{figure}

To address this, we leverage the insight that a NN search algorithm that \textit{guarantees} one metric, either precision or recall, while \textit{maximizing} the other can effectively navigate this trade-off. We build on two approximate FRNN algorithms: PQE-PT and PQE-RT \cite{DujianPQA}, which, given a query \(q\), a proxy model \(P\), an oracle model \(O\), and a predefined precision (resp. recall) target \(t\), return a neighbor set that satisfies \(t\) with high probability while maximizing the complementary\footnote{Precision is complementary w.r.t. recall and vice versa.} metric. Both algorithms achieve statistical guarantees by thresholding proxy distances and validating candidates using oracle embeddings. In our \sprint framework, the pilot sample \(S_p\), for which oracle embeddings are already available, allows PQE-PT and PQE-RT to run without incurring additional oracle cost. We adopt PQE-PT in \sprintv for its implementation stability and consistent empirical performance across datasets. Moreover, we observe empirically that the \(\text{F1}_S\) achieved by PQE-PT is unimodal w.r.t. the precision target \(t\) (see Fig.~\ref{fig:f1-empirical}). This unimodality enables an efficient search for the optimal precision target \(t^*\) which maximizes the F1 score via ternary search.

\begin{algorithm}[t]
\caption{\sprintv}
\label{algo:sprint_v}
\begin{algorithmic}[1] 
\State \textbf{Input:} \( S \), \( S_p \), \( O \), \( P \), \( q \), \( r \), \( \omega_V \)
\State \textbf{Output:} \(\sprint_S\)

\State \(t_{\text{min}} \leftarrow 0\), \(t_{\text{max}} \leftarrow 1\) 

\While{$t_{\text{max}} - t_{\text{min}} > \omega_V$}
    \State \(t_1 \leftarrow t_{\text{min}} + \frac{t_{\text{max}} - t_{\text{min}}}{3}\)
    \State \(t_2 \leftarrow t_{\text{max}} - \frac{t_{\text{max}} - t_{\text{min}}}{3}\)
    
    \State \(NN(t_1) \leftarrow \text{PQE-PT}(S_p, O, P, q,r, t_1)\)
    \State \(NN(t_2) \leftarrow \text{PQE-PT}(S_p, O, P, q,r, t_2)\)
    
    \State \(\text{F1}_{S_p}(t_1) \leftarrow \frac{2 \text{P}_{S_p}(t_1)  \text{R}_{S_p}(t_1)}{\text{P}_{S_p}(t_1) + \text{R}_{S_p}(t_1)}\)
    \State \(\text{F1}_{S_p}(t_2) \leftarrow\frac{2\text{P}_{S_p}(t_2)  \text{R}_{S_p}(t_2)}{\text{P}_{S_p}(t_2) + \text{R}_{S_p}(t_2)}\)

    \If{\(\text{F1}_{S_p}(t_1) > \text{F1}_{S_p}(t_2)\)}
        \State \(t_{\text{max}} \leftarrow t_2\)
    \Else
        \State \(t_{\text{min}} \leftarrow t_1\)
    \EndIf
\EndWhile
\State \(t^* \leftarrow (t_{\text{min}} + t_{\text{max}})/2\)

\State \Return \(\text{PQE-PT}(S, O, P, q, r, t^*)\)
\end{algorithmic}
\end{algorithm}

As shown in Algorithm~\ref{algo:sprint_v}, \sprintv takes as inputs the sample \(S\), pilot sample \(S_p\), oracle \(O\), proxy \(P\), query target \(q\), query radius \(r\) and a tolerance \(\omega_V\) for terminating the ternary search, and outputs the \sprint neighbors in \(S\). The algorithm initializes the precision target range with \(t_{\text{min}} = 0\) and \(t_{\text{max}} = 1\) (Line 3) and iteratively narrows the range until \(t_{\text{max}}-t_{\text{min}} \leq \omega_V\) (Line 4). In each iteration, two candidate targets \(t_1\) and \(t_2\) are selected by dividing the range into thirds (Line 5-6). The PQE-PT subroutine is then called with each target to retrieve neighbors in \(S_p\) (Lines 7-8). Based on the selected neighbors and the oracle-labeled ground truth in \(S_p\), we compute precision, recall, and F1 scores at both targets in \(S_p\) (Lines 9-10). The more promising region of the precision range is retained for the next iteration (Lines 11-15). Once the search converges, the optimal precision target \(t^*\) is set as the midpoint of the final interval (Line 17). PQE-PT is then called one last time on \(S\) to identify \(\sprint_S\) (Line 18), which will later be used to compute the approximate aggregate.

\begin{algorithm}[t!]
\caption{\sprintc}
\label{algo:sprint_c}
\begin{algorithmic}[1]
\State \textbf{Input:} \( S \), \( S_p \), \( O \), \( P \), \( q \), \( r \), \( \omega_C \)
\State \textbf{Output:} \(\sprint_S\)

\State \(t_{\text{min}} \leftarrow 0\), \(t_{\text{max}} \leftarrow 1\) 
\While{$|\text{R}_{S_p} - \text{P}_{S_p}| > \omega_C$}
    \State \( t^* \leftarrow (t_{\text{min}} + t_{\text{max}})/2 \)
    \State \(NN \leftarrow \text{PQE-PT}(S_p, O, P, q, r, t^*)\)
    \State \(\text{R}_{S_p} \leftarrow \frac{|NN \cap ON_{S_p}|}{|ON_{S_p}|}\)
    \State \(\text{P}_{S_p} \leftarrow \frac{|NN \cap ON_{S_p}|}{|NN|}\)
    \If{$\text{P}_{S_p} \leq \text{R}_{S_p}$}
        \State \( t_{\text{min}} \leftarrow t^* \)
    \Else
        \State \( t_{\text{max}} \leftarrow t^* \)
    \EndIf
\EndWhile
\State \Return \(\text{PQE-PT}(S, O, P, q, r, t^*)\)
\end{algorithmic}
\end{algorithm}

\subsubsection{\sprintc} \label{subsec:sprint_c}
For count-sensitive AQNNs, we propose \sprintc to find neighbors in \(S\). Similar to \sprintv, \sprintc leverages PQE-PT to iteratively find the optimal precision target such that the selected neighbors in \(S\) achieve equal precision and recall. The intuition is that balancing the false positives and false negatives in the retrieved neighbor set leads to an unbiased estimate of the true neighbor count. We formalize this insight using \(\mathtt{PCT}\) as an example, by expressing its approximate estimate as:
\begin{equation}
\widetilde{\mathtt{PCT}_S} = \frac{|\sprint_S|}{s} = \frac{|(\sprint_S \cap ON_S) \cup (\sprint_S \setminus ON_S)|}{s}.
\label{eq:approx_PCT_S}
\end{equation}
where $\sprint_X$ (resp. $ON_X$) denotes the NNs found by \sprint (resp. by the oracle, i.e., the true NNs) in dataset $X$, $X \in \{S, D\}$. When, in \(S\), the number of false positives \((\sprint_S \setminus ON_S)\) equals the number of false negatives \((ON_S \setminus \sprint_S)\), \(\widetilde{\mathtt{PCT}_S}\) can accurately reflect \(\mathtt{PCT}_S\). This condition holds when precision equals recall, which implies a balance between false positives and false negatives. Formally, the precision and recall of the \sprint neighbors in \(S\) are:
    $\text{P}_S = \frac{|\sprint_S \cap ON_S|}{|\sprint_S|}$, and
    $\text{R}_S = \frac{|\sprint_S \cap ON_S|}{|ON_S|}$.
By setting \(\text{P}_S = \text{R}_S\), we can infer 
    $|\sprint_S \setminus ON_S|  = |ON_S \setminus \sprint_S|$,
i.e., the numbers of false positives and false negatives are equal. As a result, when \(\text{P}_S = \text{R}_S\), \(\widetilde{\mathtt{PCT}_S}\) is an unbiased estimator of \(\mathtt{PCT}_S\). 

As shown in Algorithm \ref{algo:sprint_c}, \sprintc takes similar inputs as \sprintv and outputs the \sprint neighbors in \(S\). Since there is an inverse relationship between precision and recall, where a higher precision leads to a lower recall, a binary search is employed to find the optimal precision target \(t^*\) which can equalize precision and recall in \(S\). Specifically, \sprintc initializes the search range with \(t_{\text{min}} = 0\) and \(t_{\text{max}} = 1\) (Line 3) and iteratively narrows this range until the difference between the recall \(R_{S_p}\) and precision \(P_{S_p}\) in the pilot sample is within a tolerance \(\omega_C\) (Line 4). In each iteration, a candidate precision target is set to the midpoint of the current range (Line 5). The PQE-PT subroutine is then invoked to retrieve the NNs in \(S_p\) (Line 6). The corresponding recall and precision \(R_{S_p}\) and \(P_{S_p}\) are calculated based on the oracle-determined true NNs and the selected NNs in \(S_p\) (Line 7-8), and the search range is updated accordingly (Lines 9–13). Finally, once the optimal \(t^*\) is determined, PQE-PT is called on \(S\) with \(t^*\) to retrieve \(\sprint_S\) (Line 15). The neighbor set will then be used in the aggregation step. 

\stitle{Time Complexity.} The time complexity of both algorithms is dominated by the invocation of PQE-PT, which has a complexity of \(O(s^2\log(s))\) \cite{DujianPQA}, $s = |S|$ being the sample size. During the search for \(t^*\), \sprintv employs a ternary search, requiring \(O(\log(1/\omega_V))\) iterations, while \sprintc uses a binary search with \(O(\log(1/\omega_C))\) iterations\footnote{Ternary search uses \(\log_3\), while binary search uses \(\log_2\). However, in asymptotic analysis, these differences are absorbed into constant factors.}. Assuming \(\omega_C=\omega_V =\omega^*\), the overall time complexity is \(O(\log(1/\omega^*) s^2 \log(s))\). Evaluating precision and recall over \(S_p\) introduces an additional linear term, \(O(s_p)\), but since \(s_p \ll s\), its impact is negligible in asymptotic analysis. Therefore, \sprintv and \sprintc scale quadratically with \(s\) (up to logarithmic factors), while the number of iterations needed for convergence depends only logarithmically on \(\omega^*\).

\subsection{Theoretical Analysis} \label{sec:theory}
In this section, we derive probabilistic error bounds of \sprint for value- and count-sensitive queries. 
Also, we establish the minimum sample and pilot sample sizes required to achieve the error bounds. The detailed proofs are provided in the Appendix. 

\stitle{Key insight.} The total approximation error arises from two components: 
(i) the sampling error \(\omega_S\) from aggregating over NNs in \(S\) instead of the full dataset \(D\), and 
(ii) the NN selection error \(\omega_{NN}\) introduced by selecting NNs based on a mixture of oracle and 
proxy embeddings in \(S\). These two sources of error are bounded separately in our analysis.

To help interpret the theoretical bounds, let us unpack the NN selection error \(\omega_{NN}\) at a high level. For value-sensitive queries, \(\omega_{NN}\) depends on how well the F1 score in the sample \(S\) is optimized. Using \sprintv, we maximize \(\text{F1}_{S_p}\) and leverage it as an estimator for \(\text{F1}_S\). By ensuring that \(\text{F1}_{S_p}\) closely approximates \(\text{F1}_S\), i.e.  \(|\text{F1}_S-\text{F1}_{S_p}| \leq \lambda\), the neighbor selection strategy derived from \(S_p\) also performs well on \(S\), keeping \(\omega_{NN}\) small. For count-sensitive queries, \(\omega_{NN}\) is determined by the precision-recall gap \(|P_S - R_S|\), which reflects the imbalance between false positives and false negatives in the retrieved neighbor set in \(S\). \sprintc controls this gap via \(|P_{S_p} - R_{S_p}| \leq \omega_C\) in the pilot sample \(S_p\), which ensures with high probability that \(\omega_{NN}\) is small.


\begin{theorem}[Aggregation Error Bound for Value-Sensitive Queries]
\label{thm:sprint-v}
Let \(\widetilde{\mathrm{agg}}_S\) denote the approximate aggregate returned by \sprintv for a value-sensitive query (e.g., \(\mathtt{AVG}\), \(\mathtt{VAR}\)). Given user-specified tolerances \(\omega_S\) and \(\omega_{NN}\), and confidence level \(1-\alpha\), the approximation error satisfies
\[
\Pr\left[
    \left| \widetilde{\mathrm{agg}}_S - \mathrm{agg}_D \right| \leq \omega_S + \omega_{NN}
\right] \geq 1-\alpha,
\]
provided that the sample size \(s\) and pilot sample size \(s_p\) satisfy
\[
s \geq \kappa_1\cdot\frac{1}{\omega_S^2},\quad
s_p \geq \kappa_2\cdot\frac{1}{\lambda^2},
\]
where \(\lambda\) is a configurable tolerance on the  difference between \(\mathrm{F1}_S\) and \(\mathrm{F1}_{S_p}\) 
and
\(
\omega_{NN} = C_1\cdot\lambda + C_2\cdot\lambda^2.
\)
Here, \(\kappa_1, \kappa_2, C_1,\) and \(C_2\) depend on the specific aggregation function as follows:
\begin{itemize}
    \item For \(\mathtt{AVG}\),
    \[
    \begin{aligned}
    \kappa_1 &= \frac{(b-a)^2\ln(2/\alpha)}{2\rho},\quad
    \kappa_2=32\ln(2/\alpha),\\
    C_1 &= \frac{2(b-a)}{|ON_S|},\quad
    C_2=0.
    \end{aligned}
    \]
    \item For \(\mathtt{VAR}\),
    \[
    \begin{aligned}
    \kappa_1 &= \frac{(b-a)^4\ln(2/\alpha)}{2\rho},\quad
    \kappa_2=32\ln(2/\alpha),\\
    C_1&=\frac{3(b-a)^2}{|ON_S|},\quad
    C_2=\frac{(b-a)^2}{|ON_S|}.
    \end{aligned}
    \]
\end{itemize}
where \(a\) and \(b\) are the lower and upper bounds of the target attribute \footnote{In clinical settings, adult patients’ heart rates are commonly observed in the range of 50-120 bpm, although the typical resting range is 60-100 bpm.}; \(\rho = |ON_S|/|S|\) is the the neighborhood density.
\end{theorem}

\stitle{Practical Implications.} 
To illustrate the practical significance of Theorem~\ref{thm:sprint-v}, consider the value-sensitive query in Fig. \ref{fig:framework}. Suppose we set \(\alpha = 0.05\) and choose a radius \(r\) such that \(\rho = 0.8\). Here, \(\rho\) is estimated from the dataset by the proportion of objects in \(D\) within radius \(r\). Setting \(\omega_S = 5\) bpm and \(\omega_{NN} = 0.2\), which are clinically reasonable tolerances, results in a minimum sample size of \(s \geq 452\) and pilot sample size of \(s_p \geq 111\). These correspond to 10.6\% and 2.6\% of the population size (\(|D|=4{,}245\)), respectively. By applying these bounds, practitioners can configure parameters to balance computational cost and aggregation accuracy in real-world settings.

\begin{theorem}[Aggregation Error Bound for Count-Sensitive Queries]
\label{thm:sprint-c}
Let \(\widetilde{\mathrm{agg}}_S\) denote the approximate aggregate returned by \sprintc for a count-sensitive query (e.g., \(\mathtt{PCT}\), \(\mathtt{COUNT}\)). Given user-specified tolerances \(\omega_{S}\) and \(\omega_{NN}\), and confidence level \(1-\alpha\), the approximation error satisfies
\[
\Pr\left[
    \left| \widetilde{\mathrm{agg}}_S - \mathrm{agg}_D \right| \leq \omega_S + \omega_{NN}
\right] \geq 1-\alpha,
\]
provided that the sample size \(s\) and pilot sample size \(s_p\) satisfy
\[
s \geq \kappa_1\cdot\frac{1}{\omega_S^2},\quad
s_p \geq \kappa_2\cdot\frac{1}{\left(\omega_{NN}-\rho\cdot\omega_C\right)^2},
\]
where \(\rho = |ON_S|/|S|\) is the the neighborhood density.
Here, \(\kappa_1\) and \(\kappa_2\) depend on the specific aggregation function as follows:
\begin{itemize}
    \item For \(\mathtt{PCT}\),
    \[
    \kappa_1 = \frac{1}{2}\ln\left(\tfrac{2}{\alpha}\right),\quad
    \kappa_2 = 2\ln\left(\tfrac{2}{\alpha}\right)\cdot\rho^{-2}.
    \]
    \item For \(\mathtt{COUNT}\),
    \[
    \kappa_1 = \frac{|D|^2}{2}\ln\left(\tfrac{2}{\alpha}\right),\quad
    \kappa_2 = 2\ln\left(\tfrac{2}{\alpha}\right)\cdot\rho^{-2}.
    \]
\end{itemize}
\end{theorem}

\stitle{Practical Implications.} 
To illustrate the practical significance of Theorem~\ref{thm:sprint-c}, consider a count-sensitive query such as computing the proportion of neighbors in the population of a target insomnia patient. Suppose we set \(\omega_S = 0.05\), \(\omega_{NN} = 0.1\), and \(\omega_C = 0.0001\). These tolerances yield a minimum sample size of \(s \geq 738\) and pilot sample size of \(s_p \geq 739\), which both correspond to about 17.4\% of (\(|D|=4{,}245\)). Notably, the lower bound for \(s_p\) may exceed that of \(s\) because \(s\) and \(s_p\) serve distinct roles. \(s\) controls how well the aggregate computed over \(S\) approximates the true aggregate over \(D\). In contrast, \(s_p\) ensures the precision and recall estimates obtained from \(S_p\) are sufficiently accurate to represent those in \(S\), which is essential for selecting a stable precision target \(t^*\). Hence, to ensure both accuracy and stability, we set \(s = \max(s, s_p)\) and the smaller one be the pilot sample. 


\subsection{Practical Algorithm for \(\mathtt{SUM}\) Queries}\label{sec:SUM_algo}

\(\mathtt{SUM}\) query differs from the value-sensitive or count-sensitive queries because its error depends on both the cardinality of the selected neighborhood and their target attribute values. Hence, we propose a practical strategy, called \textit{Two-Phase}, to balance these two aspects.

\stitle{Two-Phase Strategy.} 
To handle \(\mathtt{SUM}\)'s sensitivity to both count and value of NNs, we propose Two-Phase: (1) apply \sprintc to identify a precision target \(t\) that balances precision and recall, aligning the neighbor counts; (2) refine it within a small window (i.e., \(t\pm0.05\)) and use \sprintv to maximize the F1-score, ensuring that the selected neighbors are value-consistent. 

\begin{theorem}[Aggregation Error Bound for \(\mathtt{SUM}\) Queries]
\label{thm:sprint-sum}
Let \(\widetilde{\mathrm{SUM}}_S\) denote the approximate aggregate returned by Two-Phase for the \(\mathtt{SUM}\) query. Given user-specified tolerances \(\omega_S\) and \(\omega_{NN}\), and confidence level \(1-\alpha\), the approximation error satisfies
\[
\Pr\left[
    \left|\widetilde{\mathrm{SUM}}_S - \mathrm{SUM}_D\right| \leq \omega_S + \omega_{NN}
\right] \geq 1-\alpha,
\]
provided that the sample size \(s\) and pilot sample size \(s_p\) satisfy
\[
s \geq \max\left(s^{count},\ s^{avg}\right),\quad
s_p \geq \max\left(s_p^{count},\ s_p^{avg}\right).
\]
Here:
\squishlist
    \item \(s^{count}= \frac{2|D|^2|\mathtt{AVG}_S|^2\ln(4/\alpha)}{\omega_S^2}\) bounds the sampling error on the neighbor count.
    \item \(s^{avg}= \frac{2(b-a)^2|ON_D|^2\ln(4/\alpha)}{\omega_S^2}\) bounds the sampling error on the attribute mean.
    \item \(s_p^{count}= \kappa_2^{\mathtt{COUNT}}\cdot\frac{1}{\left(\omega_{NN}^{\mathtt{SUM}}-\rho\cdot\omega_C\right)^2}\) bounds the count selection error,
    where \(\omega_{NN}^{\mathtt{SUM}} = \frac{\omega_{NN}|S|}{2|D||\widetilde{\mathtt{AVG}}_S|}\) and \(\rho = \tfrac{|ON_S|}{s}\) denotes the neighborhood density in \(S\).
    \item \(s_p^{avg}= \kappa_2^{\mathtt{AVG}}\cdot\frac{1}{\lambda^2}\) bounds the mean selection error,
    where \(|\text{F1}_S-\text{F1}_{S_p}|\leq\lambda\).
\squishend
\end{theorem}

\begin{table}[t!]
\caption{Datasets, oracle, and proxy models.}
\label{tab:dataset}
\centering
\small
\begin{tabular}{lcc}
\toprule
\textbf{Datasets} & \textbf{Oracle} & \textbf{Proxy} \\
\midrule
Mimic-III, eICU & Expert Notations & LIG-Doctor~\cite{DBLP:journals/isci/RodriguesGSBA21} \\
Jigsaw & User Labels & GLiClass~\cite{zaratiana2023gliner} \\
Yelp, Amazon-E
 & MiniLM-L12~\cite{DBLP:conf/nips/WangW0B0020} & MiniLM-L6~\cite{DBLP:conf/nips/WangW0B0020} \\
\bottomrule
\end{tabular}
\end{table}
\section{Experiments}\label{sec:exps}
The purpose of our experiments is to evaluate our solution for answering AQNNs w.r.t. objectives \textbf{O1} and \textbf{O2}: cost and error. \S~\ref{exp:setup} describes the datasets, baselines, and evaluation measures. \S~\ref{exp:framework} evaluates efficiency of \sprint in terms of embedding generation cost and end-to-end runtime. \S~\ref{exp:algorithm} examines the effectiveness of \sprintv and \sprintc in lowering aggregation error and query execution time. We also study the impact of dataset size (scalability), sample size, pilot sample size, and NN sparsity (controlled by the query radius \(r\)) on the aggregation error. Finally, \S~\ref{exp:HT} demonstrates one-sample hypothesis testing as a downstream application of AQNNs. 




\subsection{Experimental Setup}\label{exp:setup}
In our experiments, we use static datasets to simulate settings where embeddings must be recomputed or refreshed on demand at query time. This reflects realistic deployment scenarios, such as in healthcare or education, where data evolves rapidly, making precomputed or cached embeddings stale or infeasible. For settings with static or infrequently updated data, embeddings may be precomputed, cached, and reused for later queries; we analyze the cost of such scenarios in \S~\ref{exp:cost} by decoupling embedding generation cost and measuring only the query execution cost, assuming embeddings are precomputed.

\begin{table}[t]
\centering
\caption{Dataset size $|D|$ and default parameters for \sprint: sample size $s$ and pilot sample size $s_p$.}
\label{tab:default_proxy_oracle_calls}
\small
\begin{tabular}{lccccc}
\toprule
 & \textbf{eICU} & \textbf{MIMIC-III} & \textbf{Jigsaw} & \textbf{Yelp} & \textbf{Amazon-E} \\
\midrule
$|D|$     & 8,234    & 4,245     & 16,225   & 6,990,280    & 10,000    \\
$s$       & 1,000    & 500       & 1,000    & 35,000       & 2,000     \\
$s_p$     & 600      & 150       & 600      & 20,000      & 1,200     \\
\bottomrule
\end{tabular}
\end{table}

\subsubsection{Datasets}
We used five datasets from three different domains, each paired with task-appropriate proxy and oracle models (see Table~\ref{tab:dataset}). The dataset sizes are provided in Table~\ref{tab:default_proxy_oracle_calls}, top row.

\stitle{Medical:} MIMIC-III \cite{johnson2016mimic} and eICU \cite{pollard2018eicu} are publicly available clinical datasets containing de-identified ICU records from U.S. hospitals. We adopt LIG-Doctor \cite{DBLP:journals/isci/RodriguesGSBA21}, a lightweight RNN-based model for patient trajectory prediction, to generate proxy embeddings from ICU admission sequences. The oracle embeddings are derived from ground-truth physician diagnoses. Each patient record has demographic and physiological attributes (e.g., heart rate), which enable value-sensitive AQNNs in medical contexts. We remove records with only one ICU stay.

\stitle{Social media:} The Jigsaw toxicity classification dataset \cite{jigsaw} contains real-world user comments annotated across five toxicity types (e.g., threat, insult). We employ GLiClass \cite{zaratiana2023gliner}, a zero-shot classifier to generate proxy embeddings without domain-specific training. Oracle embeddings are constructed from the annotated toxicity labels. To support the evaluation of value-sensitive queries, we synthesize numerical engagement metrics (e.g., number of down-votes) using K-means clusters~\cite{lloyd1982least, mcqueen1967some}. 

\stitle{E-commerce:} Yelp \cite{yelp_dataset} and Amazon-Electronics (Amazon-E) \cite{hou2024bridging} are publicly available corpora of user reviews that include text, metadata, and star ratings (1-5). We generate oracle embeddings using MiniLM-L12~\cite{DBLP:conf/nips/WangW0B0020}, a transformer-based encoder known for its strong semantic fidelity, and proxy embeddings using a lighter-weight MiniLM-L6~\cite{DBLP:conf/nips/WangW0B0020}. Star ratings are used as the target attribute in value-sensitive AQNNs.

\subsubsection{Baselines}\label{exp:baseline}
\sprint is, to our knowledge, the first framework for cost-efficient and low-error approximation of AQNNs in settings where oracle embeddings are expensive and need to be recomputed on demand. We first conduct a \textit{framework-level comparison} against a Brute-Force baseline, which computes oracle embeddings for all objects and performs exact NN search, w.r.t. embedding generation and the overall end-to-end costs. We then perform an \textit{algorithm-level comparison} to evaluate our proposed NN selection algorithms, \sprintv and \sprintc, against three probabilistic ANN baselines adapted from prior work. These include:

\stitle{A. PQE-PT and PQE-RT.}
These are FRNN approximation algorithms that guarantee user-specified precision targets (PQE-PT) or recall targets (PQE-RT) with high probability~\cite{DujianPQA}. They calibrate the correlation between proxy and oracle distances using a small oracle-labeled sample from the dataset. Once calibrated, proxy distances are used to select neighbors in the full dataset. Although PQE-PT is used as a subroutine in our framework, it is designed to optimize precision only and does not account for aggregation error. We will show its suboptimality for answering AQNNs in \S ~\ref{exp:error}.

\stitle{B. SUPG-PT and SUPG-RT. } 
SUPG~\cite{DBLP:journals/pvldb/KangGBHZ20} frames FRNN as a binary classification problem: given a query target \(q\) and radius \(r\), predict whether each object \(x\) lies within the neighborhood of \(q\) using proxy-based distances. SUPG-PT and SUPG-RT are trained using oracle-labeled samples to ensure precision or recall guarantees, respectively. After training, the classifier is applied to the entire dataset using only proxy embeddings. Like PQE, SUPG incurs limited oracle cost during calibration but performs inference purely on proxy representations.

\begin{figure}[!t]
    \centering
    \begin{subfigure}{0.48\linewidth}
        \centering
        \includegraphics[width=0.8\linewidth]{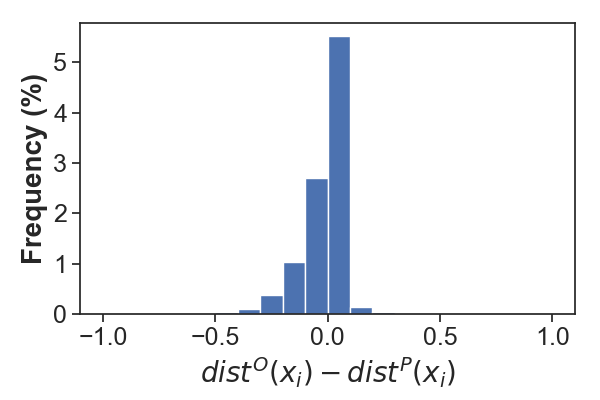}
        \caption{eICU}
        \label{fig:dist_eICU}
    \end{subfigure}
    \hfill
    \begin{subfigure}{0.48\linewidth}
        \centering
        \includegraphics[width=0.8\linewidth]{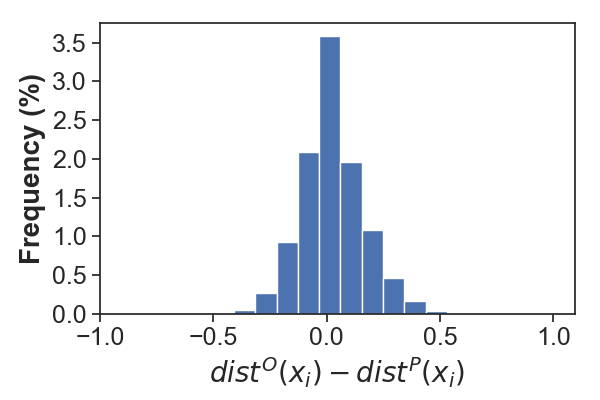}
        \caption{MIMIC-III}
        \label{fig:dist_mimic}
    \end{subfigure}
    \caption{Distribution of the difference between proxy and oracle distances.}
    \label{fig:dist}
\end{figure}

\stitle{C. Top-K.}  
Probabilistic Top-K approximates NNs by returning the top-\(K\) closest objects to the query under proxy distances, where \(K\) is set to the number of oracle-based neighbors within radius \(r\)~\cite{DBLP:conf/sigmod/LaiHLZ0K21}. It assumes that proxy distances are accurate surrogates of oracle distances. This assumption holds reasonably well in our datasets, as shown in Fig.~\ref{fig:dist}, where \(dist^X(x_i)\) represents the distance computed using \(X\) embeddings between \(x_i\) and \(q\), with \(X\in \{O,P\}\).

\subsubsection{Evaluation Measures}\label{exp:evaluation_measures}
\stitle{Relative Error (RE)} is a measure of aggregation error, defined as the proportional difference between the estimated and true aggregates:
\begin{equation}
\text{RE} = \frac{| \widetilde{\text{agg}_S} - \text{agg}_D |}{| \text{agg}_D |} \times 100\%
\end{equation}

\stitle{Cost} refers to the total computational expense incurred during AQNN evaluation. It consists of two components: (i) {\em embedding generation cost}, measured by time and also reflected by the number of oracle and proxy model calls required to generate embeddings; and (ii) {\em query execution cost}, measured by time required to answer AQNNs given computed embeddings.

\stitle{F1 score}, the harmonic mean of precision and recall, measures how accurately the selected NNs match the true NNs. It is defined as:
\begin{equation}
\text{F1} = \frac{2\cdot\text{Precision} \cdot\text{Recall}}{\text{Precision} + \text{Recall}}
\end{equation}


\subsubsection{Default Parameters}
We evaluate AQNNs using four aggregation functions that span all query types: value-sensitive (\(\mathtt{AVG}\), \(\mathtt{VAR}\)), count-sensitive (\(\mathtt{PCT}\)), and both value- and count-sensitive (\(\mathtt{SUM}\)). To use \sprint, the default values for the sample size \(s\) and pilot sample size \(s_p\) are shown in Table~\ref{tab:default_proxy_oracle_calls}. These values are derived from Theorems~\ref{thm:sprint-v}-\ref{thm:sprint-sum} by setting appropriate values for \(b\), \(a\), \(\rho\), \(\omega_S\), and \(\omega_{NN}\). Following prior work~\cite{DujianPQA}, we set the precision and recall targets to 0.95 for both PQE and SUPG. All experiments are repeated 30 times to ensure statistical reliability under the Central Limit Theorem (CLT)~\cite{kwak2017central}. Reported metrics are averaged over 10 random query targets to account for query-specific variability.

All experiments are conducted on a NVIDIA RTX 2080 GPU with batch size 32, unless otherwise specified.

\begin{figure}[t]
    \centering
    \begin{subfigure}{0.48\linewidth}
        \centering
        \includegraphics[width=0.78\linewidth]{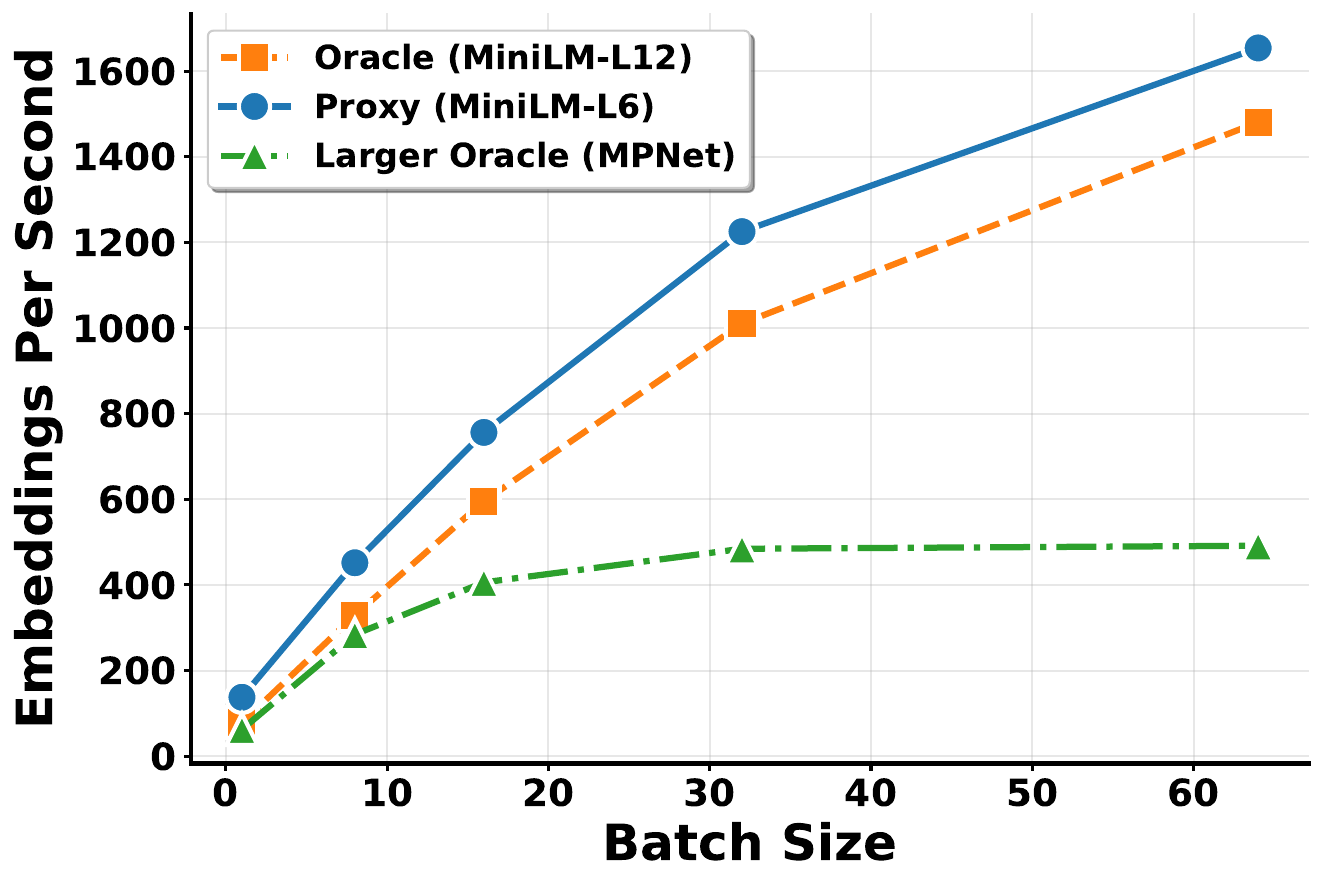}
    \end{subfigure}
    \hfill
    \begin{subfigure}{0.48\linewidth}
        \centering
        \includegraphics[width=0.78\linewidth]{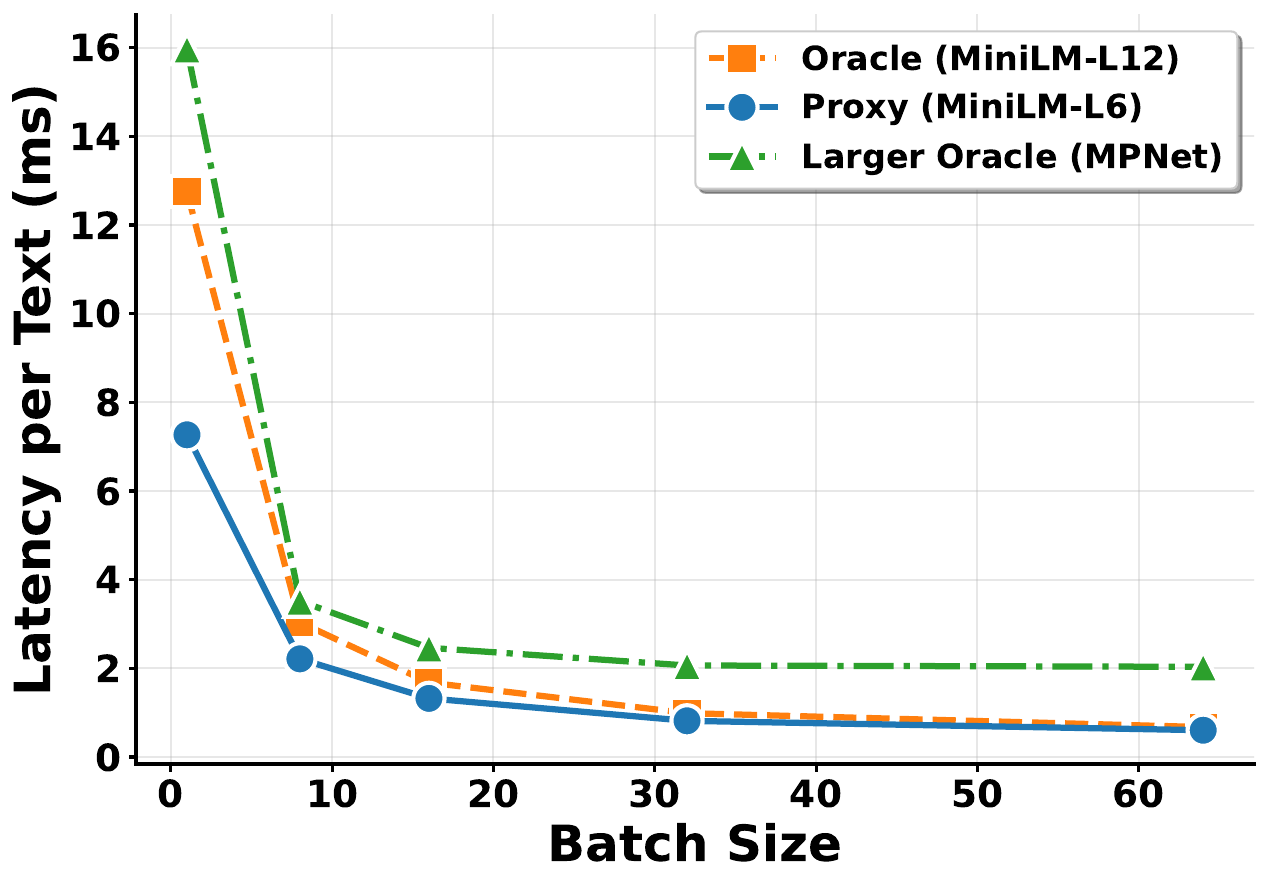}
    \end{subfigure}
    \caption{Oracle vs. Proxy embedding generation cost.}
    \label{fig:oracle_vs_proxy_emb_cost}
\end{figure}

\begin{figure*}[!t]
    \centering
    \begin{subfigure}{0.48\linewidth}
        \centering
        \includegraphics[width=0.8\linewidth]{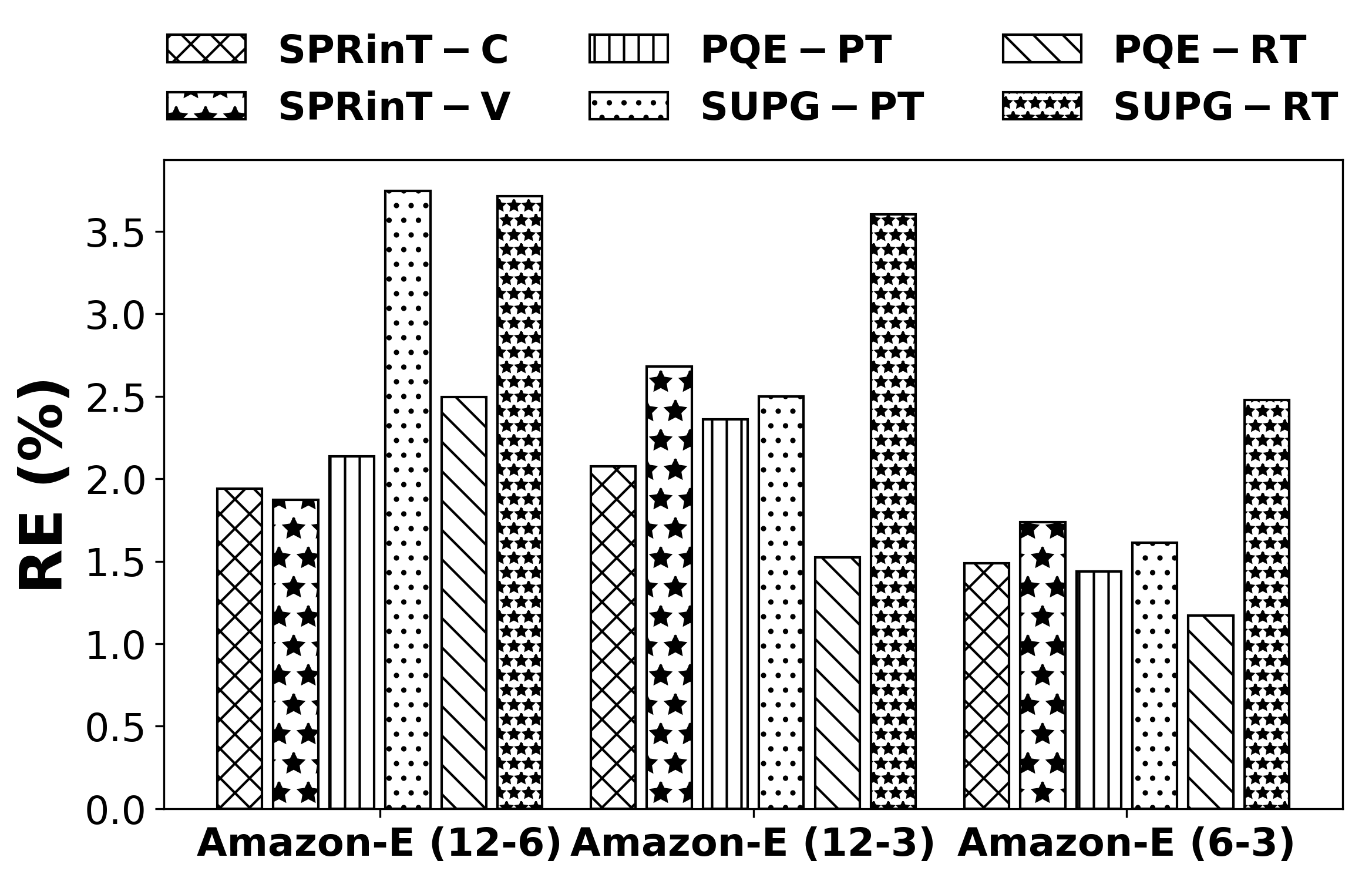}
        \caption{value-sensitive: \(\mathtt{AVG}\)}
        \label{fig:sensitivity_AVG}
    \end{subfigure}
    \hfill
    \begin{subfigure}{0.48\linewidth}
        \centering
        \includegraphics[width=0.8\linewidth]{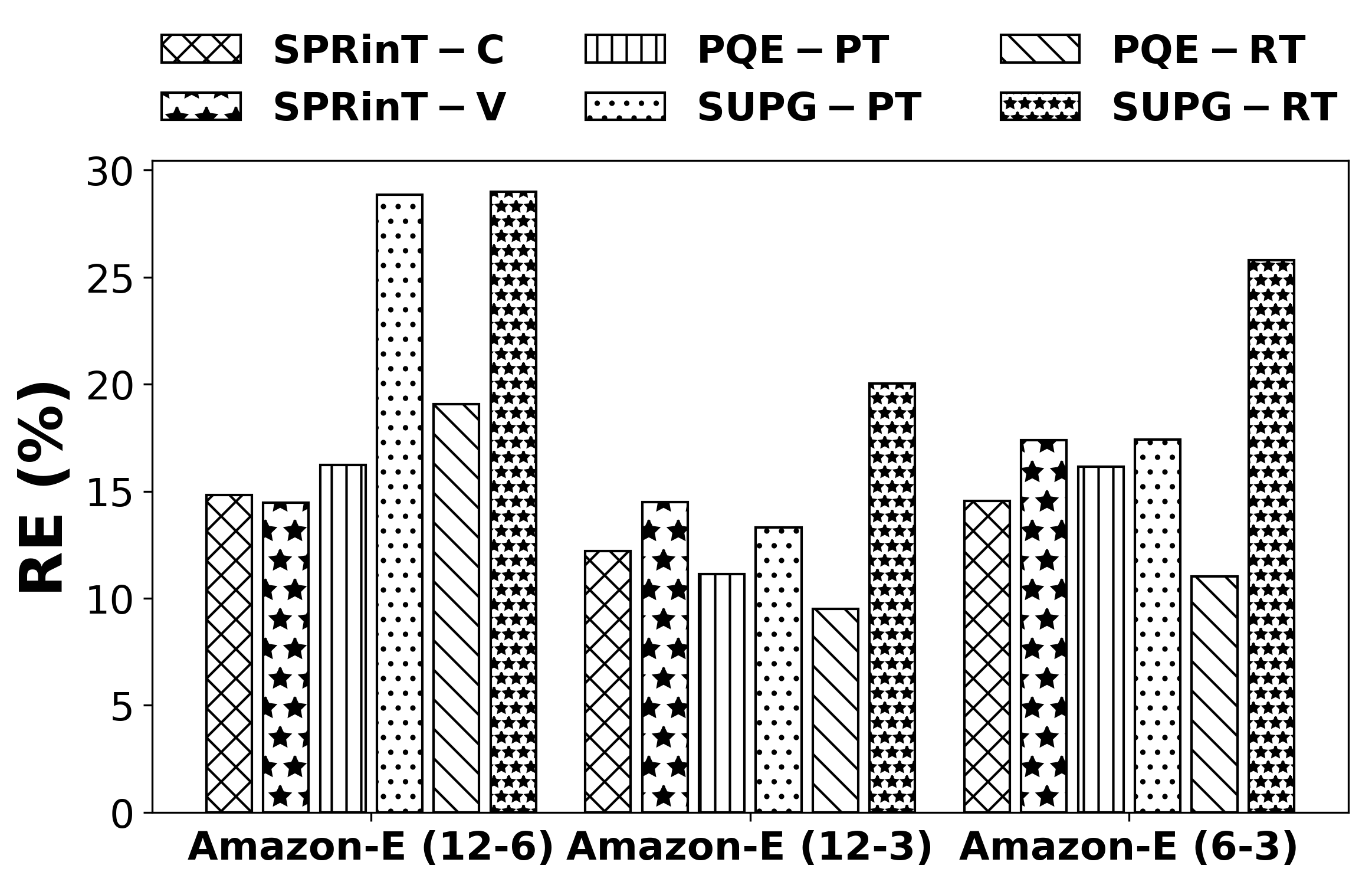}
        \caption{value-sensitive: \(\mathtt{VAR}\)}
        \label{fig:sensitivity_VAR}
    \end{subfigure}
    \centering
    \begin{subfigure}{0.48\linewidth}
        \centering
        \includegraphics[width=0.8\linewidth]{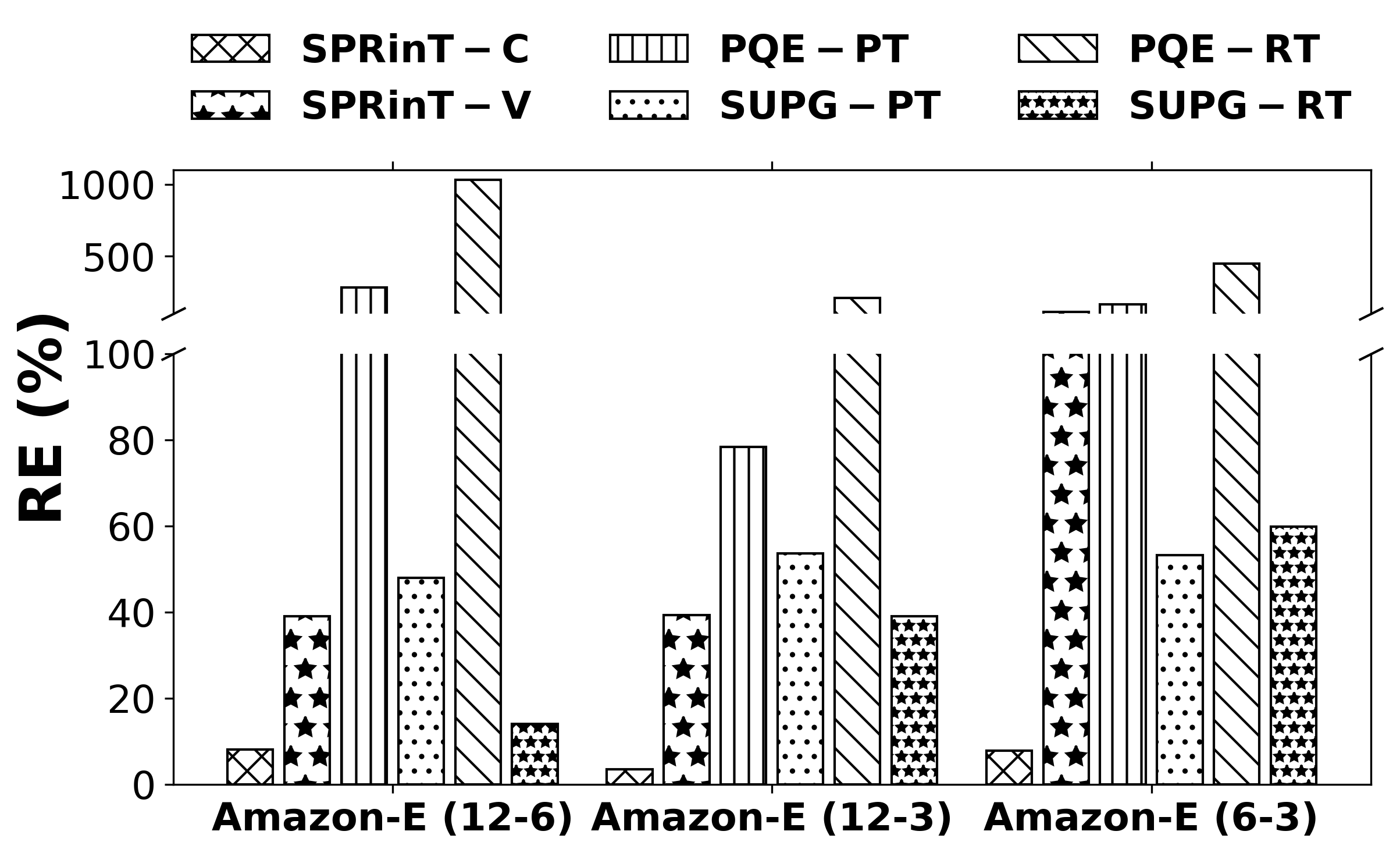}
        \caption{count-sensitive: \(\mathtt{PCT}\)}
        \label{fig:sensitivity_PCT}
    \end{subfigure}
    \hfill
    \begin{subfigure}{0.48\linewidth}
        \centering
        \includegraphics[width=0.8\linewidth]{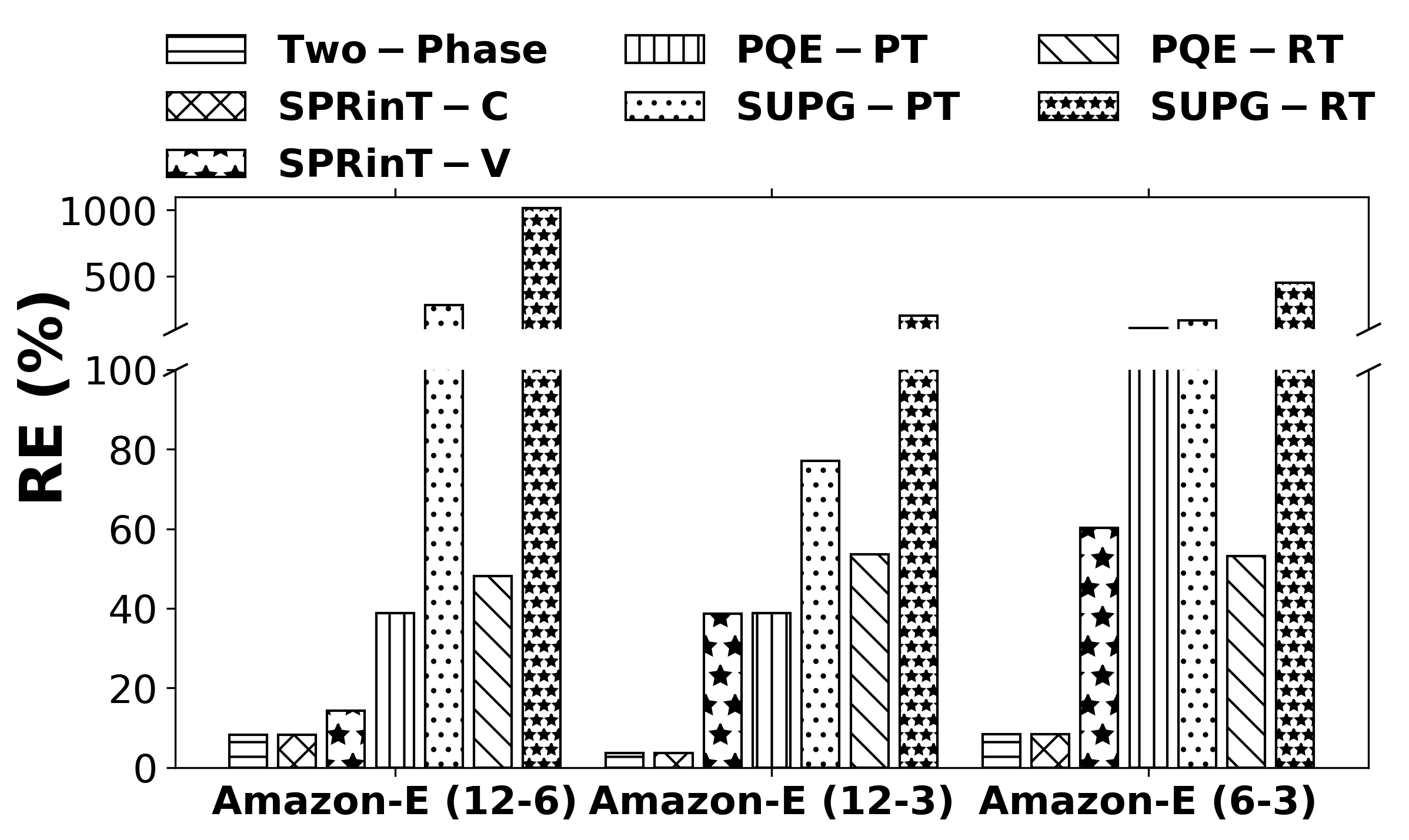}
        \caption{value- and count-sensitive: \(\mathtt{SUM}\)}
        \label{fig:sensitivity_SUM}
    \end{subfigure}
    \caption{RE performance under different choices of oracle and proxy models on Amazon-E.}
    \label{fig:proxy_oracle_variants}
\end{figure*}

\subsection{Framework-Level Evaluation}\label{exp:framework}

We evaluate the embedding generation cost and overall end-to-end runtime of answering AQNNs by comparing \sprint against Brute-Force. Aggregation error is not compared when evaluating cost, as Brute-Force trivially achieves zero error. Also, we conduct a sensitivity analysis on the choice of oracle-proxy pair to evaluate the robustness of \sprint to this choice.

\noindent \textbf{Empirical Evidence: Oracle vs Proxy Embedding Generation Cost.} To quantify the computational gap between oracle and proxy models, we measure embedding generation latency and throughput across batch sizes for three text-embedding models, MiniLM-L6 (proxy), MiniLM-L12 (oracle), and MPNet ~\cite{DBLP:conf/nips/WangW0B0020, song2020mpnet} using 1,000 Amazon-E reviews. Fig.~\ref{fig:oracle_vs_proxy_emb_cost} shows that oracle models are approximately \(2\times\) slower in per-text latency than proxy models at small to medium batch sizes, consistent with the benchmark results reported in~\cite{DBLP:conf/nips/WangW0B0020}. At large batch sizes, per-text latency converges due to GPU amortization, but proxy models still achieve higher throughput and remain significantly more efficient for query-time embedding generation.

\subsubsection{Embedding Generation Cost}\label{exp:framework:embedding_generation_cost}
Table~\ref{tab:embedding_cost_comparison} compares the embedding generation costs for all datasets. We assume oracle calls are $2\times$ slower than proxy calls, which is conservative, as prior work reports $2$–$10\times$ gaps~\cite{DBLP:conf/nips/WangW0B0020,reddi2020mlperf}. \sprint achieves $4.5$–$186.4\times$ speedup by avoiding the majority of oracle calls. For instance, on Jigsaw, \sprint uses proxy models for $\sim$6\% of objects to avoid $>$96\% of oracle calls.

\begin{table}[t]
\caption{Embedding generation cost in terms of oracle and proxy calls. \textbf{Bold} indicates the best performance.}
\label{tab:embedding_cost_comparison}
\centering
\small
\resizebox{0.6\columnwidth}{!}
{ 
\begin{tabular}{lcccc}
\toprule
\textbf{Dataset} & \textbf{Framework} & \textbf{Oracle Calls}~$\downarrow$ & \textbf{Proxy Calls}~$\downarrow$ & \textbf{Speedup}~$\uparrow$ \\
\midrule
\multirow{2}{*}{eICU} 
    & Brute-Force  & 8,234 & \textbf{0}    & 1.0× \\
    & \sprint & \textbf{600}  & 1,000 & \textbf{7.5×} \\
\midrule
\multirow{2}{*}{MIMIC-III} 
    & Brute-Force  & 4,245 & \textbf{0}    & 1.0× \\
    & \sprint & \textbf{150}  & 500 & \textbf{10.6×} \\
\midrule
\multirow{2}{*}{Jigsaw} 
    & Brute-Force  & 16,225 & \textbf{0}    & 1.0× \\
    & \sprint & \textbf{600}  & 1,000 & \textbf{14.8×} \\
\midrule
\multirow{2}{*}{Yelp} 
    & Brute-Force  & 6,990,280 & \textbf{0}    & 1.0× \\
    & \sprint & \textbf{20,000}  & 35,000 & \textbf{186.4×} \\
\midrule
\multirow{2}{*}{Amazon-E} 
    & Brute-Force  & 10,000 & \textbf{0}    & 1.0× \\
    & \sprint & \textbf{1,200}  & 2,000 & \textbf{4.5×} \\
\bottomrule
\end{tabular}
}
\end{table}

\begin{table}[t]
\centering
\caption{End-to-end cost (s). \textbf{Bold} indicates the best performance.}
\label{tab:end_to_end_cost}
\small
\resizebox{0.6\columnwidth}{!}
{ 
\begin{tabular}{l c c c c c}
\toprule
\textbf{Dataset} & \textbf{Framework} & \makecell{\textbf{Embedding}\\ \textbf{Generation}\\ \textbf{Time (s)}~$\downarrow$} & \makecell{\textbf{Query}\\ \textbf{Execution}\\ \textbf{Time (s)}~$\downarrow$} & \makecell{\textbf{Total}\\\textbf{Time (s)}~$\downarrow$} & \textbf{Speedup}~$\uparrow$ \\
\midrule
\multirow{2}{*}{Amazon-E} 
    & Brute-Force & 10.88 & \textbf{0.16} & 11.04 & 1.0× \\
    & \sprint     & \textbf{3.01}  & 0.94 & \textbf{3.95} & \textbf{2.8×} \\
\midrule
\multirow{2}{*}{Yelp} 
    & Brute-Force & 8996.49 & \textbf{0.16} & 8996.65 & 1.0× \\
    & \sprint     & \textbf{65.54}  & 207.84 & \textbf{273.38} & \textbf{32.9×} \\
\bottomrule
\end{tabular}
}
\end{table}

\subsubsection{End-to-End Runtime}\label{exp:framework:end_to_end_cost}
We measure end-to-end latency by running both frameworks on Yelp and Amazon-E, the two datasets for which both oracle and proxy models are executable on GPU. Table~\ref{tab:end_to_end_cost} shows that \sprint reduces total query latency by $32.9\times$ on Yelp and $2.8\times$ on Amazon-E. While \sprint's query execution time is higher due to its additional logic, this overhead is negligible compared to the savings from reducing the cost of oracle embedding generation, which dominates the total cost when data evolves and the embeddings must be refreshed on query time.

\subsubsection{Impact of Oracle and Proxy Choices.}\label{exp:framework:oracle_proxy_choices}
To evaluate the robustness of \sprint under different choices of oracle and proxy models, we conduct a sensitivity analysis on Amazon-E with the following three model pairs: 
\begin{enumerate}
    \item \textbf{MiniLM-L12 (oracle)} and \textbf{MiniLM-L6 (proxy)};
    \item \textbf{MiniLM-L6 (oracle)} and \textbf{MiniLM-L3 (proxy)};
    \item \textbf{MiniLM-L12 (oracle)} and \textbf{MiniLM-L3 (proxy)}.
\end{enumerate} 

Since all variants use the same default sample and pilot sample sizes under \sprint, their end-to-end speedup over Brute-Force remains comparable (see Table~\ref{tab:end_to_end_cost}). In terms of aggregation error, Fig.~\ref{fig:proxy_oracle_variants} shows that for \(\mathtt{PCT}\) and \(\mathtt{SUM}\) queries, our proposed methods \sprintc and Two-Phase consistently outperform baselines across all datasets. Interestingly, we observe that \sprintc performs nearly identically to Two-Phase in most cases. This similarity arises because, in these datasets, the attribute values within neighborhoods exhibit low variance, making the benefit over \sprintc from the second-phase refinement of the precision target in Two-Phase modest.
For value-sensitive queries like \(\mathtt{AVG}\) and \(\mathtt{VAR}\), different oracle-proxy variants may lead to varying \sprintv performance. Particularly, \sprintv outperforms all baselines when the oracle is MiniLM-L12 and the proxy is MiniLM-L6. However, precision-targeted methods (e.g., PQE-PT and SUPG-PT) perform better than \sprintv for other oracle-proxy pairs. We attribute this to the highly concentrated value distribution of Amazon-E star ratings. When attribute values are similar in the neighborhood, even a small, high-precision set can provide accurate estimates. This highlights an important insight: the optimal neighbor selection strategy for AQNNs may depend not only on the aggregation function but also on the distribution of neighbor attribute values. We defer a deeper analysis of value-sensitive AQNNs and the impact of query targets on aggregation error to \S~\ref{exp:error}-A.


\begin{figure*}[!t]
    \centering
    \begin{subfigure}{0.48\linewidth}
        \centering
        \includegraphics[width=0.8\linewidth]{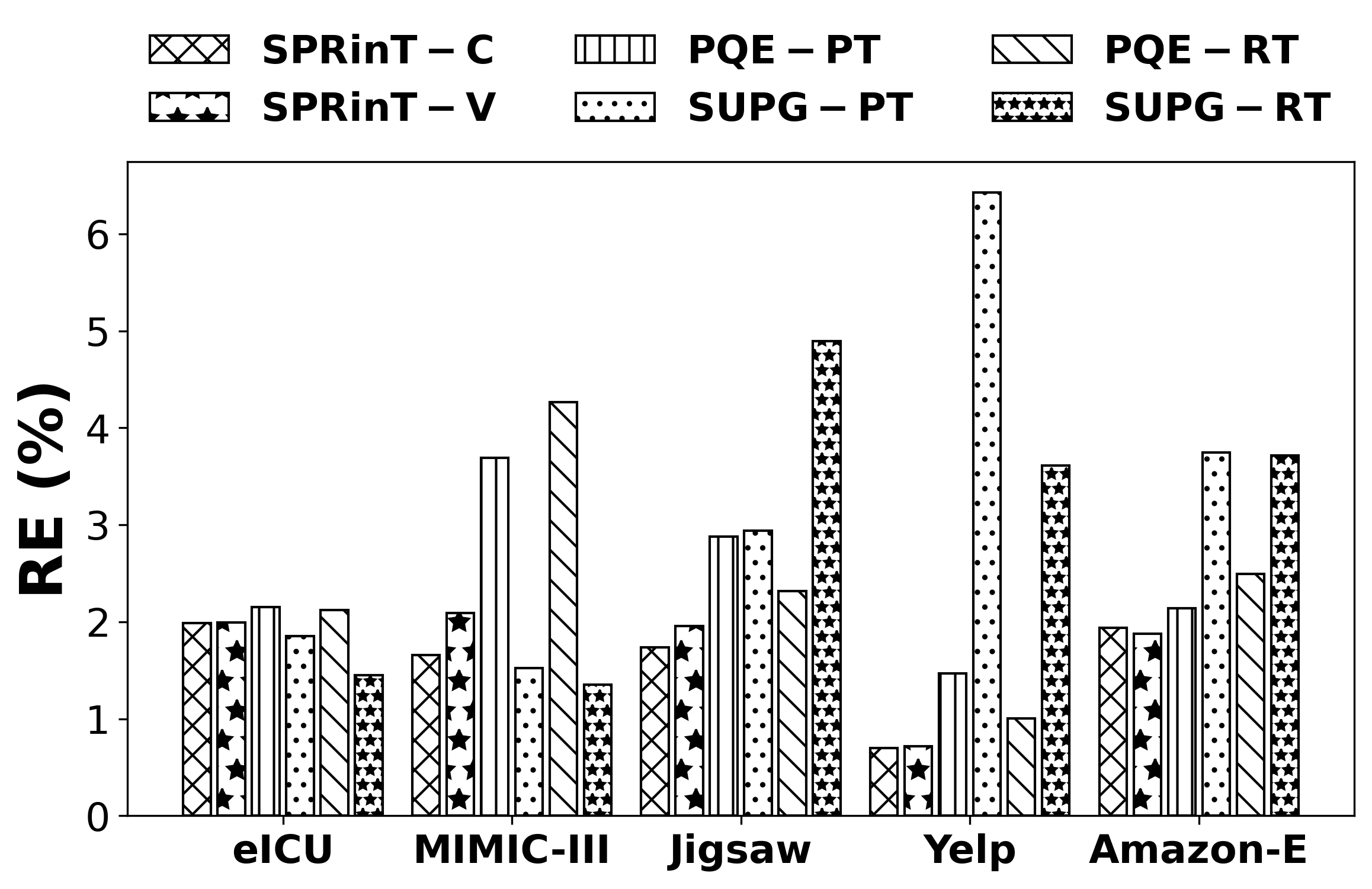}
        \caption{value-sensitive: \(\mathtt{AVG}\)}
        \label{fig:RE_PQE_AVG}
    \end{subfigure}
    \hfill
    \begin{subfigure}{0.48\linewidth}
        \centering
        \includegraphics[width=0.8\linewidth]{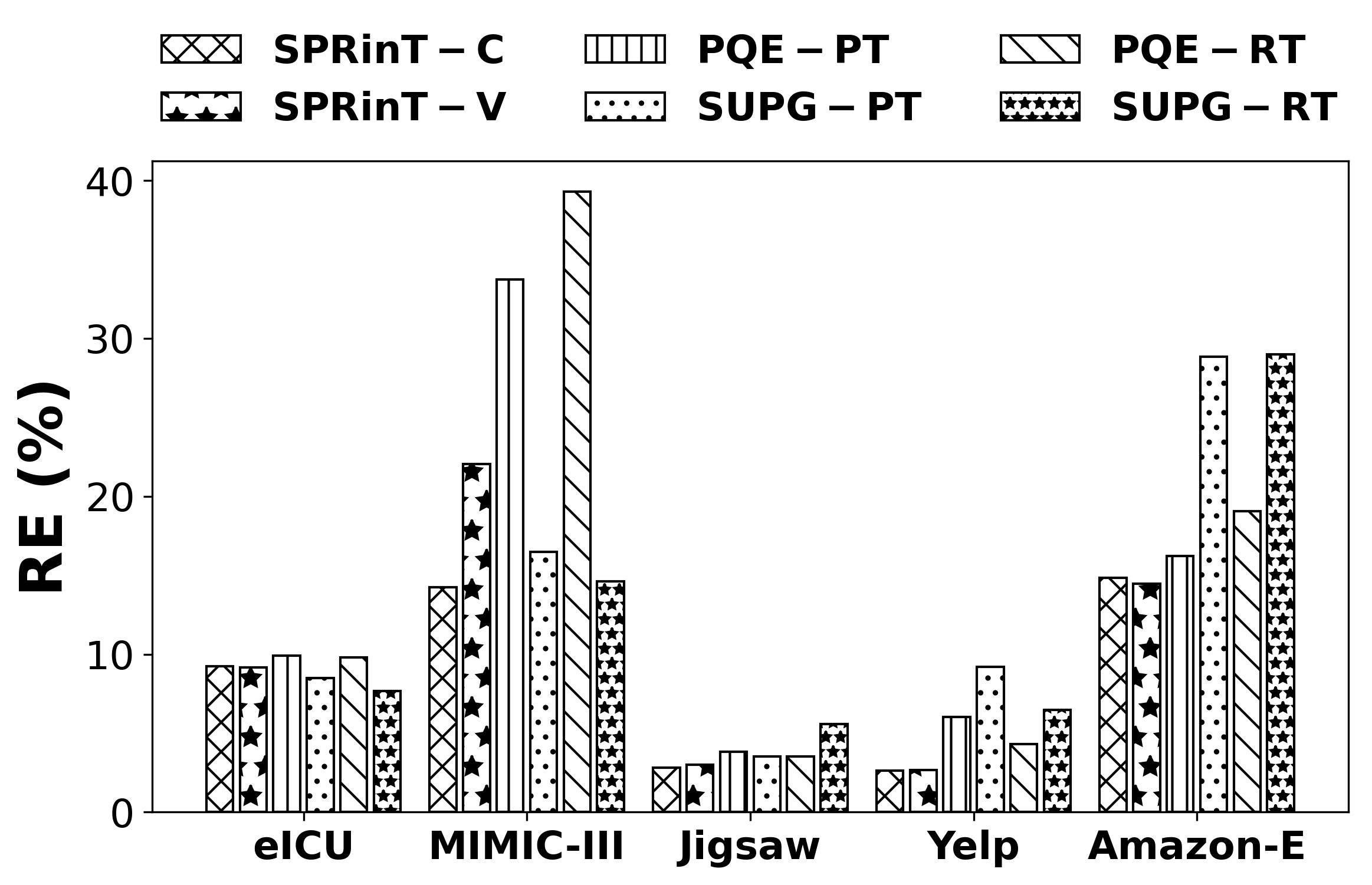}
        \caption{value-sensitive: \(\mathtt{VAR}\)}
        \label{fig:RE_PQE_VAR}
    \end{subfigure}
    \centering
    \begin{subfigure}{0.48\linewidth}
        \centering
        \includegraphics[width=0.8\linewidth]{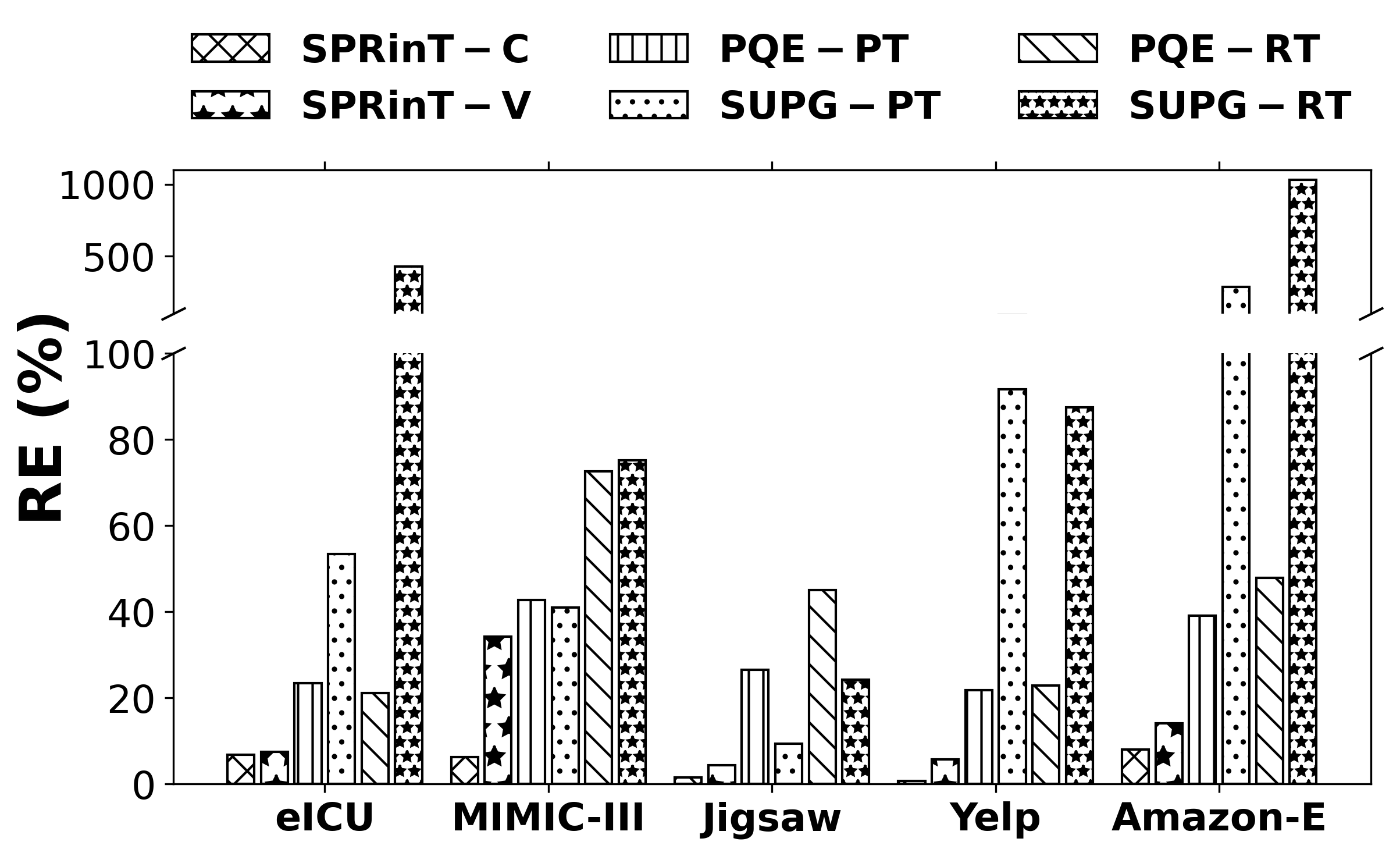}
        \caption{count-sensitive: \(\mathtt{PCT}\)}
        \label{fig:RE_PQE_PCT}
    \end{subfigure}
    \hfill
    \begin{subfigure}{0.48\linewidth}
        \centering
        \includegraphics[width=0.8\linewidth]{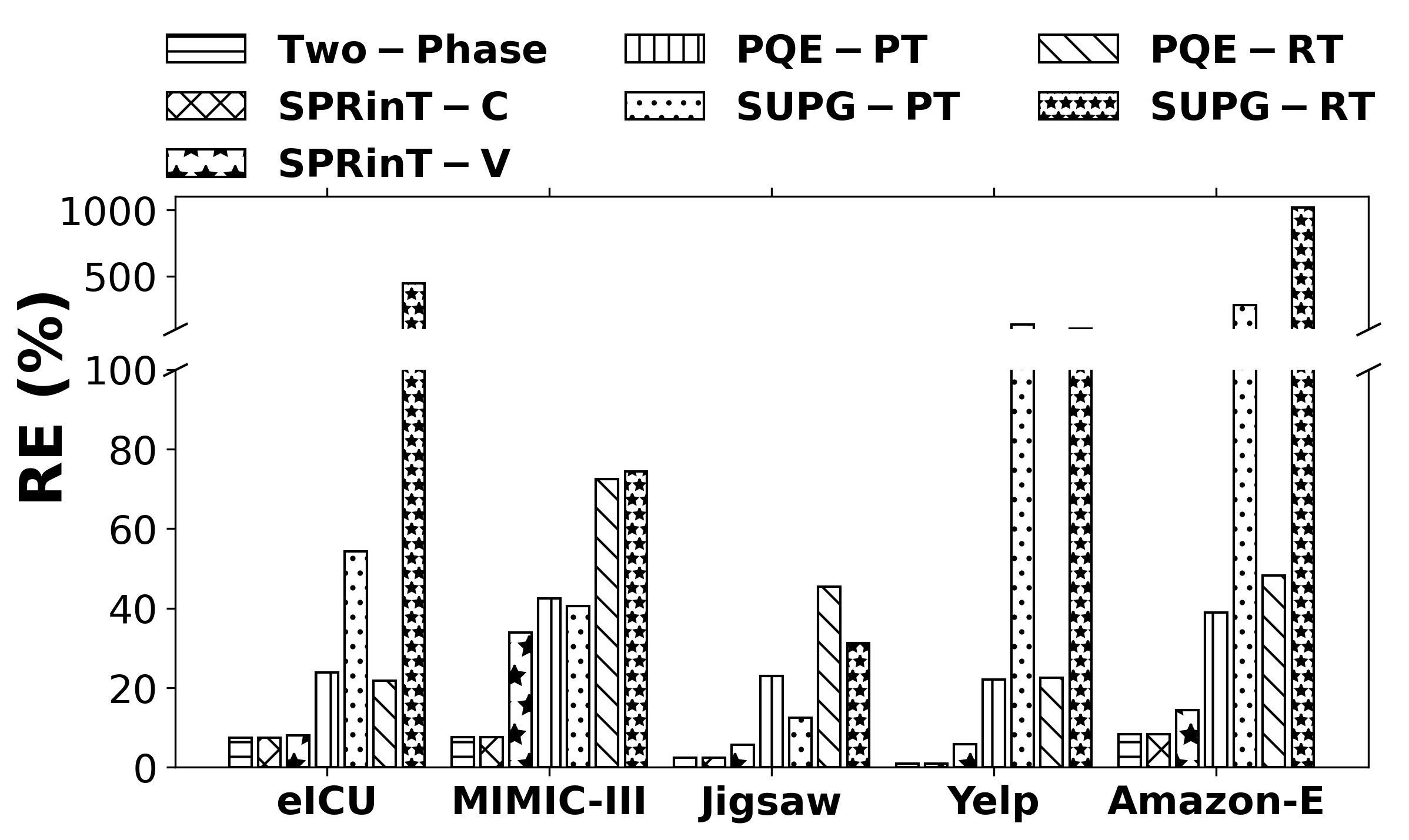}
        \caption{value- and count-sensitive: \(\mathtt{SUM}\)}
        \label{fig:RE_PQE_SUM}
    \end{subfigure}
    \caption{RE performance for different AQNN queries.}
    \label{fig:RE}
\end{figure*}

\subsection{Algorithm-Level Evaluation}\label{exp:algorithm}
We compare the aggregation error in terms of RE and query execution runtime of our proposed neighbor selection algorithms, including \sprintv, \sprintc, and Two-Phase, against ANN baselines in \S~\ref{exp:baseline} within the same \sprint framework. Also, we examine scalability, the impact of sample and pilot sample sizes, and the effect of query radius on RE.

\subsubsection{Examining Aggregation Error}\label{exp:error}
Fig.~\ref{fig:RE} reports the RE of all algorithms except Top-K, which is compared separately in Table~\ref{tab:topk}. This is for a fair comparison as Top-K dynamically incurs oracle calls and does not follow a fixed oracle budget.


\stitle{A. Value-sensitive AQNNs.}
In Fig.~\ref{fig:RE_PQE_AVG} and ~\ref{fig:RE_PQE_VAR}, on \(\mathtt{AVG}\) and \(\mathtt{VAR}\), \sprintv performs best on Amazon-E. On other datasets, recall-targeted methods (e.g., SUPG-RT, PQE-RT) occasionally outperform \sprintv. To explain this, we compare the quality of neighbor selection using \(\text{F1}_S\) scores (Table~\ref{tab:f1_score_AVG}). SUPG-RT and PQE-RT retrieve low-quality neighbors yet achieve lower RE. For instance, on eICU, SUPG-RT attains the lowest \(\text{F1}_S\) but surprisingly also the lowest RE. Further investigation reveals that the query target chosen and the target attribute value distribution in its neighborhood can play a decisive role here. 

\begin{table}[t]
\caption{\(\text{F1}_S\) performance (\(\uparrow\)) for \(\mathtt{AVG}\) and \(\mathtt{VAR}\). \textbf{Bold} indicates the best per dataset.}
\small
\label{tab:f1_score_AVG}
\resizebox{0.7\columnwidth}{!}
{%
\begin{tabular}{lcccccc}
\toprule
\textbf{Dataset}& \textbf{\sprintv} & \textbf{\sprintc} & \textbf{PQE-PT} & \textbf{PQE-RT} & \textbf{SUPG-PT} & \textbf{SUPG-RT} \\ \midrule
eICU        & \textbf{0.95} & \textbf{0.95} & 0.87   & 0.84   & 0.89    & 0.57    \\ 
MIMIC-III   & \textbf{0.82} & 0.79          & 0.67   & 0.81   & 0.43    & 0.75    \\ 
Jigsaw    & \textbf{0.96} & \textbf{0.96} & 0.81   & 0.95   & 0.69    & 0.89    \\ 
Yelp        & \textbf{0.93} & \textbf{0.93} & 0.89   & 0.83   & 0.83    & 0.80    \\ 
Amazon-E & 0.84          & \textbf{0.85} & 0.79   & 0.47   & 0.68    & 0.30    \\ 
\bottomrule
\end{tabular}
}
\end{table}

\begin{figure}[t]
    \centering
    \begin{subfigure}{0.48\linewidth}
        \centering
        \includegraphics[width=0.8\linewidth]{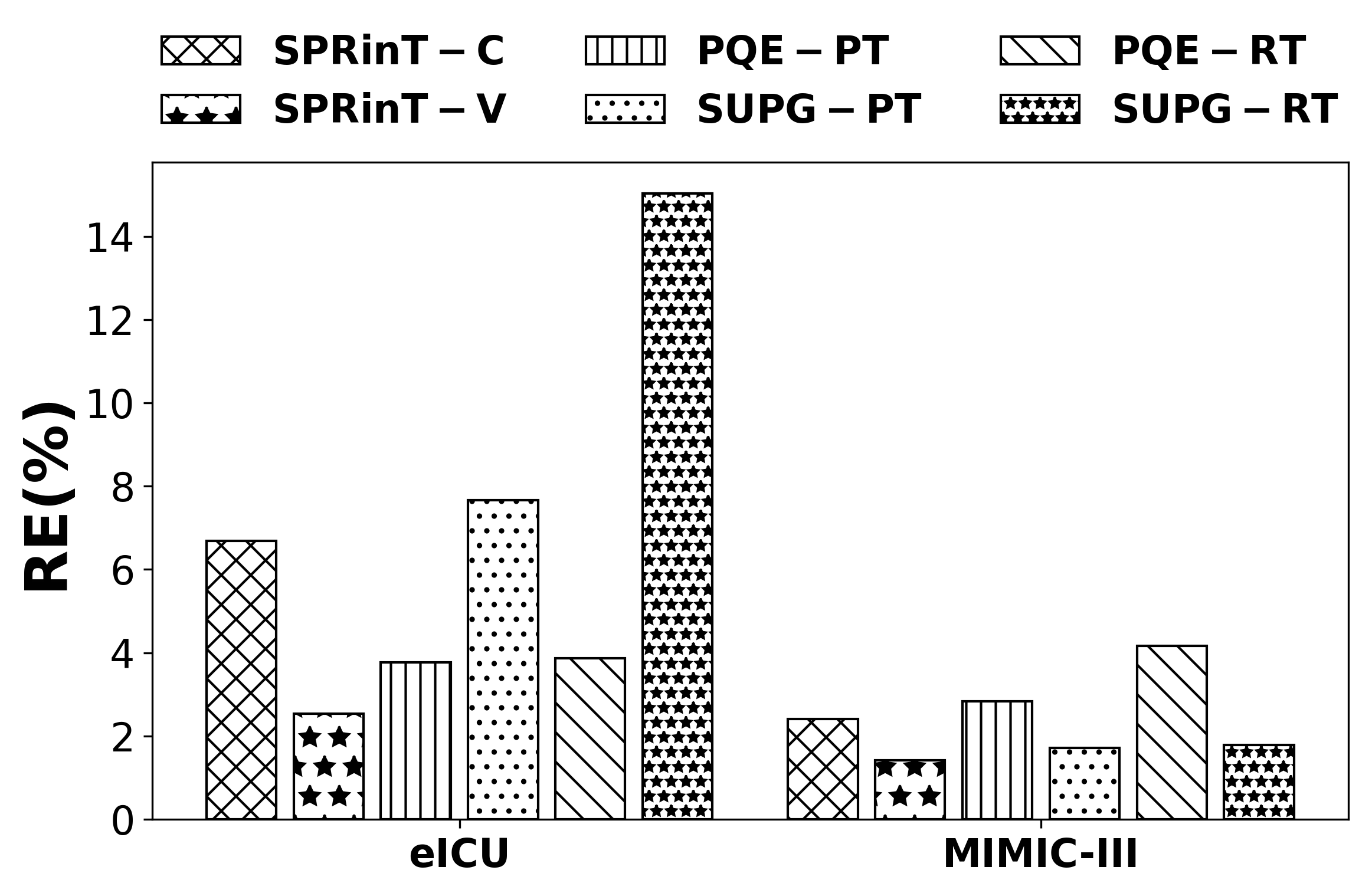}
        \caption{value-sensitive: \(\mathtt{AVG}\)}
        \label{fig:RE_PQE_AVG_picked}
    \end{subfigure}
    \hfill
    \begin{subfigure}{0.48\linewidth}
        \centering
        \includegraphics[width=0.8\linewidth]{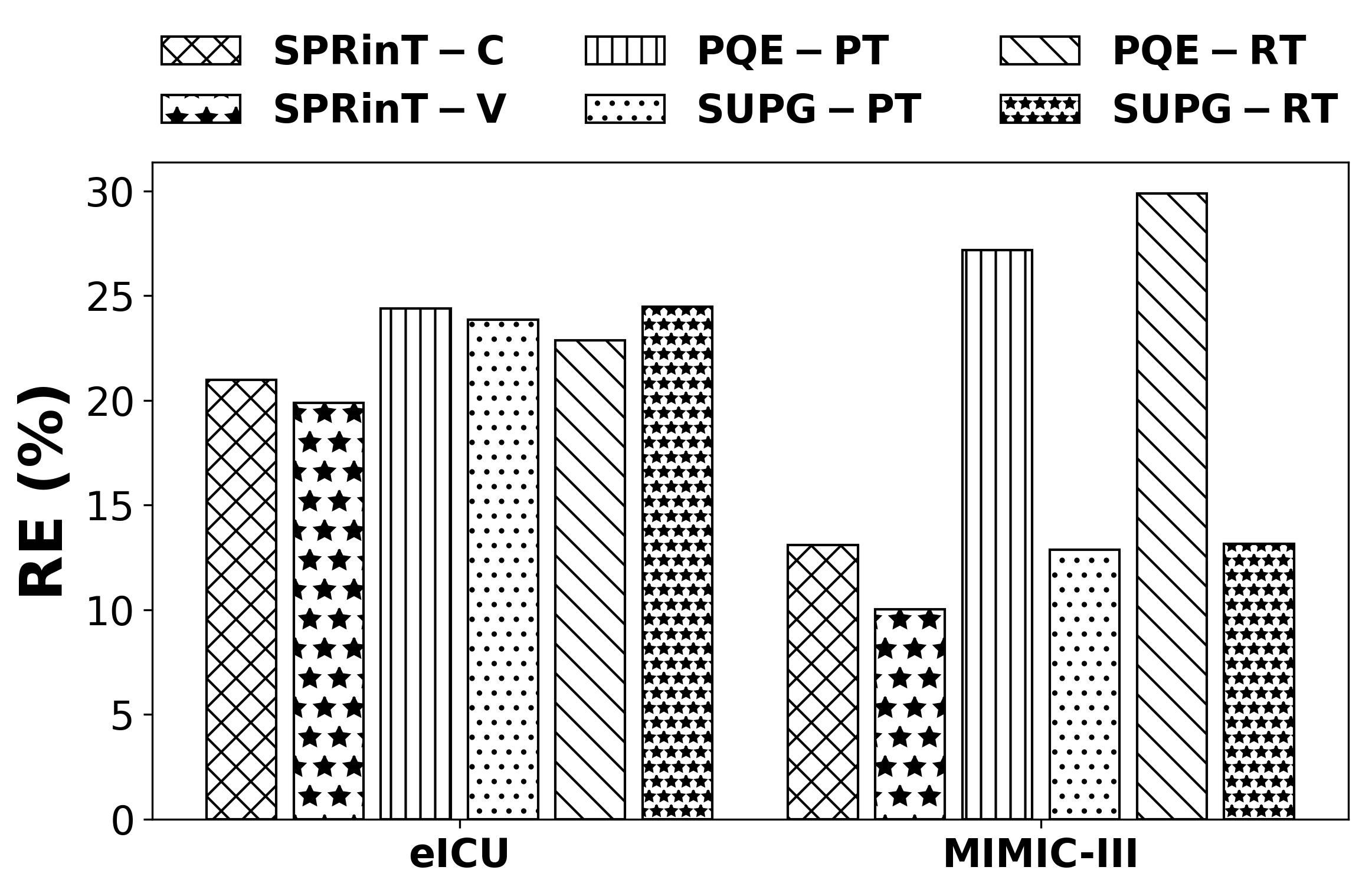}
        \caption{value-sensitive: \(\mathtt{VAR}\)}
        \label{fig:RE_PQE_VAR_picked}
    \end{subfigure}
    \caption{RE performance for representative target queries.}
    \label{fig:RE_picked}
\end{figure}

\noindent \textit{Impact of query targets.}
Value-sensitive queries are indeed a bit nuanced. When a query target's NNs have a target attribute distribution that differs from the global distribution, \sprintv consistently outperforms baselines (Fig. ~\ref{fig:RE_picked} shows representative queries from eICU and MIMIC-III). The rationale is that algorithms that achieve poorer F1 score will end up picking up ``neighbors'' which are less representative of the true neighborhood and as a result perform the subsequent aggregation on an attribute value distribution dissimilar to the neighborhood's distribution. In contrast, if the neighborhood and the global distributions are similar, the impact of a poor F1 score diminishes: this is demonstrated in the various cases in Fig. ~\ref{fig:RE} and ~\ref{fig:proxy_oracle_variants}. Indeed, in such cases, NN search may be unnecessary in the first place: directly sampling from the global distribution would already give accurate approximations of aggregates, while \sprintv may incur unnecessary overhead. A practitioner needs to know under what conditions computing AQNNs using a sophisticated approach like \sprintv is warranted. To help with this decision, we recommend first using pure proxy embeddings to compute global and neighborhood aggregates. If the difference exceeds a meaningful threshold, i.e., 10\%, selective oracle refinement using \sprint framework can significantly improve RE. Otherwise, proxy-based aggregates suffice.

\begin{table}[t]
\caption{\(|\text{P}_S-\text{R}_S|\) performance (\(\downarrow\)) for \(\mathtt{PCT}\). \textbf{Bold} indicates the best per dataset.}
\label{tab:PR_Difference_PCT}
\small
\resizebox{0.7\columnwidth}{!}
{%
\begin{tabular}{lcccccc}
\toprule
\textbf{Dataset} & \textbf{\sprintc} & \textbf{\sprintv} & \textbf{PQE-PT} & \textbf{PQE-RT} & \textbf{SUPG-PT} & \textbf{SUPG-RT} \\ \midrule
eICU        & \textbf{0.02} & 0.03     & 0.21   & 0.22   & 0.18    & 0.51    \\ 
MIMIC-III   & \textbf{0.04} & 0.22     & 0.39   & 0.25   & 0.72    & 0.38    \\ 
Jigsaw    & \textbf{0.01} & 0.04     & 0.26   & 0.08   & 0.45    & 0.15    \\ 
Yelp        & \textbf{0.00} & 0.05     & 0.16   & 0.24   & 0.23    & 0.25    \\ 
Amazon-E & \textbf{0.05} & 0.09     & 0.23   & 0.68   & 0.48    & 0.81    \\ 
\bottomrule
\end{tabular}
}
\end{table}

\begin{table}[t]
\caption{\(\text{F1}_S\) and \(|\text{P}_S-\text{R}_S|\) performance for \(\mathtt{SUM}\). \textbf{Bold} indicates the best per dataset.}
\label{tab:sum_f1_pr}
\small
\resizebox{0.8\columnwidth}{!}
{%
\begin{tabular}{lccccccc}
\toprule
\textbf{Dataset} & \textbf{Two-Phase} & \textbf{\sprintc} & \textbf{\sprintv} & \textbf{PQA-PT} & \textbf{PQA-RT} & \textbf{SUPG-PT} & \textbf{SUPG-RT} \\
\midrule
\multicolumn{8}{l}{\textbf{\(\text{F1}_S\) (\(\uparrow\))}} \\
eICU        & \textbf{0.95} & \textbf{0.95} & \textbf{0.95} & 0.87 & 0.84 & 0.89 & 0.59 \\
MIMIC-III   & 0.79          & 0.79          & \textbf{0.82} & 0.67 & 0.81 & 0.43 & 0.75 \\
Jigsaw      & \textbf{0.96} & \textbf{0.96} & \textbf{0.96} & 0.81 & 0.95 & 0.69 & 0.89 \\
Yelp        & \textbf{0.93} & \textbf{0.93} & \textbf{0.93} & 0.89 & 0.83 & 0.83 & 0.80 \\
Amazon-E    & \textbf{0.85} & \textbf{0.85} & 0.84          & 0.79 & 0.47 & 0.68 & 0.30 \\
\midrule
\multicolumn{8}{l}{\textbf{\(|\text{P}_S - \text{R}_S|\) (\(\downarrow\))}} \\
eICU        & \textbf{0.82} & \textbf{0.82} & 0.03 & 0.21 & 0.22 & 0.18 & 0.51 \\
MIMIC-III   & \textbf{0.04} & \textbf{0.04} & 0.22 & 0.39 & 0.25 & 0.72 & 0.38 \\
Jigsaw      & \textbf{0.01} & \textbf{0.01} & 0.04 & 0.26 & 0.08 & 0.45 & 0.15 \\
Yelp        & \textbf{0.00} & \textbf{0.00} & 0.05 & 0.16 & 0.24 & 0.23 & 0.25 \\
Amazon-E    & \textbf{0.05} & \textbf{0.05} & 0.09 & 0.23 & 0.68 & 0.48 & 0.81 \\
\bottomrule
\end{tabular}
}
\end{table}

\stitle{B. Count-sensitive AQNNs.} Next we turn to count-sensitive queries. We report \(\mathtt{PCT}\) results only, as \(\mathtt{COUNT}\) exhibits similar performance trends due to their proportional relationship.
As shown in Fig.~\ref{fig:RE_PQE_VAR}, \sprintc consistently achieves the lowest RE across all datasets, empirically corroborating Theorem~\ref{thm:sprint-c}. \sprintv consistently ranks second, which implies that maximizing the F1 score can be beneficial for count-sensitive queries, though it is less effective than directly controlling the neighbor count. Other baselines perform significantly worse. Particularly, the recall-targeted methods (SUPG-RT and PQE-RT) attain the highest RE because they tend to retrieve overly many false neighbors. These findings are consistent with the precision-recall gap observed in Table~\ref{tab:PR_Difference_PCT}. 

\stitle{C. Both value- and count-sensitive AQNNs.}  
Fig.~\ref{fig:RE_PQE_SUM} reports the RE for \(\mathtt{SUM}\), which is sensitive to both neighbor counts and attribute values. Two-Phase strategy consistently achieves the lowest RE across all datasets, demonstrating its ability to handle dual sensitivities effectively. A similar trend is observed here: when attribute variance within neighborhoods is low, \sprintc performs comparably to Two-Phase, consistent with our earlier observation in \S~\ref{exp:framework:oracle_proxy_choices}.
Again, we dig deeper for a better understanding by measuring the F1 score and precision-recall gaps. As shown in Table~\ref{tab:sum_f1_pr}, Two-Phase and \sprintc consistently attain the lowest \(|\text{P}_S-\text{R}_S|\) with high \(\text{F1}_S\) in most datasets. While \sprintc seems to suffice in these datasets, Two-Phase becomes critical in settings where attribute distributions are highly skewed or contain extreme outliers, as it provides an additional refinement step to control value errors.

\stitle{D. Comparison with Top-K.}
Tables~\ref{tab:topk} compares the performance of our proposed algorithms against Top-K. On all datasets, our proposed algorithms achieve comparable RE to Top-K, i.e., less than 5\% difference, on all aggregation functions, and even slightly outperform Top-K on MIMIC-III for \(\mathtt{AVG}\) and \(\mathtt{VAR}\). 


\begin{table}[t]
\caption{RE and cost comparison with Top-K. The algorithm ``ours'' refers to \sprintv for \(\mathtt{AVG}\) and \(\mathtt{VAR}\), \sprintc for \(\mathtt{PCT}\), and Two-Phase for \(\mathtt{SUM}\). \textbf{Bold} indicates the best performance.}
\label{tab:topk}
\centering
\small
\resizebox{0.62\columnwidth}{!}
{ 
\begin{tabular}{lccccccc}
\toprule
\textbf{Dataset} & \textbf{Algorithm} & 
\shortstack{\textbf{Oracle} \\ \textbf{Calls}~$\downarrow$} & 
\shortstack{\textbf{Query} \\ \textbf{Execution} \\ \textbf{Cost (s)}~$\downarrow$} & 
\shortstack{\textbf{AVG} \\ \textbf{RE}~$\downarrow$} & \shortstack{\textbf{VAR} \\ \textbf{RE}~$\downarrow$} & \shortstack{\textbf{PCT} \\ \textbf{RE}~$\downarrow$} & \shortstack{\textbf{SUM} \\ \textbf{RE}~$\downarrow$} \\
\midrule
\multirow{2}{*}{eICU} 
    & Top-K  & 991 & 47.71    & 2.01  & \textbf{8.77} & \textbf{6.15} & \textbf{6.82} \\
    & ours & \textbf{600}  & \textbf{0.24} & \textbf{1.99} & 9.15 & 6.80 & 7.37 \\
\midrule
\multirow{2}{*}{MIMIC-III} 
    & Top-K  & 500 & 1.77 & 1.60 & 14.80 & \textbf{2.87} & \textbf{4.98} \\
    & ours & \textbf{150}  & \textbf{0.09} & \textbf{1.44} & \textbf{12.92} & 6.21 & 7.55 \\
\midrule
\multirow{2}{*}{Jigsaw} 
    & Top-K  & 998 & 2.18    & \textbf{1.39} & \textbf{1.77} & \textbf{1.13} & \textbf{1.87}\\
    & ours & \textbf{600}  & \textbf{0.27} & 1.96 & 3.01 & 1.53 & 2.36\\
\midrule
\multirow{2}{*}{Yelp} 
    & Top-K  & 34,997 & 947.23 & \textbf{0.37} & \textbf{0.51} & \textbf{0.65} & 1.14 \\
    & ours & \textbf{20,000} & \textbf{125.34} & 0.71 & 2.67 & 0.77 & \textbf{0.96}\\
\midrule
\multirow{2}{*}{Amazon-E} 
    & Top-K  & 1,980 & 30.05   & \textbf{1.70} & \textbf{13.14} & \textbf{5.28} & \textbf{5.47} \\
    & ours & \textbf{1,200}  & \textbf{0.94} & 1.87 & 14.47 & 8.02 & 8.30 \\
\bottomrule
\end{tabular}
}
\end{table}

\subsubsection{Examining Cost}\label{exp:cost}
We first analyze Top-K. In Table~\ref{tab:topk}, Top-K incurs more oracle calls, leading to higher embedding generation cost, and takes significantly higher query execution runtime than \sprint. While it sometimes yields lower RE, this comes at the expense of considerably more computational overhead. For other baselines executed within the \sprint framework, Table~\ref{tab:runtime} shows the query execution cost across datasets. By comparing query execution cost only, we simulate settings with static or infrequently updated data, where oracle and proxy embeddings can be precomputed, cached, and reused for later queries. This is the \textit{least favorable setting} for our framework, yet our algorithms remain competitive with baselines in most cases, except on Yelp, whose large sample size (see Table~\ref{tab:default_proxy_oracle_calls}) leads to higher runtimes.

Overall, though our proposed algorithms may sometimes spend more time on NN selection, \sprint, analyzed in \S~\ref{exp:framework}, saves a significant amount of embedding generation cost and thus reduces end-to-end cost, demonstrating the efficiency of our solution.




\begin{table}[t] 
\caption{Query execution cost (s). \textbf{Bold} indicates the lowest cost per dataset.}
\small 
\label{tab:runtime} 
\resizebox{0.7\columnwidth}{!} {%
\begin{tabular}{lcccccc}
\toprule
\textbf{Dataset} & \textbf{\sprintv} & \textbf{\sprintc} & \textbf{PQE-PT} & \textbf{PQE-RT} & \textbf{SUPG-PT} & \textbf{SUPG-RT} \\ \midrule 
eICU & 0.24 & 0.09 & \textbf{0.01} & 0.09 & 0.02 & 0.02 \\ 
MIMIC-III & 0.09 & 0.03 & \textbf{0.00} & 0.02 & 0.01 & 0.01 \\  
Jigsaw & 0.27 & 0.14 & 0.06 & 0.11 & \textbf{0.02} & \textbf{0.02} \\ 
Yelp & 207.84 & 119.12 & 9.07 & 120.42 & 0.45 & \textbf{0.44} \\ 
Amazon-E & 0.94 & 0.54 & 0.12 & 0.42 & \textbf{0.02} & \textbf{0.02} \\ \bottomrule 
\end{tabular} 
} 
\end{table}


\subsubsection{Examine Scalability}
\begin{figure}[!t]
    \centering
    \begin{subfigure}{0.48\linewidth}
        \centering
        \includegraphics[width=0.8\linewidth]{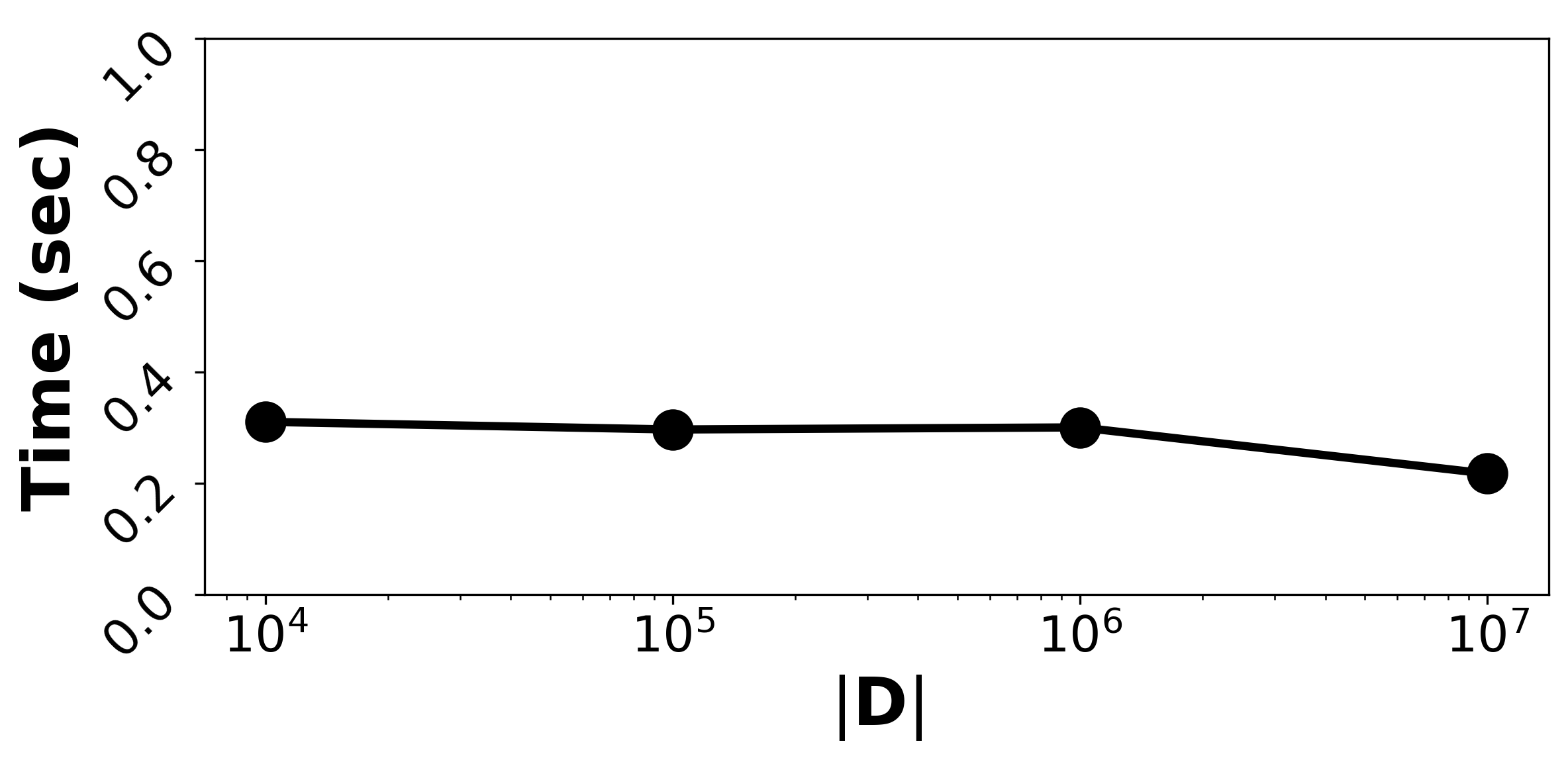}
        \caption{\sprintv for \(\mathtt{AVG}\)}
        \label{fig:avg_yelp_scalability}
    \end{subfigure}
    \hfill
    \begin{subfigure}{0.48\linewidth}
        \centering
        \includegraphics[width=0.8\linewidth]{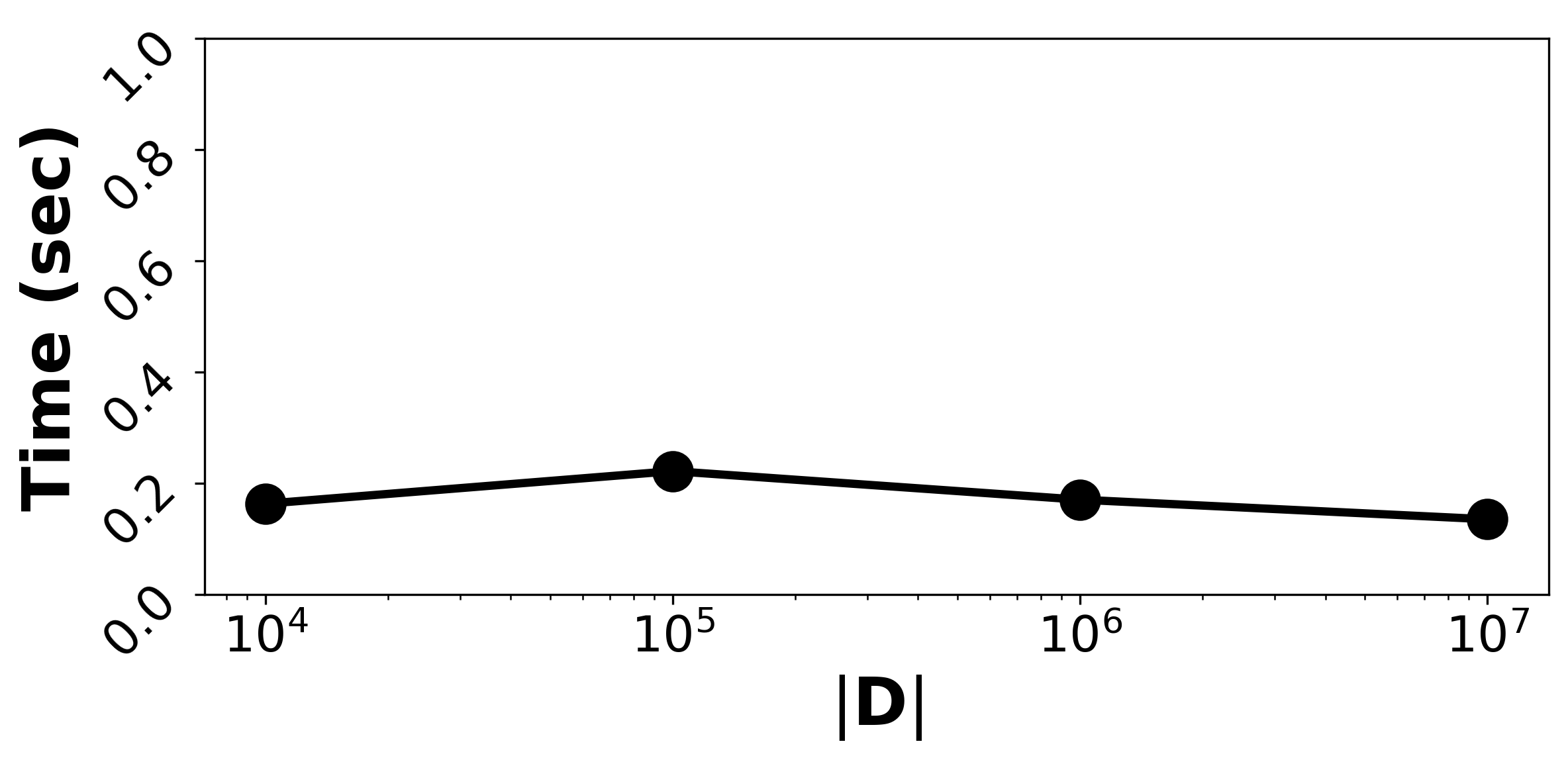}
        \caption{\sprintc for \(\mathtt{PCT}\)}
        \label{fig:pct_yelp_scalability}
    \end{subfigure}
    \centering
    \begin{subfigure}{0.48\linewidth}
        \centering
        \includegraphics[width=0.8\linewidth]{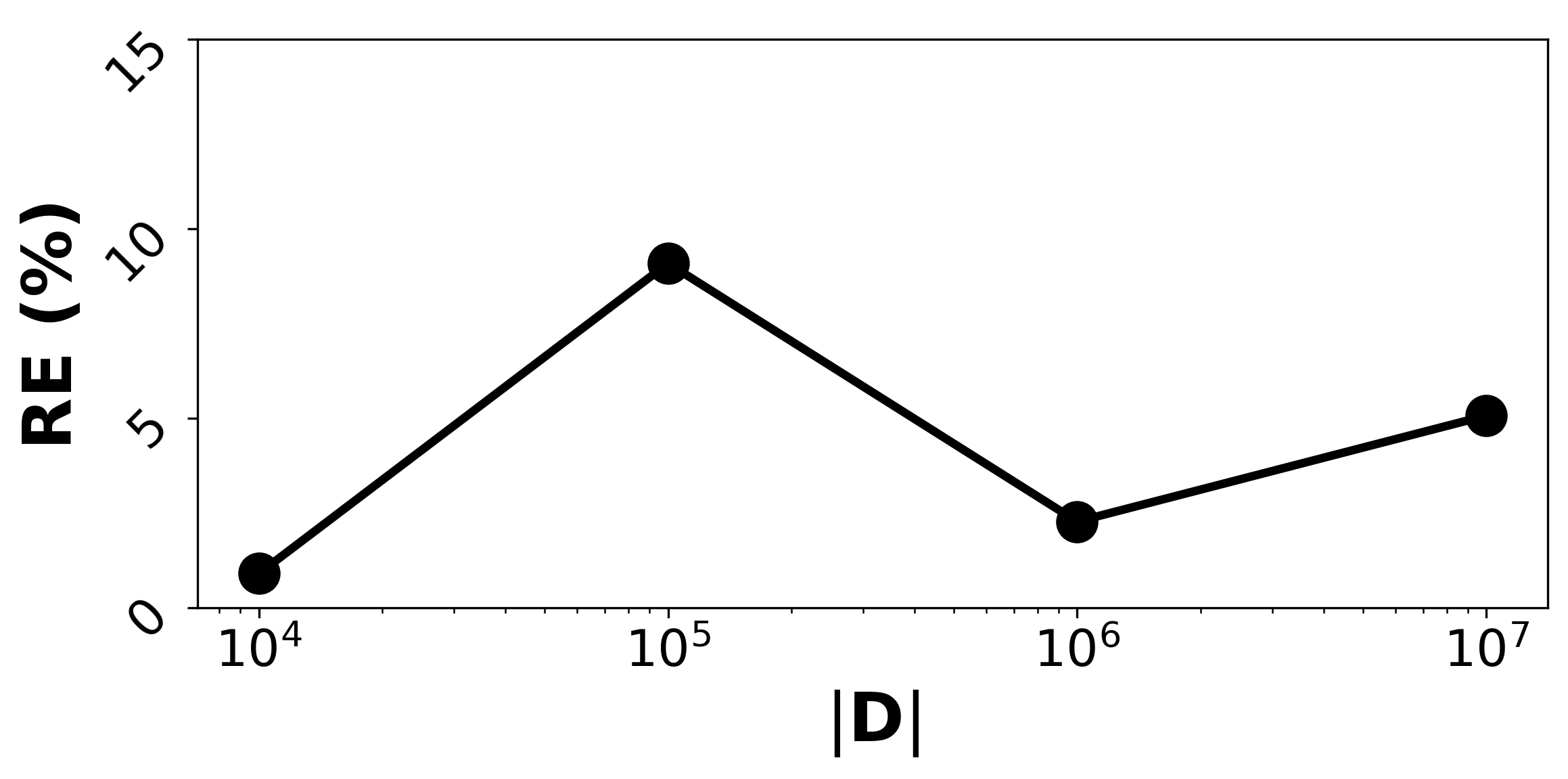}
        \caption{\sprintv for \(\mathtt{AVG}\)}
        \label{fig:avg_yelp_scalability_re}
    \end{subfigure}
    \hfill
    \begin{subfigure}{0.48\linewidth}
        \centering
        \includegraphics[width=0.8\linewidth]{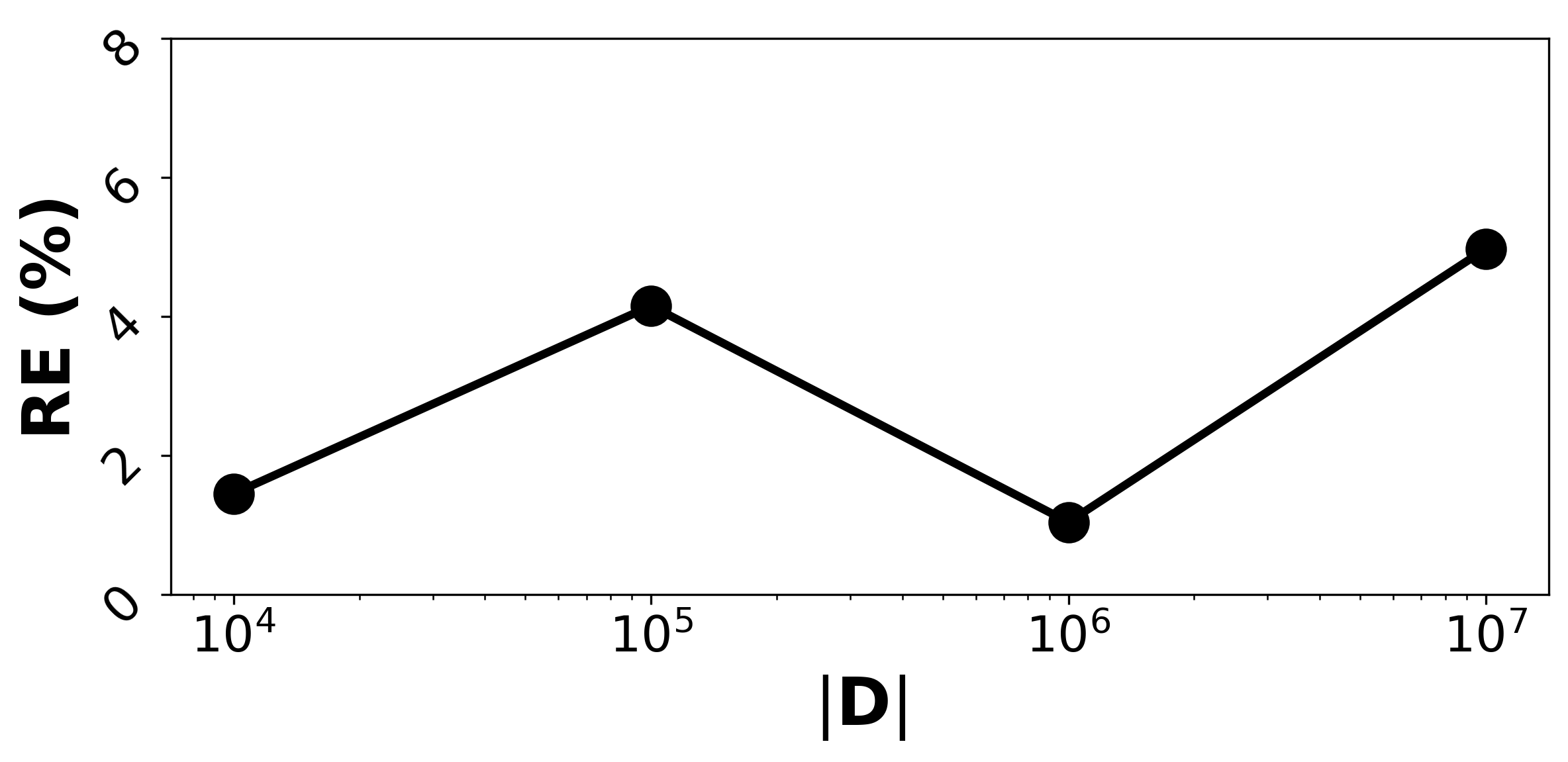}
        \caption{\sprintc for \(\mathtt{PCT}\)}
        \label{fig:pct_yelp_scalability_re}
    \end{subfigure}
    \caption{Impact of dataset size \(|D|\) on the execution time and RE using \sprintv for \(\mathtt{AVG}\) (a, c) and \sprintc for \(\mathtt{PCT}\) (b, d) on the Yelp dataset.}
    \label{fig:scalability}
\end{figure}

We evaluate scalability using a semi-synthetic version of the Yelp dataset. Specifically, we start with a sampled subset of \(|D|=10^4\) objects and scale \(|D|\) up to \(10^7\) by duplicating and slightly perturbing records to preserve the original attribute value distribution. The sample size \(s=1000\) and pilot sample size \(s_p=600\) remain fixed as \(|D|\) increases. Fig. ~\ref{fig:avg_yelp_scalability} and~\ref{fig:pct_yelp_scalability} report runtime. Fig. ~\ref{fig:avg_yelp_scalability_re} and~\ref{fig:pct_yelp_scalability_re} show RE performance.

Results confirm that both \sprintv and \sprintc scale very well w.r.t. \(|D|\), validating that our sampling-based framework’s cost is independent of the dataset size. The RE also remains fairly stable. The slight fluctuations are due to changes in neighborhood density. This aligns with our theoretical analysis: once \(s\) and \(s_p\) are sufficiently large, the approximation error and computational cost remain relatively unaffected by \(|D|\). 


\begin{figure}[!t]
    \centering
    \begin{subfigure}{0.48\linewidth}
        \centering
        \includegraphics[width=0.8\linewidth]{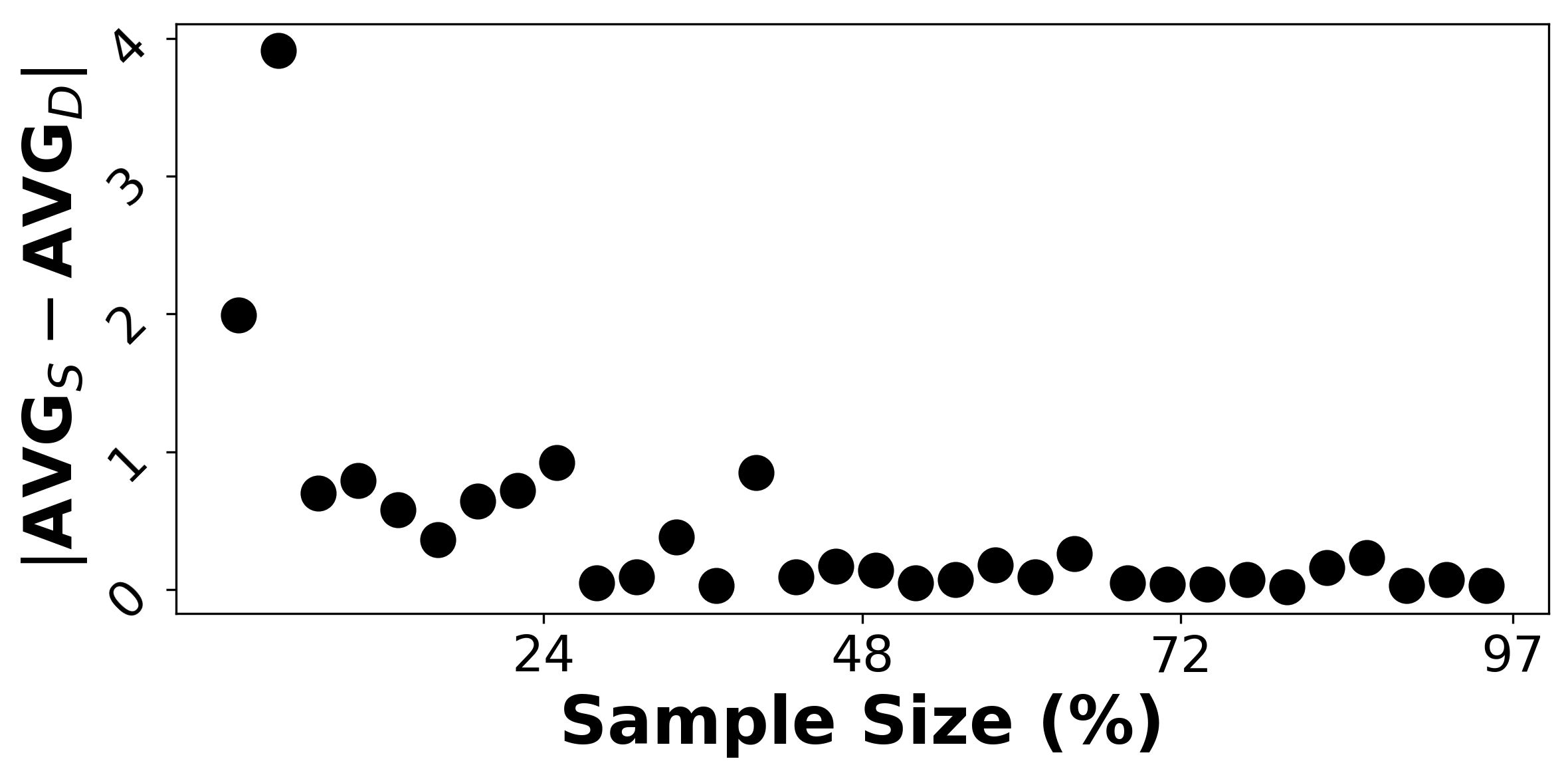}
        \caption{\(s\) vs \(\mathtt{AVG}_S\)}
        \label{fig:avg_s}
    \end{subfigure}
    \hfill
    \begin{subfigure}{0.48\linewidth}
        \centering
        \includegraphics[width=0.8\linewidth]{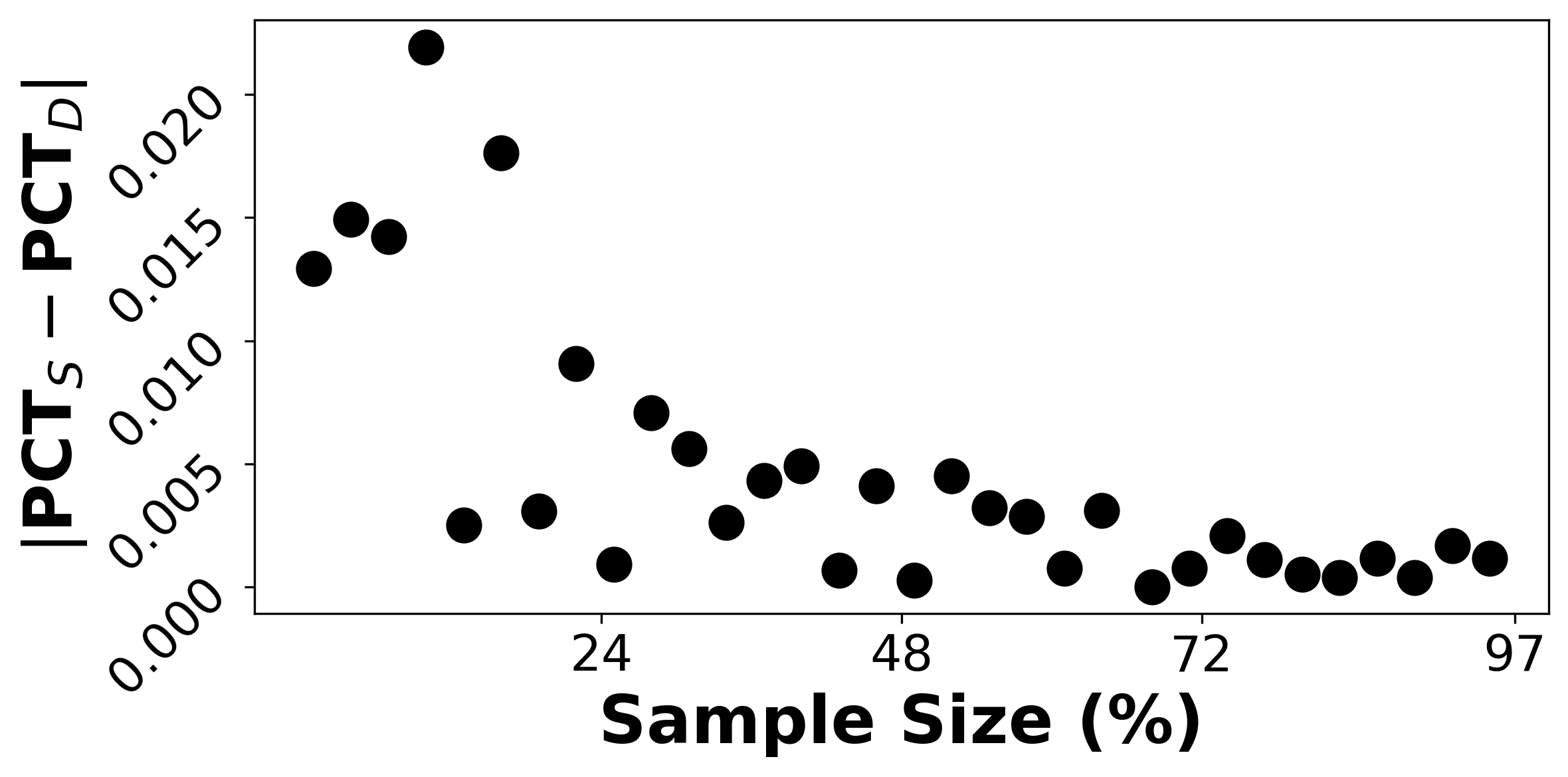}
        \caption{\(s\) vs \(\mathtt{PCT}_S\)}
        \label{fig:pct_s}
    \end{subfigure}
    \centering
    \begin{subfigure}{0.48\linewidth}
        \centering
        \includegraphics[width=0.8\linewidth]{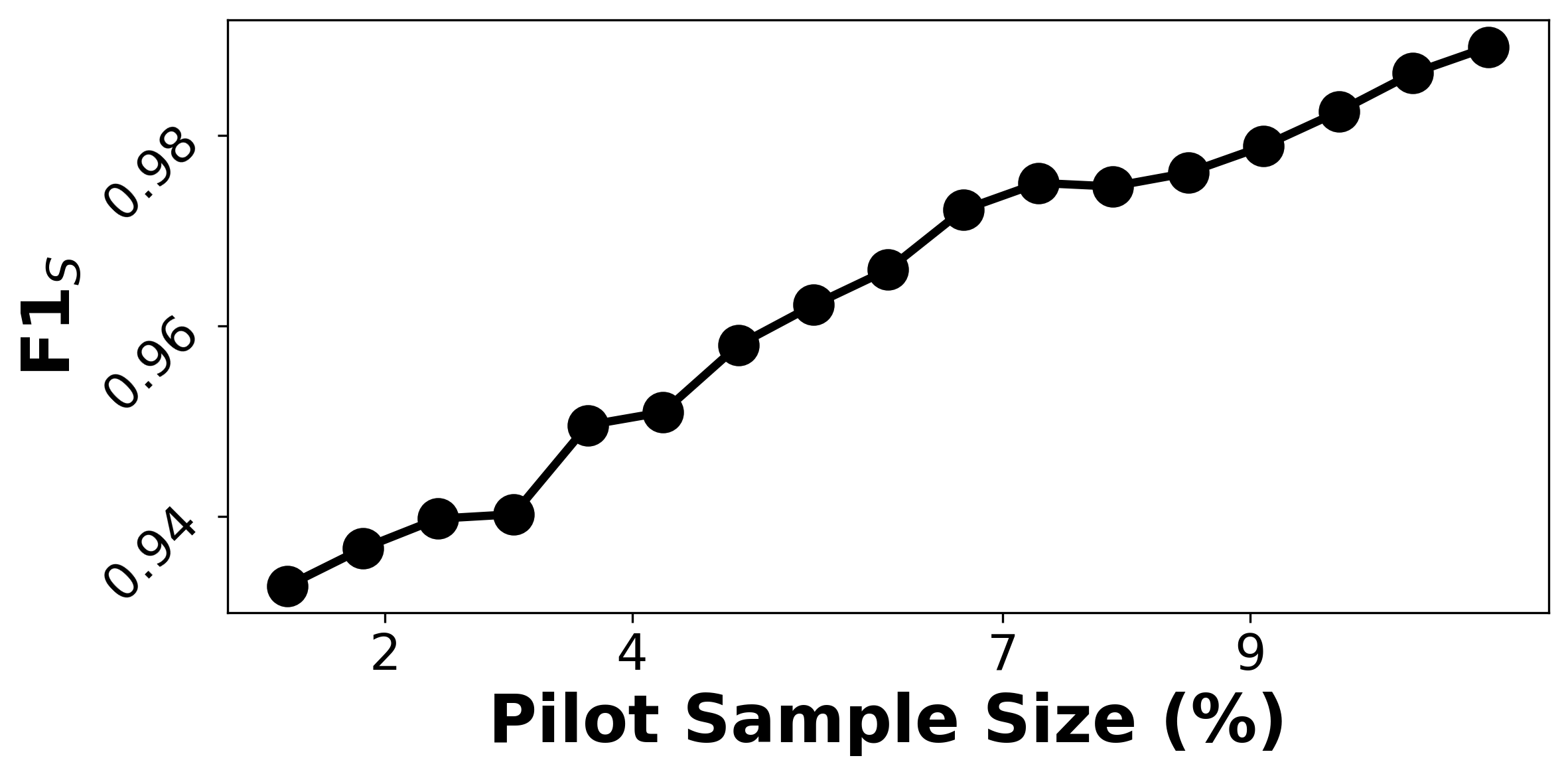}
        \caption{\sprintv}
        \label{fig:avg_s_p_f1}
    \end{subfigure}
    \hfill
    \begin{subfigure}{0.48\linewidth}
        \centering
        \includegraphics[width=0.8\linewidth]{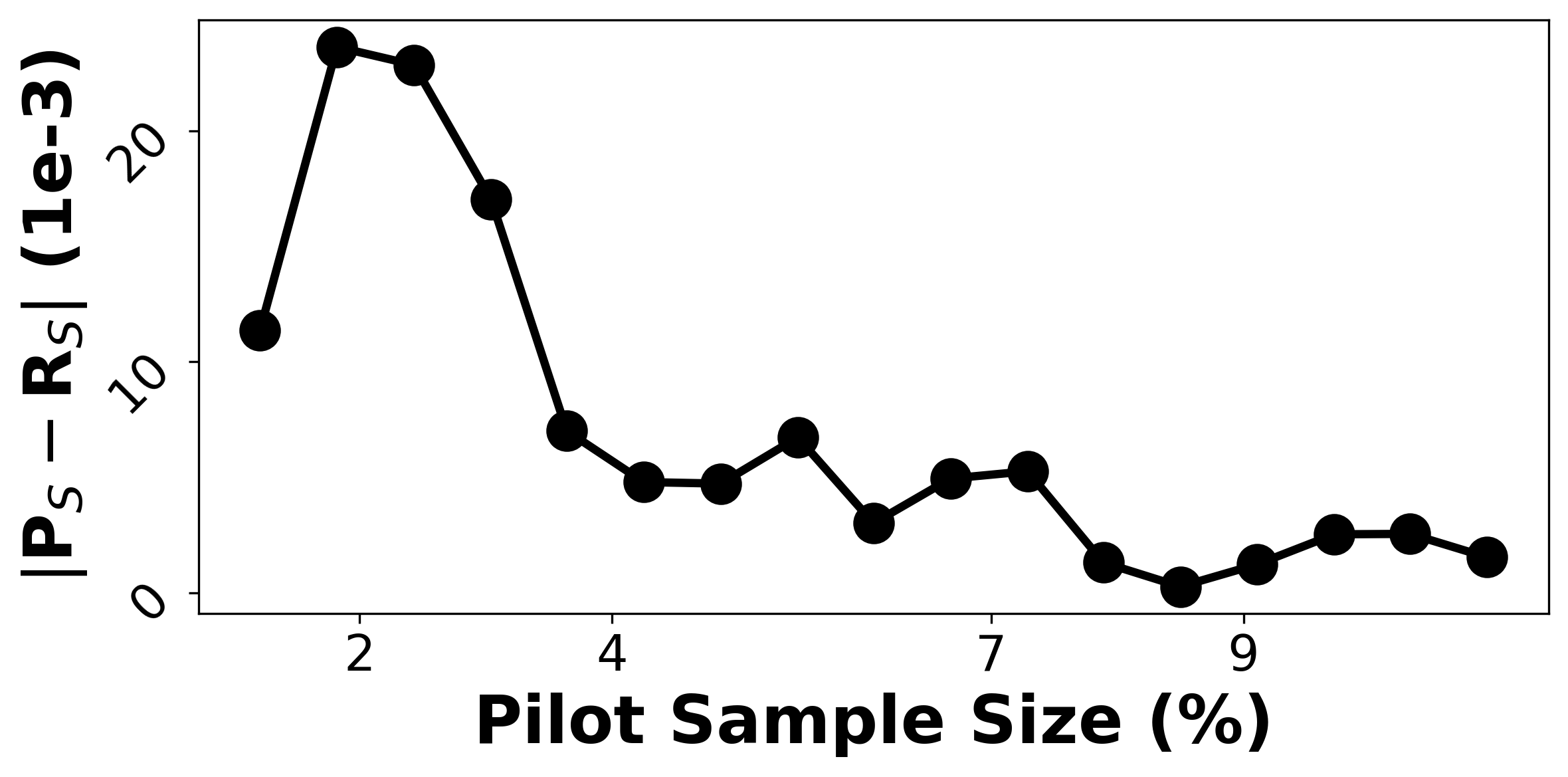}
        \caption{\sprintc}
        \label{fig:pct_s_p_pr}
    \end{subfigure}
    \centering
    \begin{subfigure}{0.48\linewidth}
        \centering
        \includegraphics[width=0.8\linewidth]{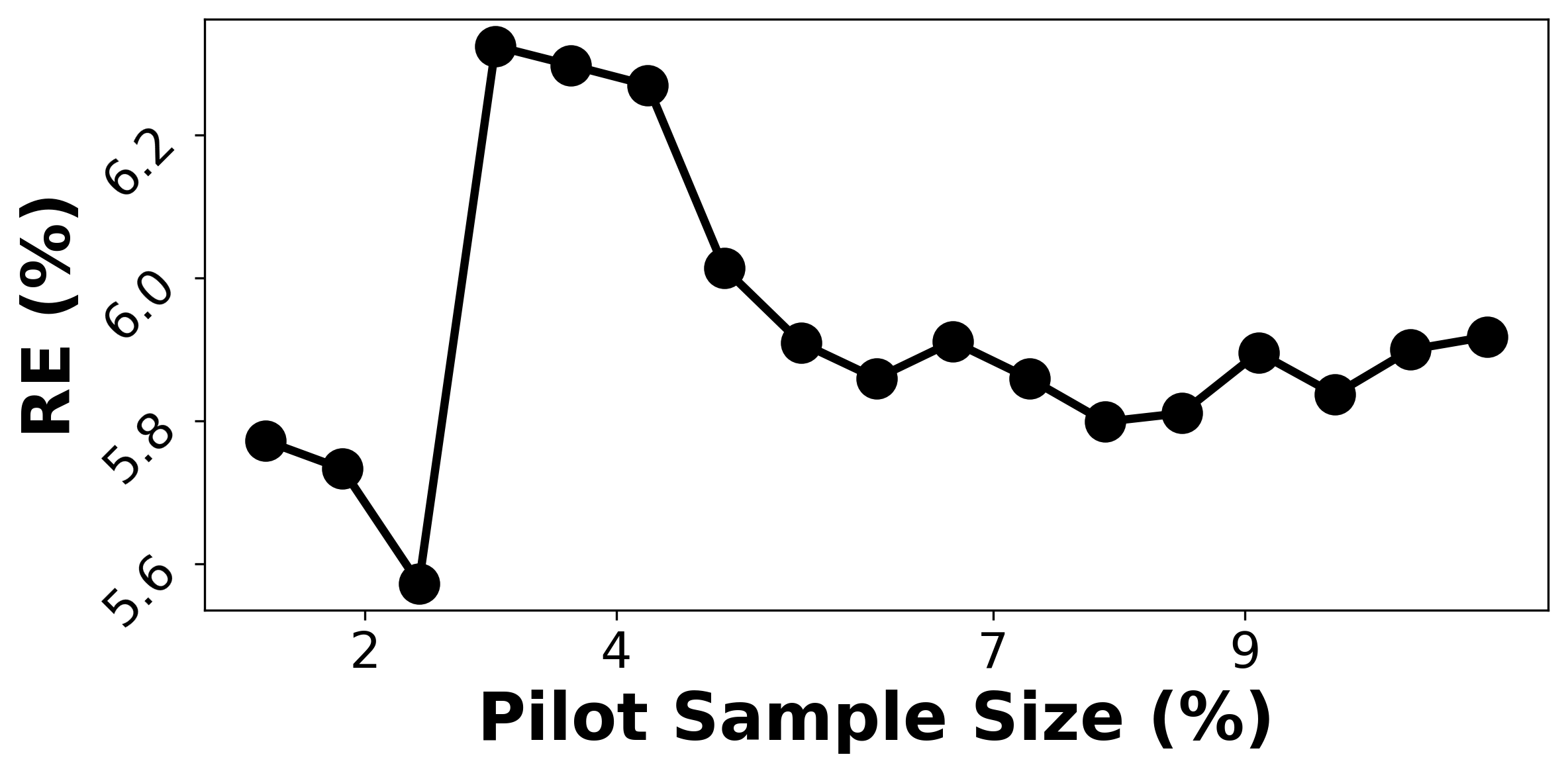}
        \caption{\sprintv}
        \label{fig:avg_s_p_re}
    \end{subfigure}
    \hfill
    \begin{subfigure}{0.48\linewidth}
        \centering
        \includegraphics[width=0.8\linewidth]{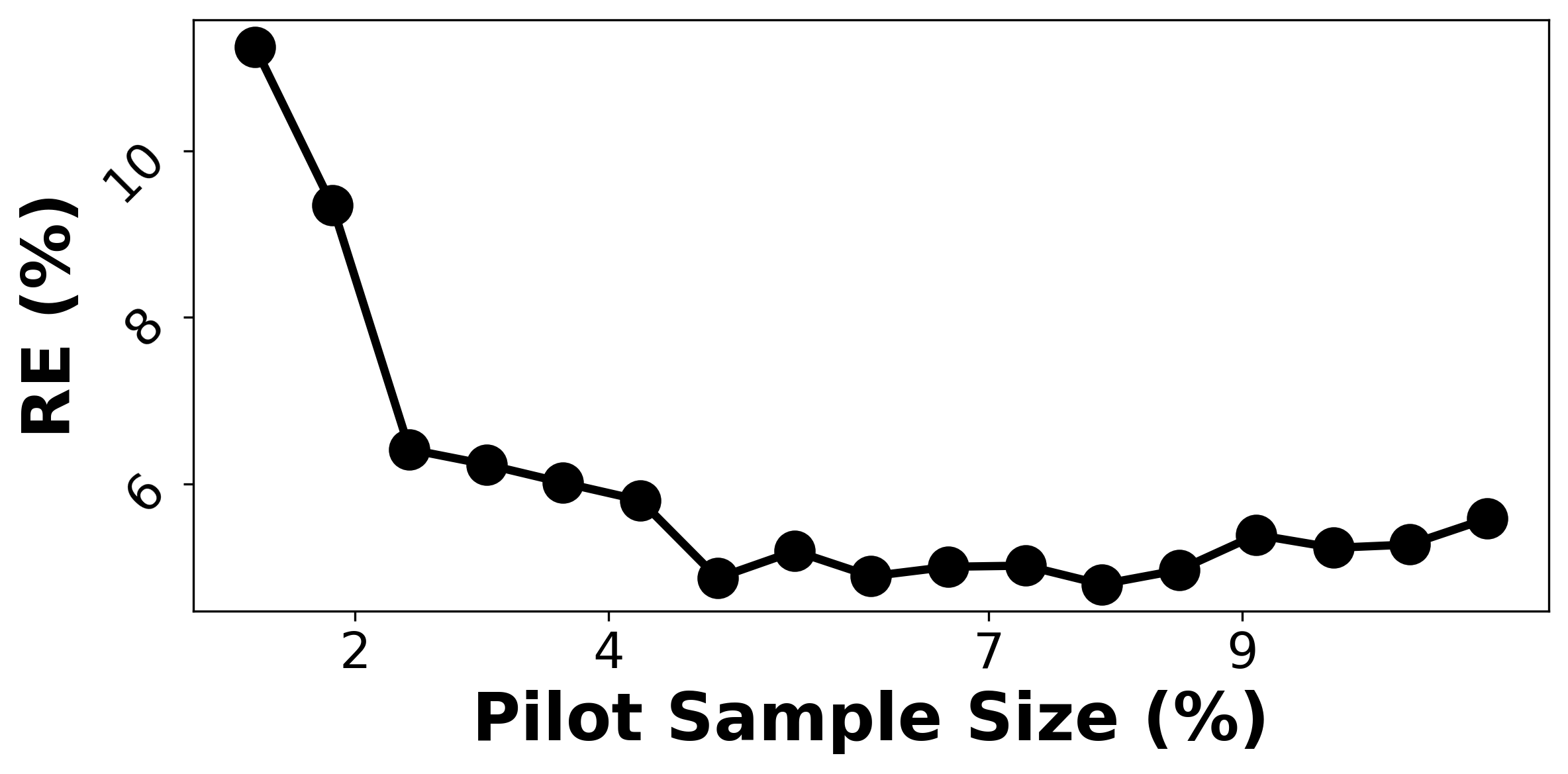}
        \caption{\sprintc}
        \label{fig:pct_s_p_re}
    \end{subfigure}
    \caption{Impact of \(s_p\) and \(s\) on RE using \sprintv for \(\mathtt{AVG}\) (a, c, e) and \sprintc for \(\mathtt{PCT}\) (b, d, f) on the eICU dataset.}
    \label{fig:sample_complexity}
\end{figure}

\subsubsection{Impact of Sample and Pilot Sample Sizes}\label{subsubsec:exp_s_s_p}

In \S~\ref{sec:algo}, we present the theoretical minimum \(s\) and \(s_p\) required to guarantee a certain aggregation error at \(1-\alpha\) confidence level. Here, we empirically validate how varying \(s\) and \(s_p\) affects RE using the eICU dataset for \(\mathtt{AVG}\) and \(\mathtt{PCT}\) in Fig.~\ref{fig:sample_complexity}. 

\stitle{A. Effect of Sample Size.}
Fig.~\ref{fig:avg_s} and~\ref{fig:pct_s} plot the absolute difference \(|\mathtt{agg}_S - \mathtt{agg}_D|\) against the sample size as a percentage of the dataset. As \(s\) increases, both the absolute difference and its variance decrease, as the aggregate over \(S\) better approximates \(D\). Notably, using around 25\% of the data already provides a close approximation to \(\mathtt{agg}_D\). Beyond a certain threshold (e.g., \(s > 25\%\)), the improvement plateaus, indicating diminishing returns. This suggests that low RE in AQNN evaluation can be achieved with a moderate sample size \(s\), which also means fewer proxy calls and less overhead.

\stitle{B. Effect of Pilot Sample Size.}
Fig.~\ref{fig:avg_s_p_f1} and~\ref{fig:avg_s_p_re} show that for \sprintv, increasing \(s_p\) improves \(\text{F1}_S\) and consequently lowers RE. For \sprintc, Fig.~\ref{fig:pct_s_p_pr} and~\ref{fig:pct_s_p_re} indicate that increasing \(s_p\) narrows the precision-recall gap in \(S\), which in turn reduces RE. Early instabilities are observed across all plots when \(s_p\) is very small, i.e., lacking true neighbors to reliably approximate aggregates. 

Overall, sufficiently large \(s\) and \(s_p\) ensure stable and low-error aggregation. But interestingly, using as little as 5\% of the data as the pilot sample already achieves stable RE in most cases. The key message is not simply that larger samples are better, but that even small samples, once above a certain threshold, suffice to yield accurate estimates. This phenomenon is not directly predicted by the theory and it reflects the efficiency of our framework.

\begin{figure}[!t]
    \centering
    \begin{subfigure}{0.48\linewidth}
        \centering
        \includegraphics[width=0.8\linewidth]{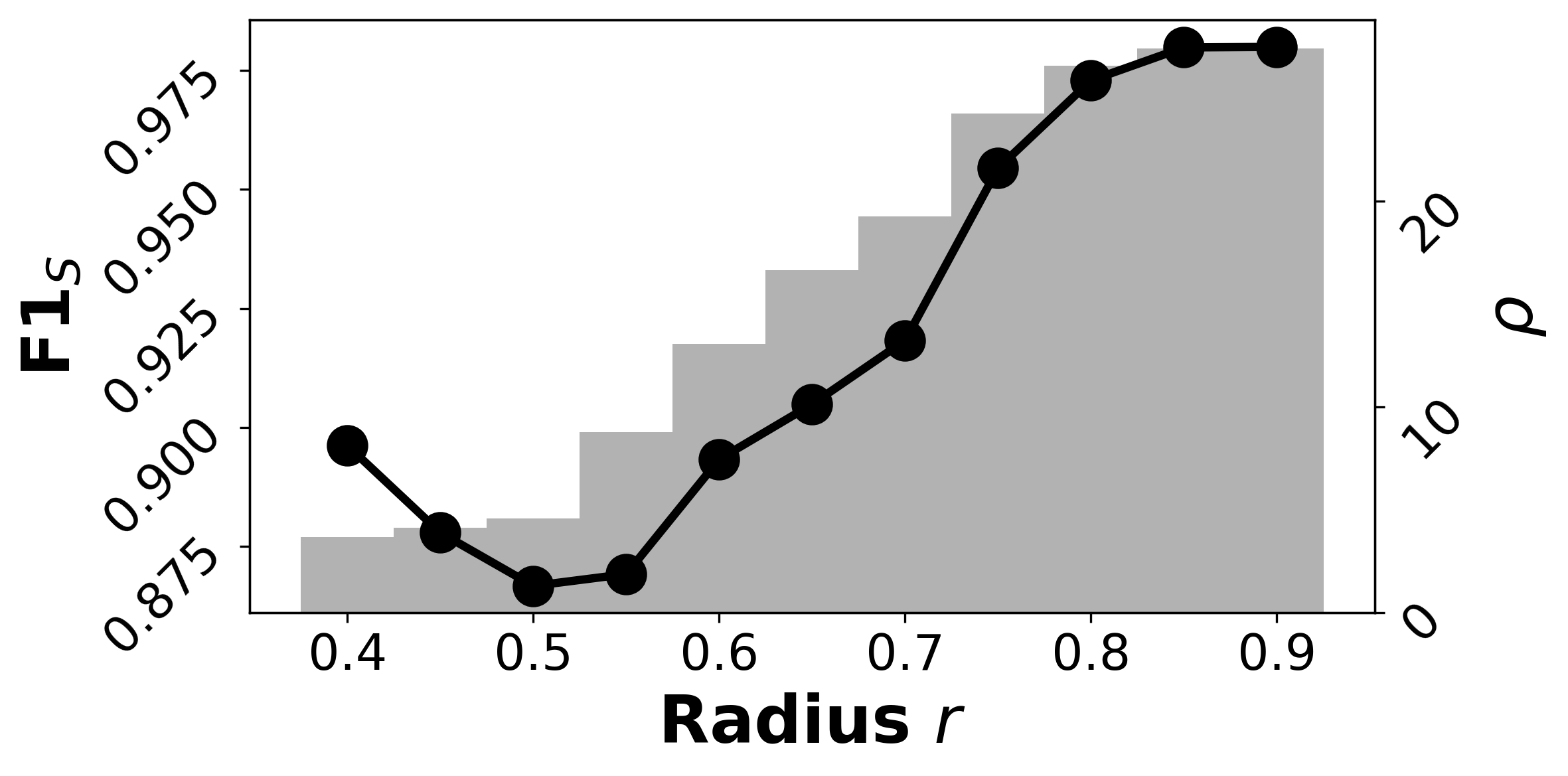}
        \caption{\sprintv: \(r\) vs \(\text{F1}_S\)}
        \label{fig:avg_rho_f1}
    \end{subfigure}
    \hfill
    \begin{subfigure}{0.48\linewidth}
        \centering
        \includegraphics[width=0.8\linewidth]{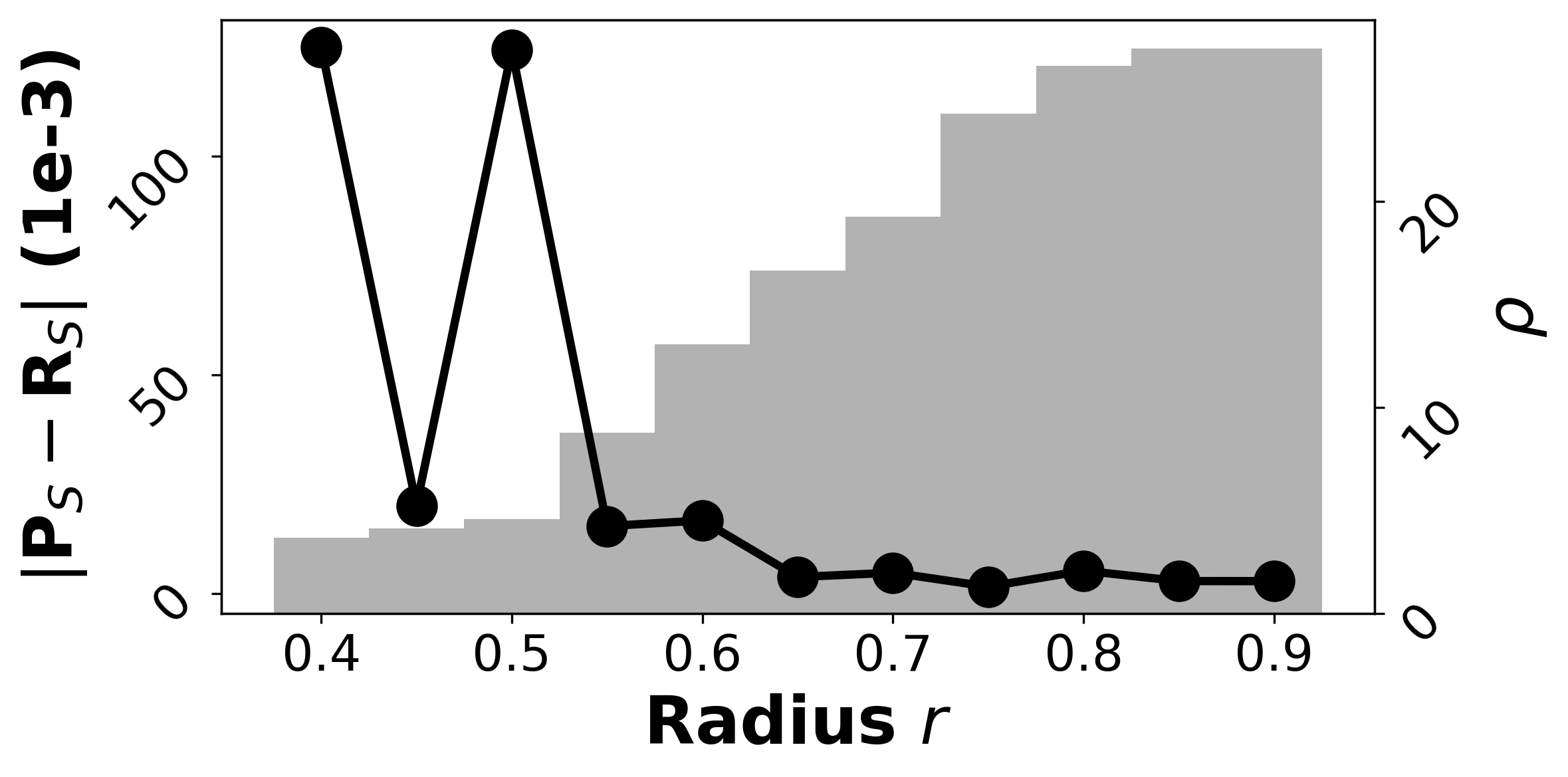}
        \caption{\sprintc: \(r\) vs \(|\text{P}_S-\text{R}_S|\)}
        \label{fig:pct_rho_pr}
    \end{subfigure}
    \centering
    \begin{subfigure}{0.48\linewidth}
        \centering
        \includegraphics[width=0.8\linewidth]{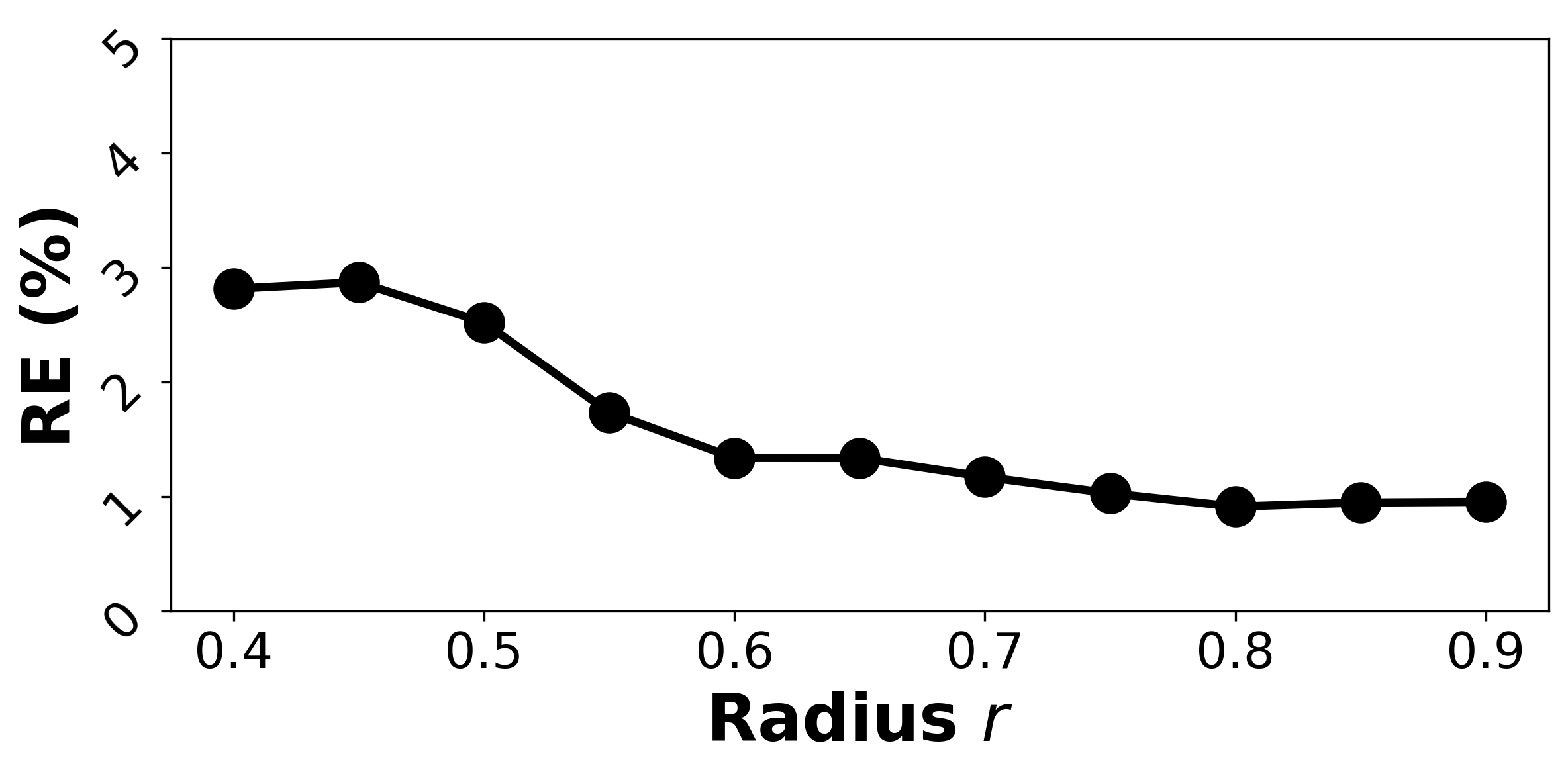}
        \caption{\sprintv: \(r\) vs RE}
        \label{fig:avg_rho_re}
    \end{subfigure}
    \hfill
    \begin{subfigure}{0.48\linewidth}
        \centering
        \includegraphics[width=0.8\linewidth]{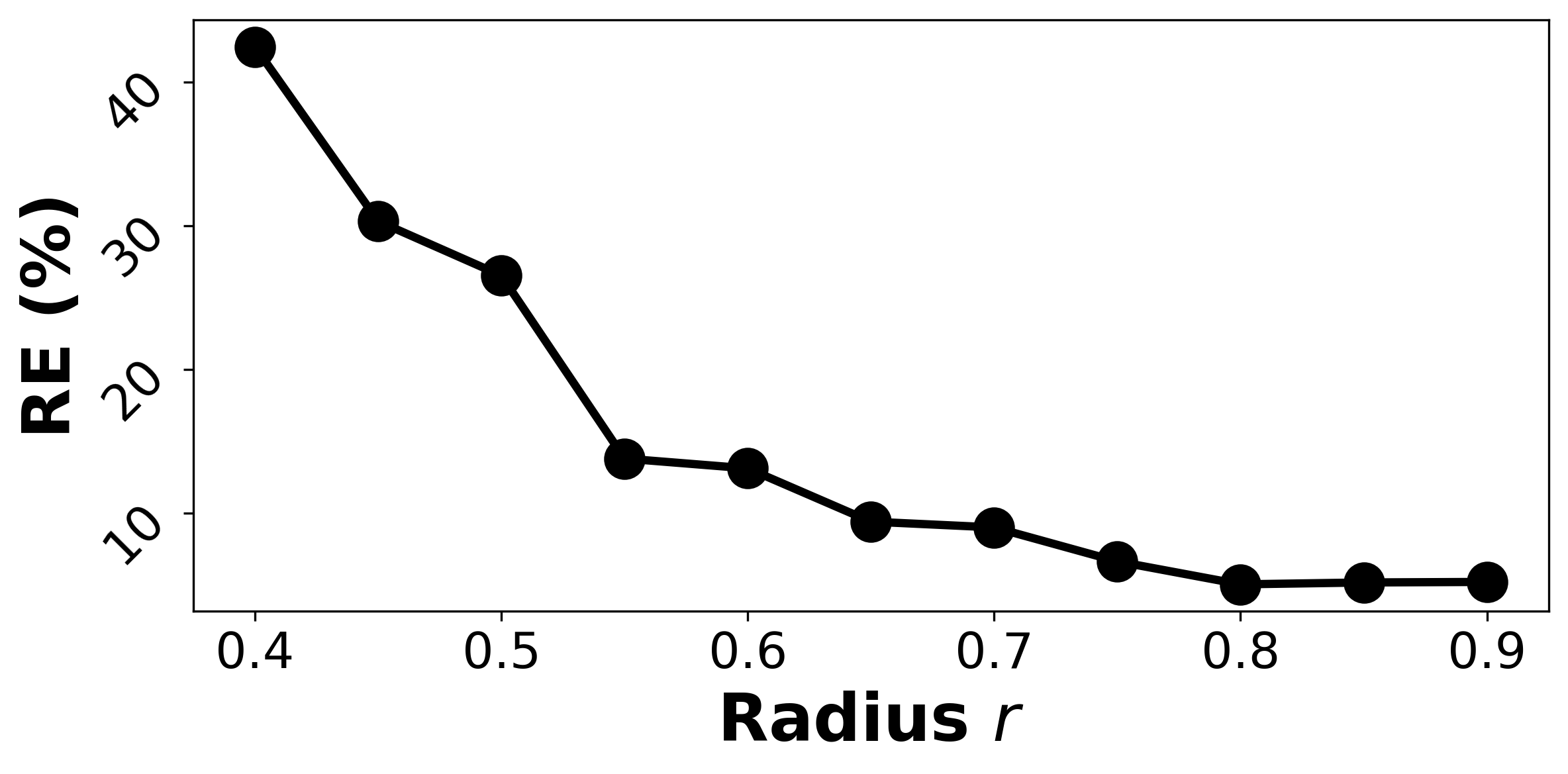}
        \caption{\sprintc: \(r\) vs RE}
        \label{fig:pct_rho_re}
    \end{subfigure}
    \caption{Impact of radius on RE using \sprintv for \(\mathtt{AVG}\) (a, c) and \sprintc for \(\mathtt{PCT}\) (b, d) on the eICU dataset.}
    \label{fig:level_of_difficulty}
\end{figure}

\subsubsection{Impact of Radius}\label{subsubsec:radius}
The radius controls the sparsity of the set of NNs for a given query target. A smaller radius makes it challenging to compute AQNN with a low RE. We conduct an experiment to assess this impact on \(\mathtt{AVG}\) and \(\mathtt{PCT}\). In Fig.~\ref{fig:level_of_difficulty}, we report RE, \(\text{F1}_S\), and precision-recall gap across different radii \(r \in [0.4, 0.9]\). We denote \(\rho\) as the effective neighborhood size ratio, i.e., the proportion of true neighbors in \(S\), which increases with the radius. It is visualized as bars in Fig.~\ref{fig:avg_rho_f1} and~\ref{fig:pct_rho_pr}.

For both \(\mathtt{AVG}\) and \(\mathtt{PCT}\), RE is high at small radii. This can be explained by the fluctuating \(\text{F1}_S\) and precision-recall gap. As \(r\) increases, the neighborhood size grows and these metrics stabilize, i.e., lower RE, higher \(\text{F1}_S\), and smaller precision-recall gaps. These observations demonstrate that aggregation error is sensitive to the effective neighborhood size. Too few neighbors lead to unreliable estimates, whereas larger neighborhoods yield more stable and accurate aggregates.

\subsection{Application: Hypothesis Testing}\label{exp:HT}
AQNNs have a natural downstream application in hypothesis testing because they provide aggregate statistics over the local neighborhood of a query target, which are key inputs for statistical inference. Hypothesis testing is a statistical method used to evaluate whether there is sufficient evidence in a sample to support or reject a specific claim about a population. When the population of interest is the neighborhood of a query target, this nearest neighbor hypothesis (NNH) is formulated directly from an AQNN. Hence, accurate AQNN results are essential for accurate hypothesis testing. Testing NNHs enables practitioners to uncover meaningful patterns, validate assumptions, and make reliable and trustworthy decisions. For example, it allows clinicians to assess whether the average heart rate of patients similar to a given individual significantly deviates from normal, thereby guiding personalized treatment. Similarly, it enables instructors to evaluate whether the average grade of students with similar learning trajectories is significantly lower than the class average, prompting timely interventions.

To evaluate the effectiveness of \sprintv and \sprintc in this application, we perform one-sample t-tests and proportion z-tests, respectively, with hypotheses of the form:
\[\text{agg}(\text{NN}_D(q, r)[\texttt{attr}]) \ \text{op} \ c\] 
where agg is \(\mathtt{AVG}\) or \(\mathtt{PCT}\), \(\text{op} \in \{\geq, \leq, \neq\}\), and \(c \in \mathbb{R}\) is a constant. Each hypothesis comes as a pair of a null hypothesis (\(H_0\)) and an alternative one (\(H_a\)). For instance, \(H_0\) might state that ``\textit{the average heart rate of patients similar to \(q\) is 100 bpm}'', while \(H_a\) posits that ``\textit{it is strictly less than 100 bpm}''. We measure performance using \textbf{accuracy}, defined as the proportion of test decisions (accept/reject) based on approximate aggregates that match those based on true aggregates:
\[
\text{Accuracy} = \frac{1}{k}\sum_{i=1}^k \mathbbm{1}_{H(\text{agg}_D) == H(\widetilde{\text{agg}_S})}
\]
where \(k = 30\) is the number of samples and \(H(\cdot)\) is the accept/reject test decision. The constant \(c\) is set as a multiple of the ground-truth aggregate, i.e., \(c = \textit{factor} \times \text{ground truth}\), with the \textit{factor} ranging from 0.5 to 1.5 in steps of 0.05. Accuracy is averaged over 10 random queries under two operators (\(\geq\) and \(\leq\)).

\begin{figure}[!t]
    \centering
    \begin{subfigure}{0.48\linewidth}
        \centering
        \includegraphics[width=0.8\linewidth]{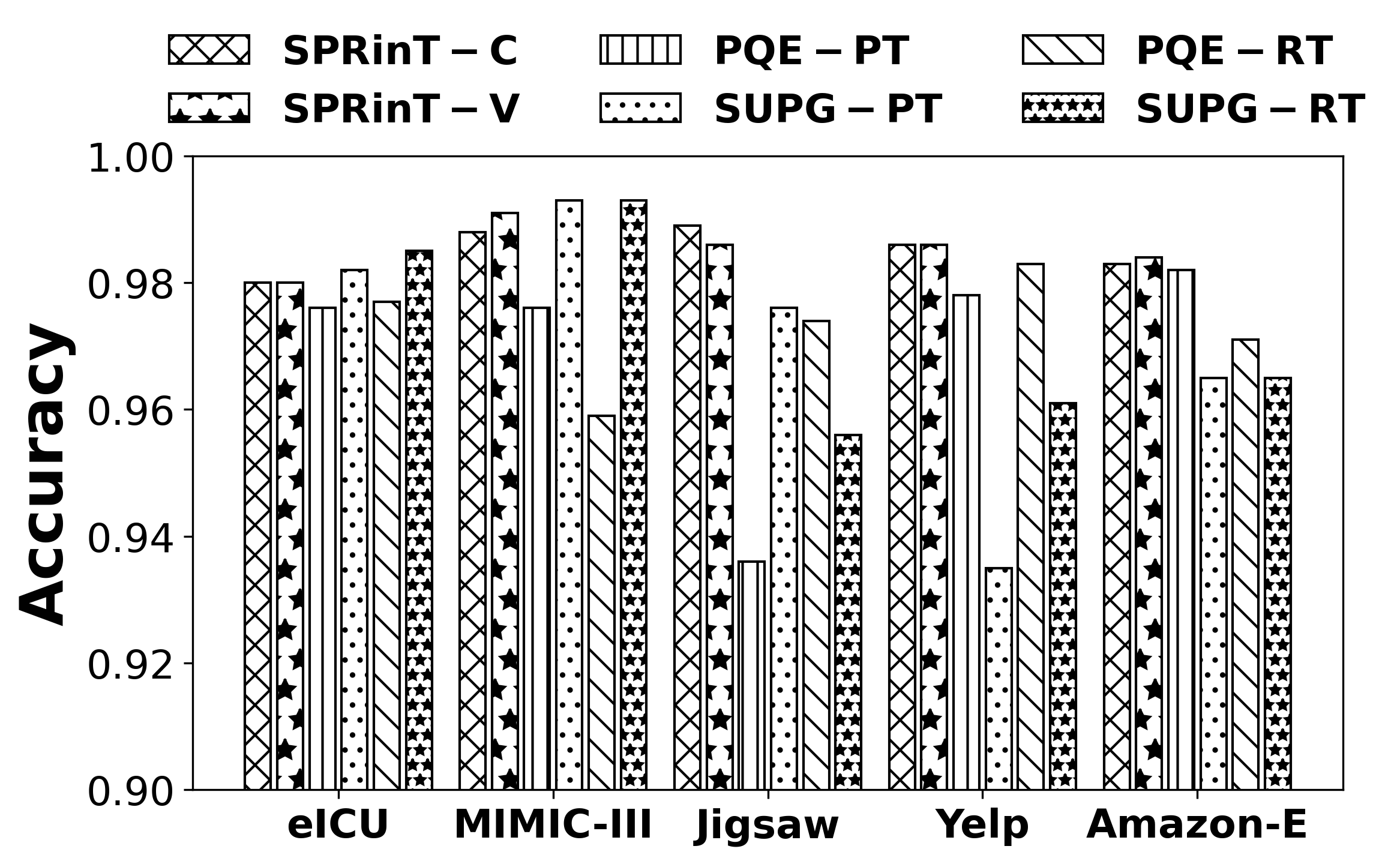}
        \caption{t-test (\(\mathtt{AVG}\))}
        \label{fig:HT_PQE_AVG}
    \end{subfigure}
    \hfill
    \begin{subfigure}{0.48\linewidth}
        \centering
        \includegraphics[width=0.8\linewidth]{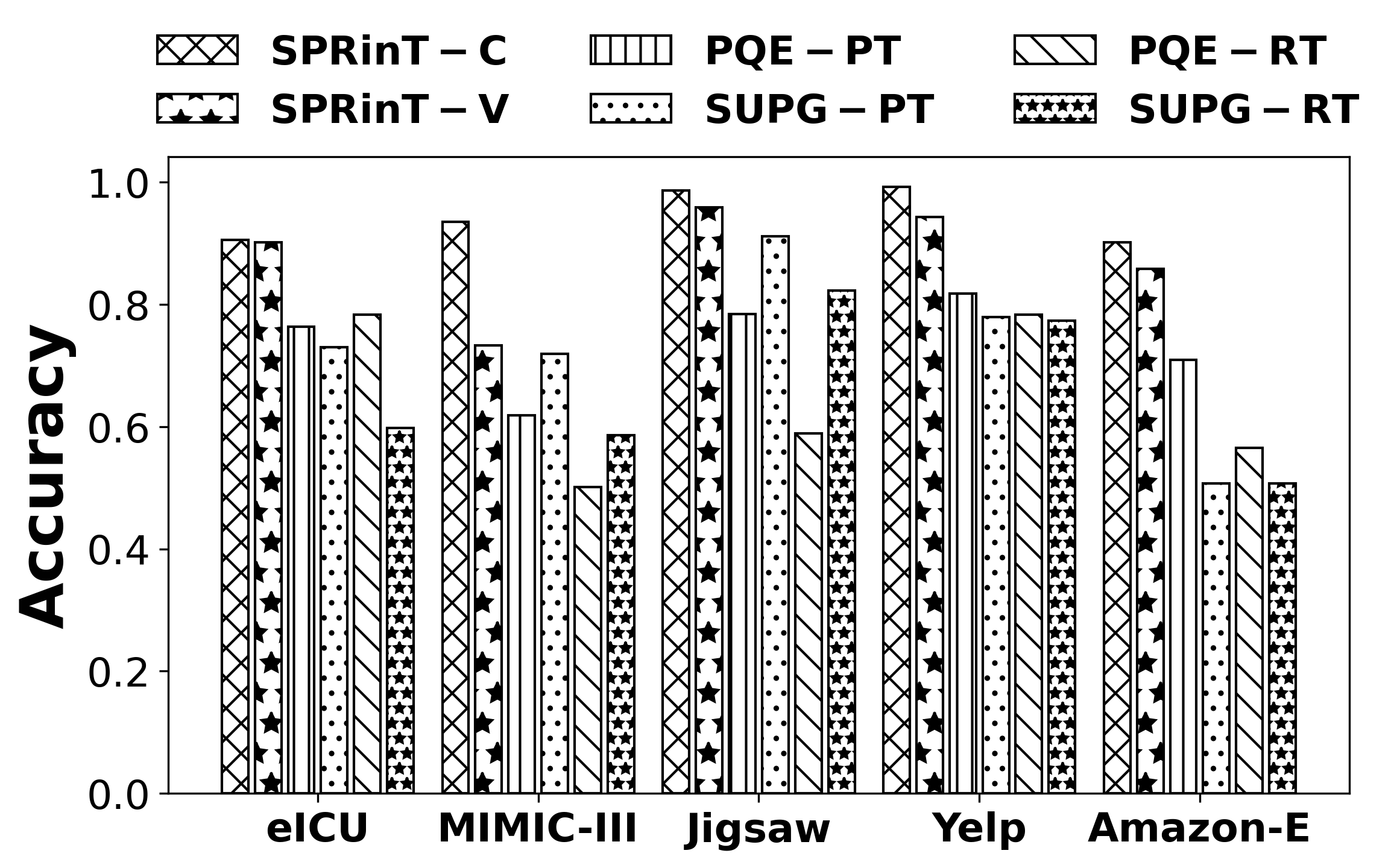}
        \caption{Proportion z-test (\(\mathtt{PCT}\))}
        \label{fig:HT_PQE_PCT}
    \end{subfigure}
    \caption{Hypothesis testing performance.}
    \label{fig:HT}
\end{figure}

Fig.~\ref{fig:HT} presents hypothesis testing accuracy across datasets and algorithms. As shown in Fig.~\ref{fig:HT_PQE_AVG} and~\ref{fig:HT_PQE_PCT}, testing accuracy closely tracks RE performance in \S~\ref{exp:error}. For \(\mathtt{PCT}\), \sprintc achieves the lowest RE and highest testing accuracy across all datasets. Similarly, for \(\mathtt{AVG}\), \sprintv outperforms baselines on Amazon-E in both RE and testing accuracy. Methods with low RE (e.g., SUPG-RT, PQE-RT) tend to yield higher testing accuracy. These results validate that low-error AQNN estimation directly enables accurate downstream hypothesis testing. Combined with \sprint's low computational cost (\S~\ref{exp:framework:embedding_generation_cost} and \S~\ref{exp:framework:end_to_end_cost}), these support our framework's practicality and scalability for real-world NNH applications.

\subsection{Deployment Considerations}
Selecting appropriate values for the sample size \(s\) and pilot sample size \(s_p\) involves balancing cost and error. The theoretical bounds in \S~\ref{sec:theory} and \S~\ref{sec:SUM_algo} provide guidelines for minimal values of \(s\) and \(s_p\) to meet target error guarantees in practice.

For deployment, \sprint can be integrated into a query engine where proxy and oracle embeddings need to be recomputed or refreshed at query time. To further optimize latency in time-sensitive applications, the search tolerance parameters (\(\omega_V\) and \(\omega_C\)) can be tuned to accelerate convergence while maintaining acceptable approximation quality.

Finally, although an AQNN computes aggregates with respect to a single designated query target, it can be extended to support SQL-style group-by semantics. Here, each group key in a group-by query may be treated as a query target, which enables an AQNN to compute aggregates over the neighborhood of each target in the embedding space. This extension is particularly useful for computing AQNNs over multiple query targets in batch.

\section{Related Work}\label{sec:relwork}


{\bf Approximate Query Processing (AQP)}. There is a rich body of prior work in the DB community on AQP methods that aim to compute approximate answers to data intensive Online Analytical Processing (OLAP) queries with high accuracy and low computational overhead. We note that AQNNs differ from OLAP queries fundamentally in both scope and operation. 

AQP techniques can generally be categorized into two types: online and offline. Online methods dynamically sample data at query time to compute approximate answers \cite{DBLP:conf/eurosys/AgarwalMPMMS13}. Offline methods, on the other hand, rely on precomputed synopses such as wavelets to answer queries efficiently \cite{DBLP:conf/vldb/ChakrabartiGRS00}. In contrast, AQNNs operate in embedding spaces where aggregation accuracy depends on both sampling and neighbor selection quality. Existing AQP methods do not address these challenges.

{\bf Nearest Neighbor Search} methods can be categorized into exact and approximate approaches. Common methods, such as brute force, k-d trees, and cell techniques, are conceptually straightforward \cite{DBLP:journals/toms/FriedmanBF77, FRNNSurvey} but impractical due to their high computational cost. Exact NN search typically relies on accurate embeddings generated by expensive oracle models and builds indexes over the entire database, leading to significant computational overhead \cite{DBLP:conf/sigmod/LiZAH20}. 

As databases and data dimensionality grow, approximation algorithms have been used to balance accuracy and efficiency. These approaches include hashing-based \cite{DBLP:journals/pvldb/HuangFZFN15}, tree-based \cite{DBLP:conf/cvpr/Silpa-AnanH08}, and index-based methods \cite{arora2018hd}. Additionally, \cite{indyk1998approximate} proposes a locality-sensitive hashing based approach and delivers sublinear query times and polynomial time processing costs, while \cite{indyk2018approximate} proposes space-efficient data structures guaranteeing compactness for Euclidean distances under specific accuracy constraints. FLANN \cite{muja2014scalable} is a library leveraging randomized kd-trees and priority search k-means trees for scalable approximate NN finding over large datasets. 

Unlike our work, these approaches assume that the ground-truth representations (aka oracle embeddings) of data objects are precomputed and can be efficiently accessed. In many real-world scenarios, however, embeddings need to be recomputed at query time because data evolves rapidly, which poses a unique challenge. To reduce the embedding generation cost, some recent works leverage cheap proxy models to approximate ground-truth oracle labels and find accurate approximations of NNs \cite{DBLP:conf/sigmod/LaiHLZ0K21, DBLP:conf/sigmod/LuCKC18, DBLP:journals/pvldb/KangGBHZ20, DBLP:journals/pvldb/KangEABZ17}. For instance, Lai et al. propose probabilistic Top-K to train proxy models to generate oracle label distribution and output approximate Top-K solutions for video analytics \cite{DBLP:conf/sigmod/LaiHLZ0K21}. In \cite{DBLP:journals/pvldb/KangGBHZ20}, the authors introduce statistical accuracy guarantees, such as meeting a minimum precision or recall target with high probability, for approximate selection queries. Recently, Ding et al. introduce a framework for approximate FRNN queries that leverages proxy models to minimize the reliance on expensive oracle models \cite{DujianPQA}. They propose four algorithms to find high-quality NNs for precision-target and recall-target queries. 

Notice that while \sprintv and \sprintc build on PQE-PT~\cite{DujianPQA} for NN selection, our goal is fundamentally different: PQE-PT focuses on retrieving a high-quality neighbor set, whereas we aim to minimize the aggregation error of the \textit{aggregate value} over neighbors. Building on this, our methods introduce two key innovations. First, unlike PQE-PT, which assumes a \textit{given} precision target and identifies neighbors that satisfy it, our algorithms \textit{dynamically optimize both precision and recall targets}. Second, whereas PQE-PT focuses on achieving a given precision target while maximizing recall, our methods explore the precision-recall space to directly optimize higher-level objectives for accurate approximation of aggregate results: \sprintv maximizes the F1 score, and \sprintc equalizes precision and recall. 
None of the prior algorithms, including PQE-PT, can support these objectives.

\section{Conclusion}\label{sec:conclusion}
We presented \sprint, a scalable framework for answering AQNNs, a new class of queries that compute aggregates over neighborhoods defined by learned representations. Exact evaluation of AQNNs using high-quality oracle embeddings is often costly, especially when data evolves rapidly. Hence, \sprint combines sampling with a judicious mix of high-quality oracle and efficient proxy embeddings to approximate aggregates with low error and computational cost. We categorized AQNNs into value- and count-sensitive queries and designed tailored strategies, \sprintv and \sprintc, to optimize F1 score or precision-recall balance. Extensive experiments as well as our theoretical analysis demonstrate that \sprint enables accurate and efficient aggregation, which also translates to high accuracy in downstream applications such as hypothesis testing.

\sprint supports aggregation functions stable under sampling, functions like \(\mathtt{MIN}\), \(\mathtt{MAX}\) and \(\mathtt{MEDIAN}\) pose unique challenges due to outlier sensitivity and rank instability. Extending \sprint to handle such queries with theoretical guarantees is an important future direction. We also plan to explore sparsity-aware sampling for diverse neighborhood structures and attribute-value-aware heuristics that directly minimize aggregation error.

\begin{acks}
This work was partially supported by DataGEMS, funded by the European Union's Horizon Europe Research and Innovation programme, under grant agreement No 101188416. Carrie Wang and Reynold Cheng were supported by the Research Grant Council of Hong Kong (RGC Project HKU 17202325), the University of Hong Kong (Project 2409100399), and the HKU Faculty Exchange Award 2024 (Faculty of Engineering). Lakshmanan's research was supported in part by a grant from the Natural Sciences and Engineering Research Council of Canada (Grant Number RGPIN-2020-05408).
\end{acks}

\bibliographystyle{ACM-Reference-Format}
\bibliography{main}

\end{document}